\documentclass{aa}  
\usepackage{natbib}
\usepackage{float}
\usepackage{graphicx}
\usepackage{txfonts}
\usepackage{color}

\begin{document}

   \title{SMC west halo: a slice of the galaxy that is being tidally
     stripped?}

   \subtitle{Star clusters trace age and metallicity gradients\thanks{Based on observations obtained at  
 the Southern Astrophysical Research (SOAR) telescope, which is a joint project of
 the Minist\'{e}rio da Ci\^{e}ncia, Tecnologia, e Inova\c{c}\~{a}o (MCTI) da Rep\'{u}blica
 Federativa do Brasil, the U.S. National Optical Astronomy Observatory
 (NOAO), the University of North Carolina at Chapel Hill (UNC), and
 Michigan State University (MSU).}\fnmsep\thanks{Tables of photometry are only available in electronic form
at the CDS via anonymous ftp to cdsarc.u-strasbg.fr (130.79.128.5)
or via http://cdsweb.u-strasbg.fr/cgi-bin/qcat?J/A+A/}}

   \author{B. Dias\inst{1,2} 
        \and L. Kerber\inst{3} 
        \and B. Barbuy\inst{2} 
        \and E. Bica\inst{4}
        \and S. Ortolani\inst{5,6} 
}

   \institute{European Southern Observatory, Alonso de C\'ordova 3107,
        Santiago, Chile\\
        \email{bdias@eso.org}
        \and 
         Instituto de Astronomia, Geof\'\i sica e Ci\^encias Atmosf\'ericas,
        Universidade de S\~ao Paulo, Rua do Mat\~ao 1226, 
        Cidade Universit\'aria, S\~ao Paulo 05508-900, SP, Brazil
        \and
        Laborat\'orio de Astrof\'{\i}sica Te\'orica e Observacional, 
        Departamento de Ci\^encias Exatas e Tecnol\'ogicas, Universidade
        Estadual de Santa Cruz,
         Rodovia Jorge Amado km 16, Ilh\'eus 45662-000, Bahia, Brazil
        \and 
        Universidade Federal do Rio Grande do Sul, IF, CP 15051, Porto Alegre,
        91501-970, RS, Brazil
        \and
        Dipartimento di Fisica e Astronomia Galileo Galilei,
        University of Padova, vicolo dell'Osservatorio 3, 35122,
        Padova, Italy.
        \and
        INAF-Osservatorio Astronomico di Padova, Vicolo dell'Osservatorio 5,
I-35122, Padova, Italy}
   \date{Received; accepted }

 
%

  \abstract
   {The evolution and structure of the Magellanic Clouds is currently under debate.
     The classical scenario in which both the Large and \object{Small
     Magellanic Clouds} (\object{LMC}, \object{SMC}) are orbiting the Milky Way has been
     challenged by an alternative in which the LMC and SMC are in their first
     close passage to our Galaxy. The clouds are close enough to us to
     allow spatially resolved observation of their stars, and detailed
     studies of stellar populations in the galaxies are expected
to be able to constrain
     the proposed scenarios. In particular, the west halo (WH) of the SMC
     was recently characterized with radial trends in age and
     metallicity that indicate tidal disruption.}
    {We intend to increase the sample of star clusters in the west
      halo of the SMC with homogeneous age, metallicity, and distance derivations
      to allow a better determination of age and metallicity gradients
      in this region. Distances and positions are compared with the
      orbital plane of the SMC depending on the scenario adopted.}
   {Comparisons of observed and synthetic V(B-V) colour-magnitude diagrams were used
     to derive age, metallicity, distance, and reddening for
     star clusters in the SMC west halo. Observations were carried out using the 4.1m SOAR
     telescope. Photometric completeness was determined through artificial
     star tests, and the members were selected   by statistical
     comparison with a control field.} 
   {We derived an age of 1.23$\pm$0.07~Gyr and [Fe/H] = -0.87$\pm$0.07 for the reference
 cluster NGC~152, compatible with literature parameters.
 Age and metallicity gradients are confirmed in the WH: 2.6$\pm$0.6~Gyr/$^{\circ}$ and -0.19$\pm$0.09~dex/$^{\circ}$, respectively.
     The age-metallicity relation for the WH has a low dispersion in metallicity and is compatible
     with a burst model of chemical enrichment. All WH clusters seem to follow the same stellar
     distribution predicted by dynamical models, with the exception of AM-3, which should belong to the
     counter-bridge. Br\"uck~6 is the youngest cluster in our sample. It is only 130$\pm$40~Myr old
     and may have been formed during the tidal interaction of SMC-LMC that  created the
     WH and the Magellanic bridge.}
   {We suggest that it is crucial to split the SMC cluster population into groups: main body,
   wing and bridge, counter-bridge, and WH. This is the way to analyse the complex star
   formation and dynamical history of our neighbour. In particular, we show that the WH has
   clear age and metallicity gradients and an age-metallicity relation that is also compatible with the dynamical
   model that claims a tidal influence of the LMC on the SMC.}

   \maketitle

%

\section{Introduction}
\label{Intro}

Hierarchical accretion of dwarf galaxies is the most likely
origin of stellar halos in large spiral galaxies such as the Milky Way
(MW), as predicted by $\Lambda$-cold dark matter ($\Lambda$CDM) models,
and early work by \cite{searle+78}. Local Group galaxies are a suitable
laboratory in which to study cosmology in very much detail by
examining the
interactions between the Milky Way and its satellites, in particular the
 Magellanic Clouds. The Large and Small Magellanic Clouds (LMC
and SMC) are the core of a complex system of gas streams with stellar
counterparts in some cases. They have an irregular shape, and complex
star formation and chemical enrichment histories. They are currently
on a close passage to the Milky Way, but there is no agreement in the
literature whether they are orbiting our Galaxy periodically (e.g. \citealp{diaz+12}) or if this
is their first close encounter (e.g. \citealp{Besla_etal07}).
It is likewise debated whether the streams were caused by
the Milky Way influence or only by the two Magellanic Clouds
themselves (e.g. \citealp{NMB08}).

Since \cite{mathewson+74} detected the trailing
gas structure called Magellanic Stream, which indicates that
both the Large and Small Magellanic Clouds are on a polar orbit around the Milky Way, many
different approaches have been implemented to understand their
history. The two most reasonable explanations would be ram pressure
stripping by an ionized gas in the Milky Way halo
(e.g. \citealp{moore+94}) or tidal stripping caused by a close
encounter of the clouds and the Milky Way about 1.5 Gyr ago
(e.g. \citealp{gardiner+96}). \cite{putman+98} ruled out the MW ram
pressure model by finding a leading arm of gas that is evidence in favour of MW tidal
stripping.

The new open question that is under debate in the
literature is whether the current passage of the SMC and LMC very close to
the Milky Way is the first (e.g. \citealp{Besla_etal07,besla+10}) or if they have
a quasi-periodic orbit around our Galaxy
(e.g. \citealp{gardiner+96}). \cite{Besla_etal07} supported a first encounter 
of the LMC and SMC with the 
Milky Way now, arguing that the SMC and LMC have entered the MW dark matter
halo about 3 Gyr ago and that the shock of encountering the MW halo gas has triggered
the recent star formation. The authors explained the formation of the Magellanic Stream
4 Gyr ago by the interaction of the SMC and LMC with each other
and without any MW influence or a stellar counterpart. \cite{diaz+12} instead suggested 
that the SMC and LMC are orbiting
the Milky Way with a period of about 2 Gyr,
arguing that this better reproduces the
observed positions and velocities for the SMC. In this case, star
formation would be triggered by the tidal forces of the MW, and there would be
stellar counterparts of the gas structures. Both scenarios are
limited to  morphology and kinematics and have their advantages
and drawbacks, but the deadlock of the
discussion might be solved with proper motions of the clouds
(e.g. \citealp{kallivayalil+13} using a space-based telescope, \citealp{vieira+10} using 
a ground-based telescope).

Close encounters of SMC and LMC, or between SMC and LMC and the Milky Way,
would trigger star formation. This means that stellar populations are very
important to understand the interaction timescales of the
galaxies.
To separate this intricate scenario for the Magellanic Cloud
evolution during the past decade, significant efforts have been made in terms
of deriving the star formation rate SFR(t), age-metallicity relation (AMR), and age and
metallicity gradients using field stars and stellar clusters.
For the SMC, several photometric works have recovered the SFR(t) and/or AMR by
means of colour-magnitude diagram (CMD) analyses. We recall the works that
were based on
very deep and small HST fields (e.g. \citealp{chiosi+07}; \citealp{cignoni+13}; \citealp{weisz+13}),
and on shallower large surveys in the visible (e.g. \citealp{harris+04}; \citealp{noel+09}; \citealp{piatti12a}; \citealp{piatti+15}) and in the near-infrared, such as the VISTA Survey for the
Magellanic Clouds (VMC; \citealp{cioni+11,rubele+15}). 
This last survey was specifically designed
to reach faint main-sequence turnoffs of the oldest clusters for 170 deg$^2$ in 
the LMC, SMC, bridge, and stream, taking the advantage that light in the near-infrared region
is almost unaffected by dust.

Spectroscopical studies in the near-infrared
involving the CaII triplet of red giant stars (\citealp{Carrera_etal08}; \citealp{dobbie+14})
or high-resolution optical spectra \citep{mucciarelli+14} have yielded fundamental
results for AMR, the metallicity distribution, and metallicity gradient for the SMC.
These photometric and spectroscopic works have concluded the
following: The SMC had an initial period of low star formation rate for ages older
than 10 Gyr.
At least two periods of enhanced SFR followed at intermediate ages,
the first around 5-8 Gyr, the second about 1-3 Gyr. These might
have been associated with close encounters with the LMC or MW.
Several bursts of star
formation in the last 500 Myr form a complex spatial pattern.
The metallicities of field stars are systematically lower than those of their
LMC and MW counterparts. The AMR is space dependent and presents a significant spread in
metallicity for intermediate ages (1 - 8 Gyr). An age gradient
is formed by the young stars that are concentrated in
the SMC bar and the old stars that form a larger and smoother ellipsoidal
structure. Finally, the spectroscopic results indicate a metallicity
gradient, but this is not consensus in the photometric works.

Additionally, the main episodes of star formation have been confirmed
by the results for stellar clusters from the analysis
of CMDs using the HST
(e.g. \citealp{glatt+08a,glatt+08b}, \citealp{girardi+13}),
the VLT \citep{parisi+14}, and 4m class telescopes
(\citealp{piatti+05b, piatti+07c, piatti+11}; \citealp{piatti11a, piatti11b}; \citealp{dias+14}),
as well as  CaII triplet spectroscopy \citep{parisi+09,parisi+15}.
These analyses also showed how difficult it is to assign a unique AMR because the metallicity for
a given age is highly dispersed. On the other hand, these works revealed no strong 
evidence of a metallicity radial gradient.
Furthermore, the metallicity distribution presented by \cite{parisi+15}
is clearly bimodal, with peaks at [Fe/H] = -1.1 and -0.8 dex.
This result do not find a counterpart in the field stars,
as demonstrated by the analysis of $\sim 3000$ red giant stars performed
by \cite{dobbie+14}, who found a single peak at [Fe/H] $\sim$ -1.0.
Equally challenging, the global AMR for both field and cluster stars
doed not present a narrow distribution in favour of a single chemical
evolution model, which suggests multiple events
during the SFH of the SMC.

The SMC complexity is also imprinted on its three-dimensional structure.
The distribution of SMC red giant branch (RGB) stars does not present any sign of
rotation \citep{HZ06}, which is the case
for the HI component \citep{stanimirovic+04}. These results suggest
that the stellar component follows a spheroidal distribution, whereas the
gas is rotating in a disc-like structure.
Furthermore, the SMC presents 
a significant line-of-sight depth, ranging from $\sim$ 6-14 kpc in the inner parts
\citep{crowl+01,subramanian+12} to $\sim$ 23 kpc
in the eastern parts along the Magellanic bridge (Nidever et al. 2013).
It is possible to find stars
$\sim 12$ kpc closer than the bulk of SMC stars in the
direction of the bridge. Bica et al. (2015) derived even shorter distances
to stellar clusters in the Magellanic bridge, raising the possibility
that they  are part of a tidal dwarf galaxy in formation.

\begin{figure}[!htb]
   \centering
   \includegraphics[width=\columnwidth]{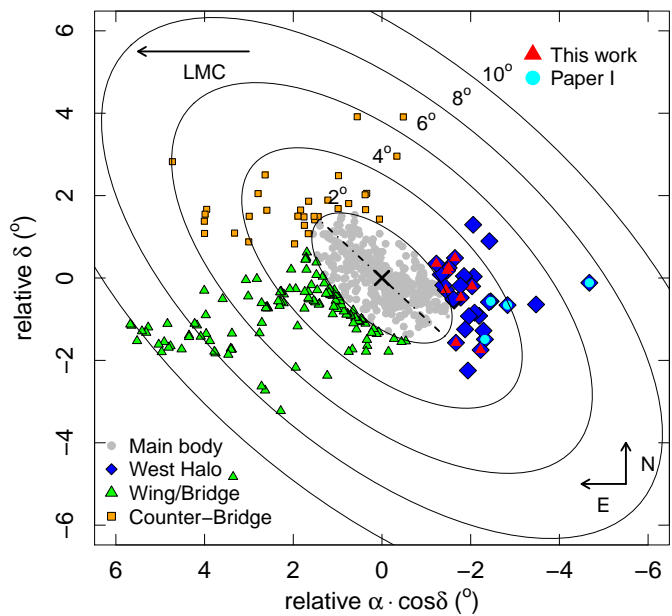}
   \caption{Distribution of SMC star clusters as catalogued by
     \cite{bica+08a}. Ellipses represent the distances from the SMC
     centre, as defined in Paper I. The different groups main body,
     wing and bridge, counter-bridge, and west halo are clearly
     identified as grey dots, green triangles, yellow squares, and
     blue diamonds. West halo clusters analysed in this work are
     highlighted with large red triangles, and those analysed in Paper I
     are displayed as large cyan circles.}
 \label{2Dpos}
   \end{figure}

The SMC has been shown to have complex structures and an involved history. We therefore 
study this galaxies in separate parts in this paper.
Considering ellipses instead of circumferences to draw cluster
distances to the SMC centre  is a better choice for this galaxy shape,
as proposed by \cite{piatti+05a}. This is shown in Fig. \ref{2Dpos}. This system
allows us to divide the clusters into an internal ($a$ < 2$^{\circ}$) and an external
($a$ > 2$^{\circ}$) group. The second group is more interesting because it
tells the tidal history of the SMC as recorded by star clusters. Finding age
and/or metallicity gradients in the external 
group is a good indicator of tidal structures, but all the clusters together show a high dispersion and no clear gradient (see
Fig. \ref{agemetgrad}). This is not the case when the external group is separated according to the different gas structures. In particular, the western
clusters (called west halo, hereafter WH) indicate some gradient in age and possibly
 in metallicity \citep[][hereafter Paper I]{dias+14}.
The WH does not correspond to
any named gas structure, it is located in the 
wing and bridge direction, but on the opposite side. The best way to
characterize the possible age and metallicity gradients is to
analyse the different groups in a homogeneous way.

In Sect. 2 the photometric data and reductions are presented. In Sect. 3
the method of isochrone fitting is described. The results are reported
in Sect. 4 and are discussed in Sect. 5. Conclusions are drawn in Sect. 6.

%

\section{Data}
\label{thedata}

%
\subsection{Selection of targets and observations}
\label{selection}

The definition of the WH region is based on the ellipses
centred on the SMC centre (P.A.=45$^{\circ}$, $e$= 0.87), as shown in
Fig. \ref{2Dpos}. All clusters located beyond $a > 2^{\circ}$
and relative $\alpha \cdot$ cos$\delta < -1^{\circ}$ are selected as
WH clusters. The catalogue of \cite{bica+08a} lists 31
clusters in this region.
Only 15 of these have at least age and metallicities available in the
literature, mostly using CMD analysis. They are  listed in Table \ref{westhalocatal}.
The remaining 16 clusters, which are fainter, have no dedicated studies so
far. We observed 9 of them that are spread through the extension of the WH.  This increased the statistics of stellar
clusters with no parameters in this region from 15 to 24 clusters
(44\% to 77\%) and offers the possibility of establishing
 gradients in age and metallicity, as previously suggested in Paper I. We
 included NGC\,152 as a reference cluster,  compared 
with the CMD analysis by \cite{rich+00} and Correnti et al. (2016, in prep.), 
both using HST images. The conversion
of magnitudes from the HST to the Johnson system may have been
problematic in \cite{rich+00}
because the CMDs of NGC~411 and NGC~419 have different colours (see their
Fig.1), while \cite{girardi+13} showed that they are similar. Moreover, 
Correnti et al. have a deeper CMD with lower photometric errors, but in different
bands than those we use in this work. We later compare our results with those
from both works, but with some reservation with respect to the
results of Rich et al.

In Fig. \ref{2Dpos} we also define wing and bridge clusters as all objects
located beyond $a > 2^{\circ}$, relative $\alpha \cdot$ cos$\delta >
-1^{\circ}$ and relative $\delta < 0.8^{\circ}$. They are located
in the
region that was originally defined as the SMC wing by \cite{shapley40}, which
extends towards the LMC. The other external group is the counter-bridge,
located beyond $a > 2^{\circ}$, relative $\alpha \cdot$ cos$\delta >
-1^{\circ}$ and relative $\delta > 0.8^{\circ}$. These clusters are in
the region that was defined by \cite{beslaPHD} and \cite{diaz+12} as the
counterpart of the bridge.

\begin{table*}[!htb]
\caption{West halo clusters with known parameters, selected as a subsample of
  \cite{bica+08a} using our definition. The semi-major axis $a$ is
  calculated as in Paper I, and the table is sorted by this parameter. References for metallicity, age, and
  distance are indicated in the last column and notes are at the bottom of
  the table.}
\label{westhalocatal}
\centering
\small
\begin{tabular}{llrrcllllll}
\hline \hline
\noalign{\smallskip}
     Cluster    &    Other names        & R.A. &   DEC. &  $a$ & [Fe/H] & Age                  &      dist.    &   ref.        \\
        &           & $^{h}$ $^{m}$ $^{s}$ &  $^{\circ}$ $^{\arcmin}$ $^{\arcsec}$ &   (deg)  &  & (Gyr)             &      (kpc)     &         \\ 
\noalign{\smallskip}
\hline
\noalign{\smallskip}
\multicolumn{9}{c}{Paper I results (homogeneous scale with this
  work)} \\
\noalign{\smallskip}
\hline
\noalign{\smallskip}
      \object{L\,3}         &    \object{ESO28SC13}          &   00:18:25 &       -74:19:07  &   2.9  &   -0.65$^{+0.18}_{-0.32}$   &   1.01$^{+0.31}_{-0.24}$     &   54.2$\pm$1.5          &  1,1,1        \\
      \object{HW\,1}        &    ---                &  00:18:27 &        -73:23:42  &   3.4  &   -1.43$^{+0.25}_{-0.66}$     &   3.90$^{+1.06}_{-0.83}$    &   58.4$\pm$1.6          &   1,1,1       \\
      \object{L\,2}         &    ---                &   00:12:55 &       -73:29:15  &   3.9  &   -1.58$^{+0.18}_{-0.31}$     &   3.54$^{+0.63}_{-0.53}$         &   56.9$\pm$1.6          &   1,1,1       \\
      \object{AM-3}       &    \object{ESO28SC4}           &   23:48:59 & -72:56:43   &   7.3  &   -0.98$^{+0.23}_{-0.54}$   &   4.88$^{+2.04}_{-1.44}$      &   63.2$^{+1.8}_{-1.7}$  &   1,1,1        \\
\noalign{\smallskip}
\hline
\noalign{\smallskip}
\multicolumn{9}{c}{Other references} \\
\noalign{\smallskip}
\hline
\noalign{\smallskip}

      \object{NGC\,152}     &    \object{K10}, \object{L15}, \object{ESO28SC24}  &   00:32:56 &     -73:06:58  &   2.0  &   -1.13$\pm$0.15    &   9.3$\pm$1.7       &   62.7$\pm$5.7          &    2,2,3     \\
      \object{K\,9}         &    \object{L13}                &   00:30:00 &       -73:22:45  &   2.2  &   -1.24                   &   4.7                        &   ---                   &   4,4,-         \\
      \object{K\,6}         &    \object{L9}, \object{ESO28SC20}       &   00:25:26 & -74:04:33 &   2.4  &   -0.7                   &   1.6$\pm$0.4        &   ---                   &  5,5,-         \\
      \object{K\,5}         &    \object{L7}, \object{ESO28SC18}       &  00:24:43 &       -73:45:18 &   2.5  &   -0.6                    &   2.0                   &   ---                   &  6,7,-          \\
      \object{K\,4}         &    \object{L6}, \object{ESO28SC17}       &  00:24:43 &       -73:45:18  &   2.7  &   -0.9                    &   3.3                      &   ---                   &   6,7,-         \\
      \object{K\,1}         &    \object{L4}, \object{ESO28SC15}       &  00:21:27 &       -73:44:55  &   2.8  &   -0.9                     &   3.3                    &   ---                   &   6,7,-         \\
      \object{L\,5}         &    \object{ESO28SC16}          &  00:21:27 &        -73:44:55 &   3.0  &   -1.2                     &   4.1                     &   ---                   &   6,7,-          \\
      \object{K\,7}         &    \object{L11}, \object{ESO28SC22}      &   00:27:46 &      -72:46:55  &   3.0  &   -0.8                        &   3.5                        &   ---                   &   8,8,-         \\
      \object{K\,3}         &    \object{L8}, \object{ESO28SC19}       &  00:24:47 &       -72:47:39  &   3.3  &   -1.12                   &   6.5                        &   56.7$\pm$1.9          &   9,9,3    \\      
      \object{NGC\,121}     &    \object{K2}, \object{L10}, \object{ESO50SC12}   &   00:26:47 &     -71:32:12  &   4.8  &   -1.2$\pm$0.12             &  10.5$\pm$0.5     &   59.6$\pm$1.8          &   8,10,3    \\
      \object{L\,1}         &    \object{ESO28SC8}           &  00:03:54 &        -73:28:19 &   5.0  &   -1.0                      &   7.5                        &   53.2$\pm$0.9          &  11,11,3   \\
\noalign{\smallskip}
\hline
\end{tabular}
\tablefoot{ 
\tablefoottext{1}{Paper I}
\tablefoottext{2}{\cite{dias+10}}
\tablefoottext{3}{\cite{crowl+01}}
\tablefoottext{4}{\cite{MSF98}}
\tablefoottext{5}{\cite{piatti+05a}}
\tablefoottext{6}{\cite{piatti+05b}}
\tablefoottext{7}{\cite{parisi+09}}
\tablefoottext{8}{\cite{DCH98}}
\tablefoottext{9}{\cite{glatt+08b}}
\tablefoottext{10}{\cite{glatt+08a}}
\tablefoottext{11}{\cite{glatt+09}}
}
\end{table*} 
\normalsize

\begin{figure*}[!htb]
   \centering
   \includegraphics[width=0.32\textwidth]{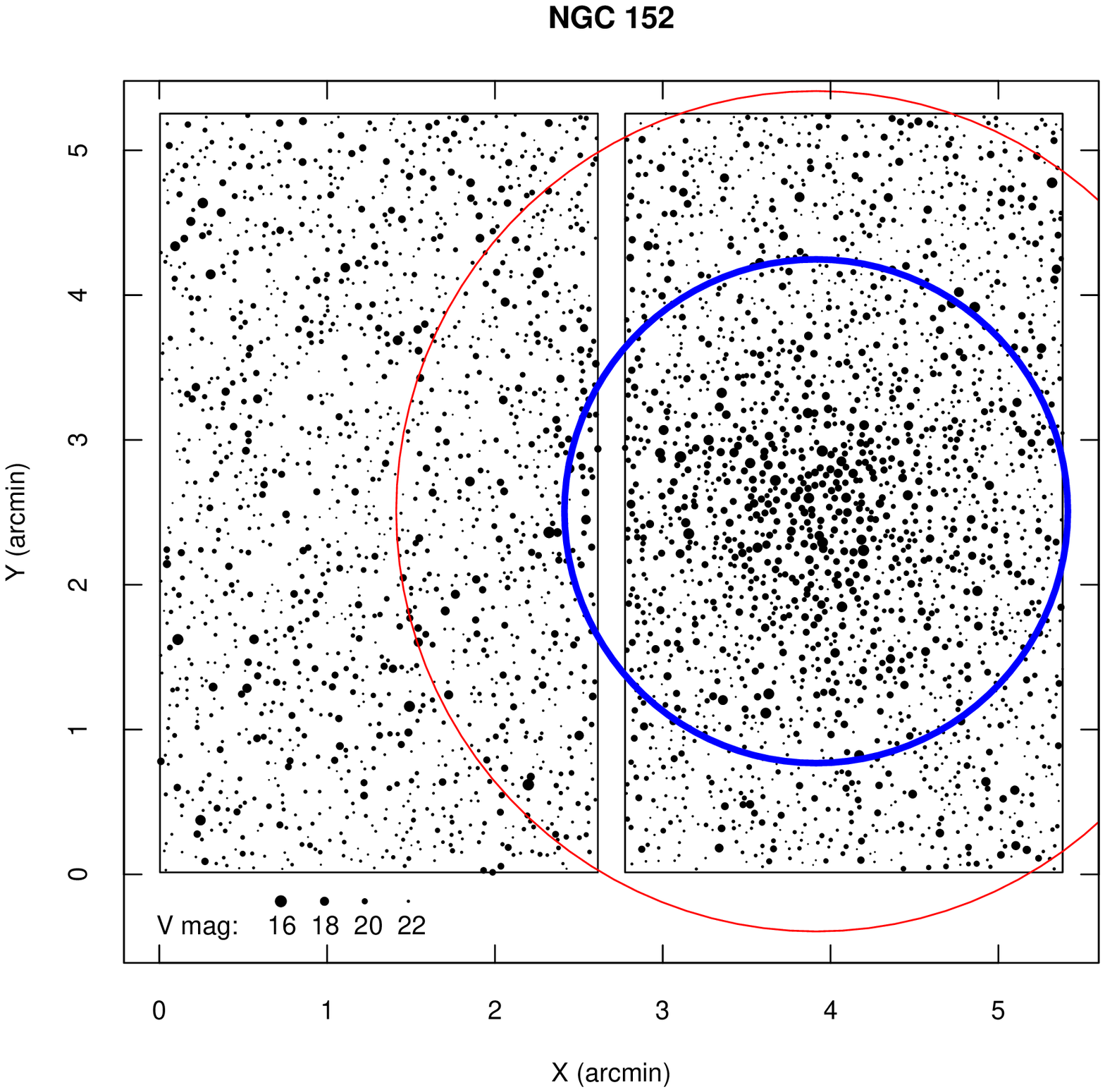}
   \includegraphics[width=0.32\textwidth]{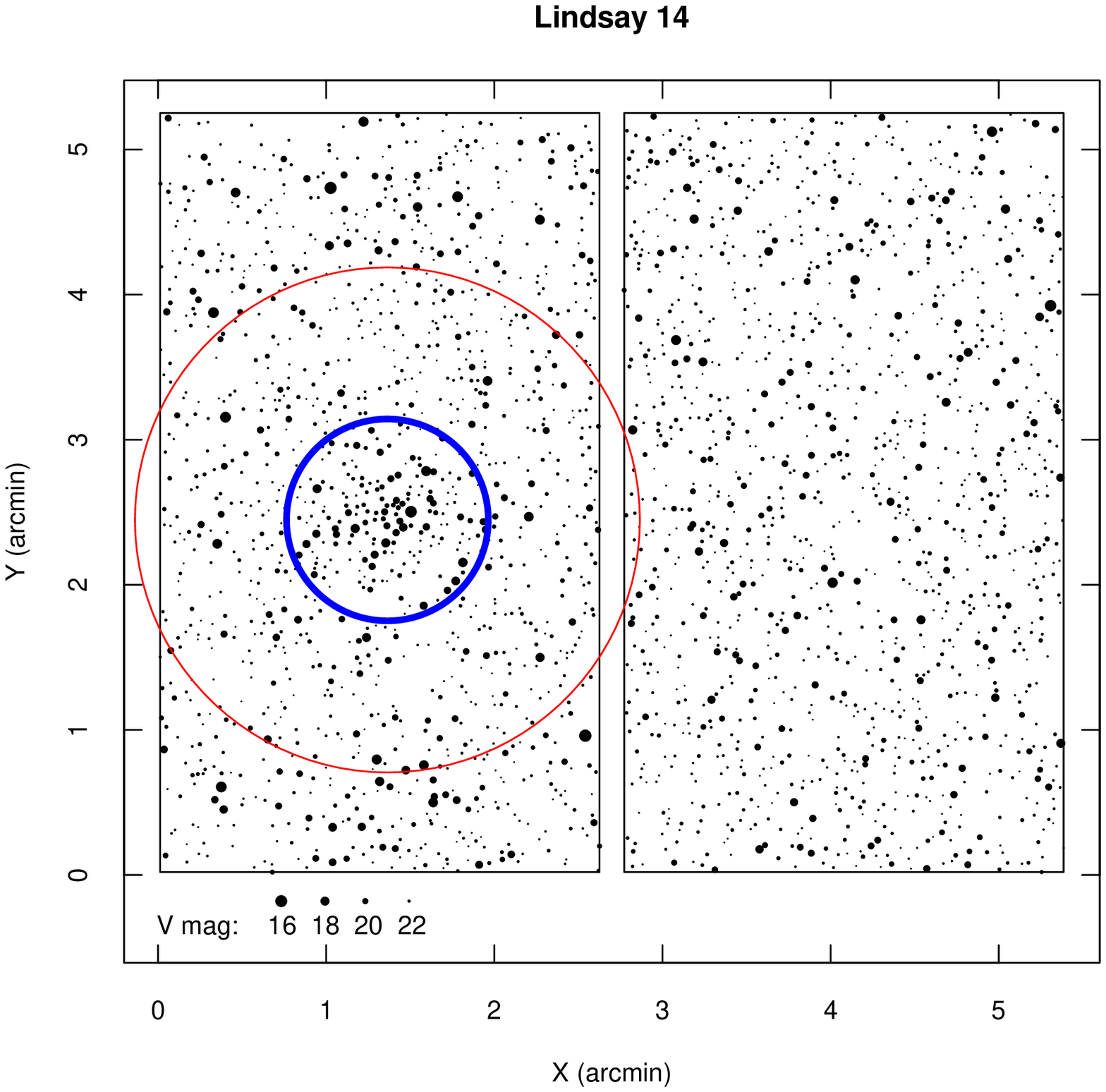}
   \includegraphics[width=0.32\textwidth]{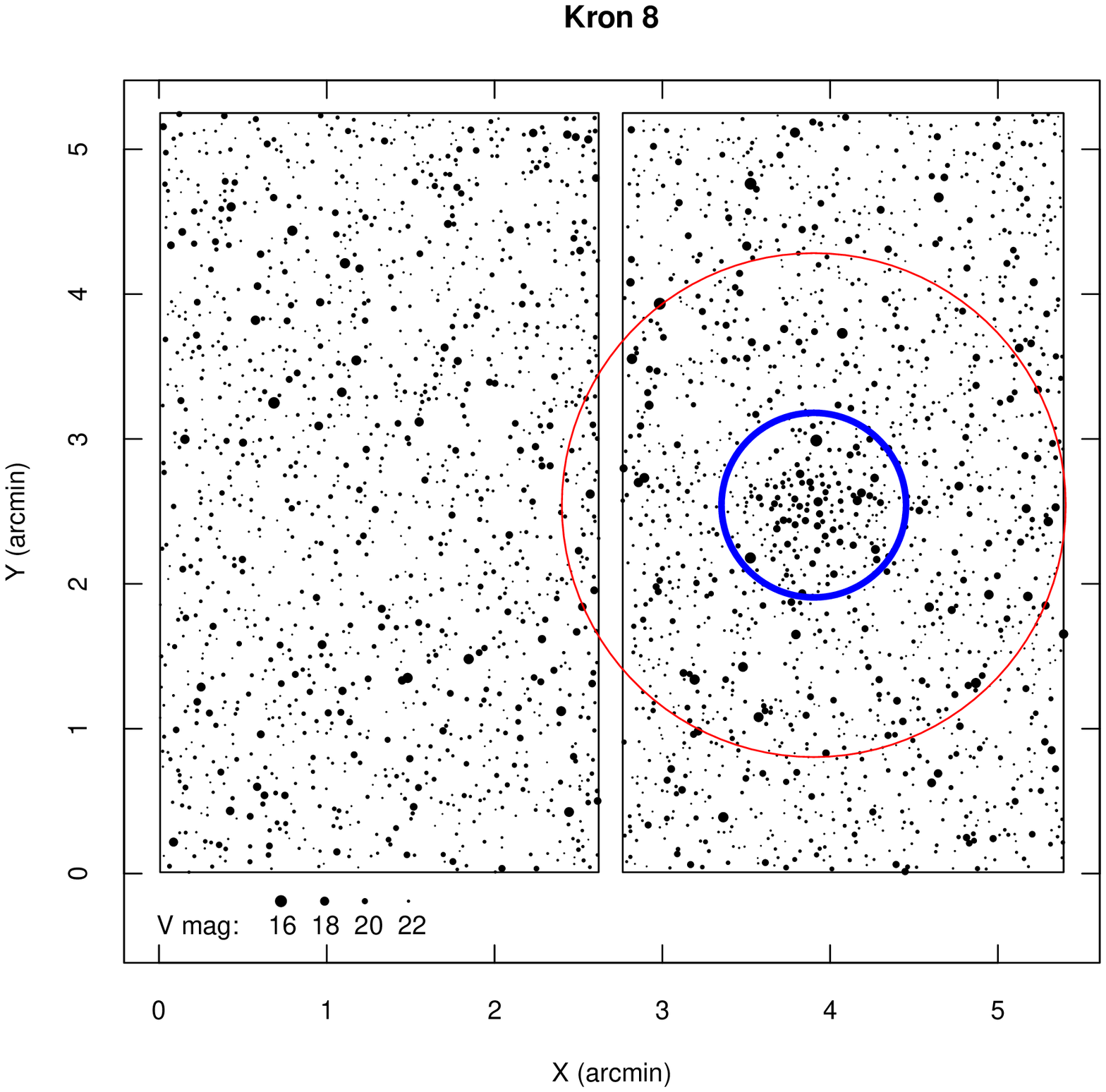}
   \includegraphics[width=0.32\textwidth]{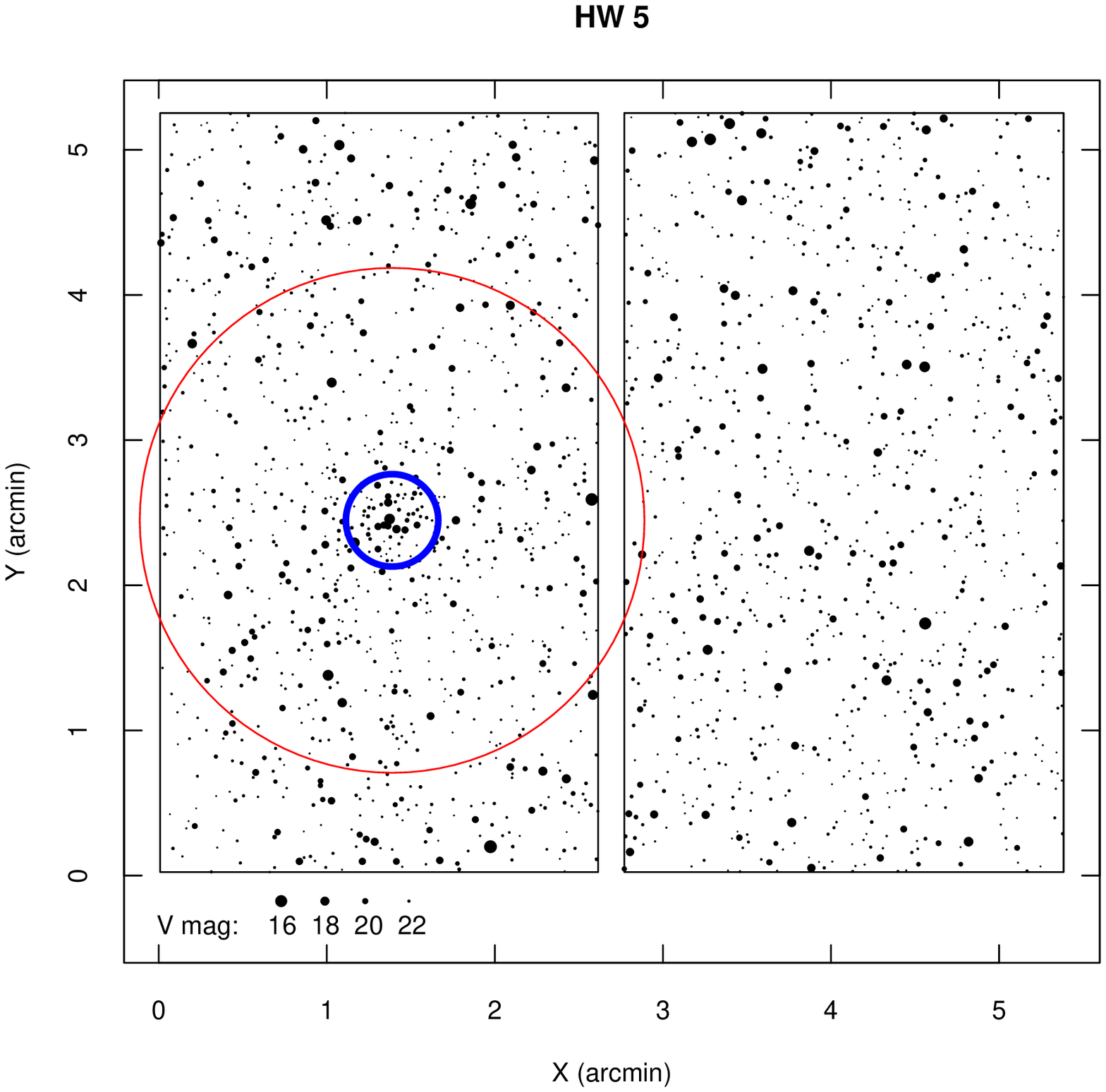}
   \includegraphics[width=0.32\textwidth]{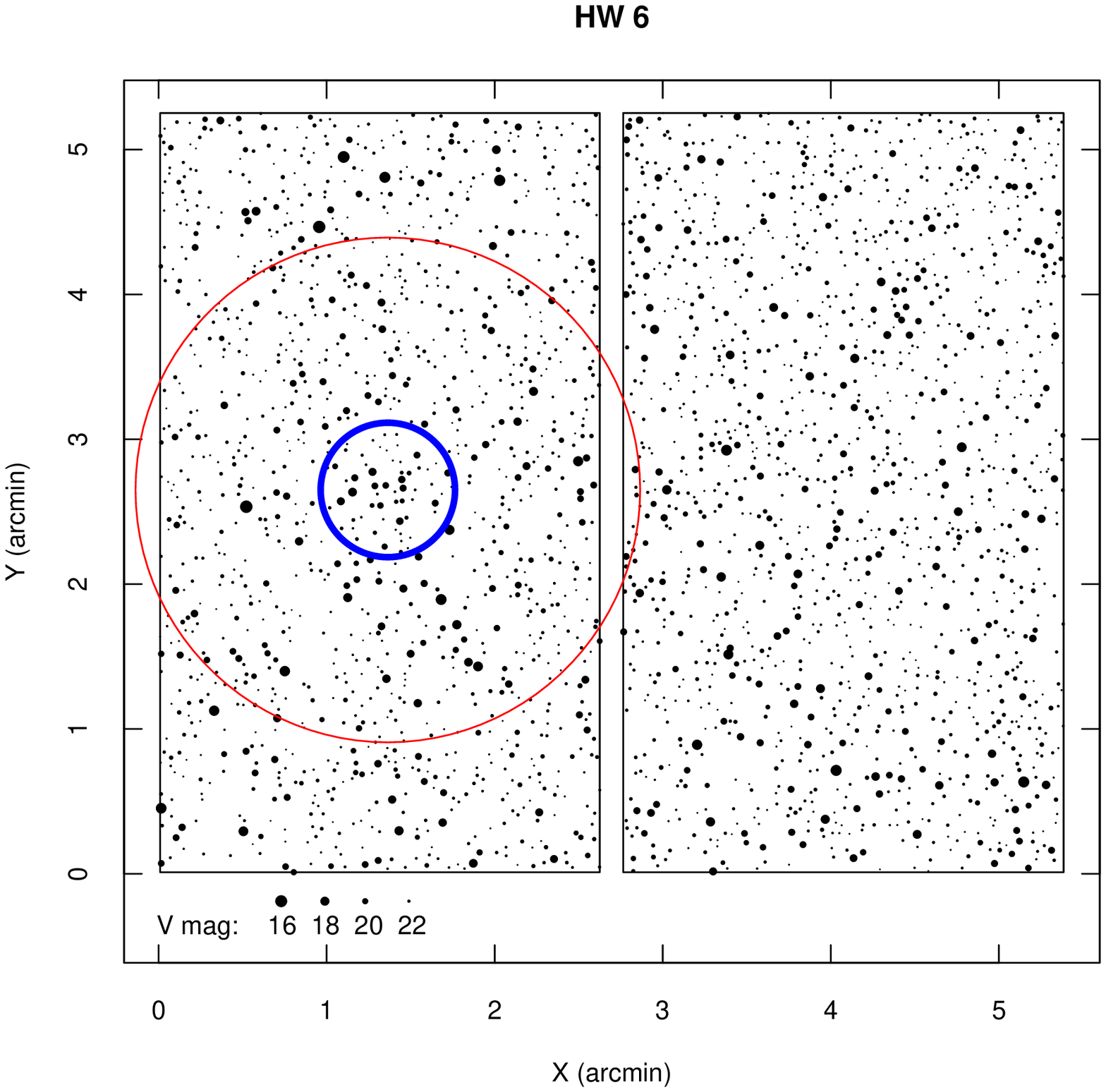}
   \includegraphics[width=0.32\textwidth]{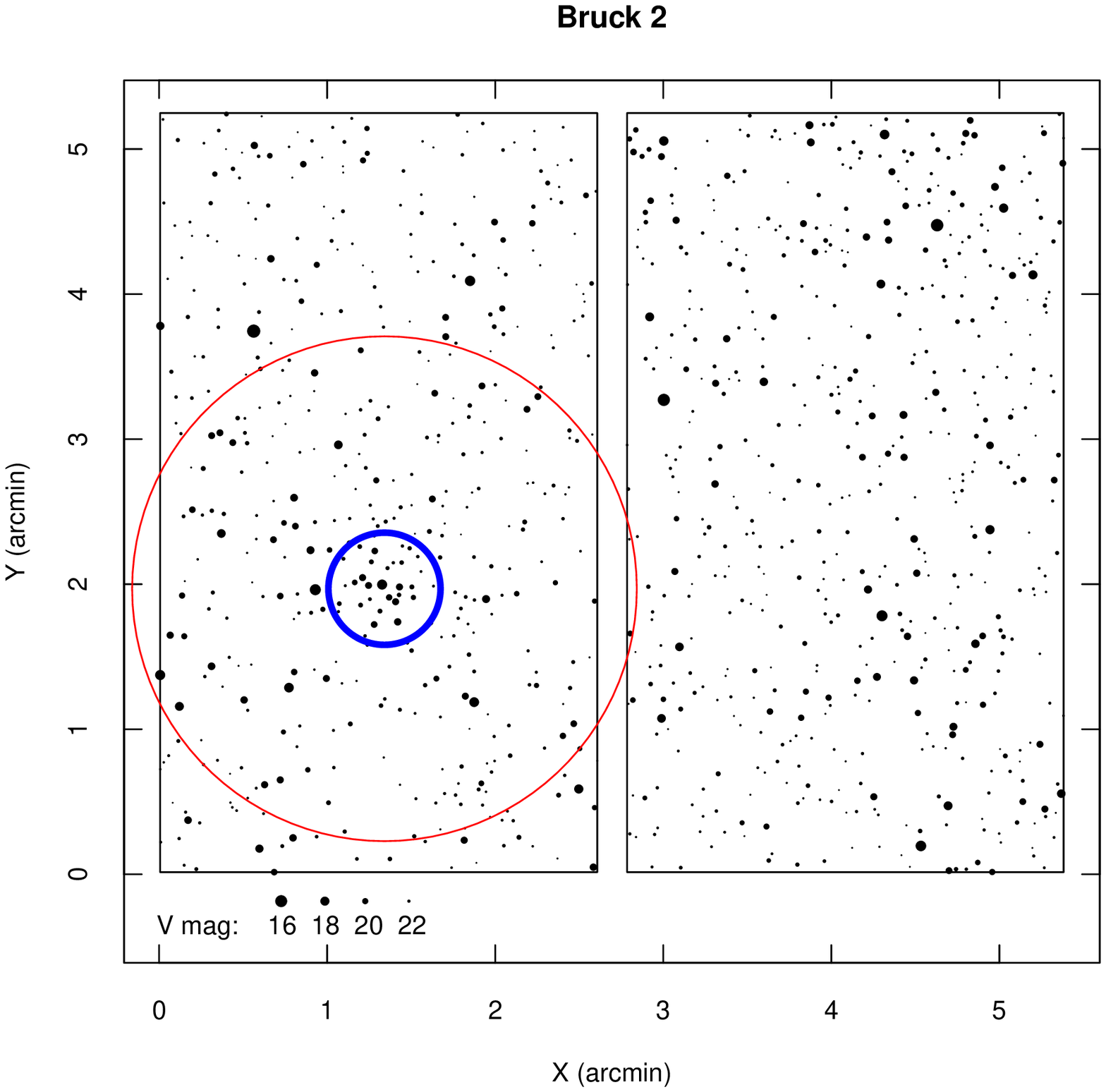}
   \includegraphics[width=0.32\textwidth]{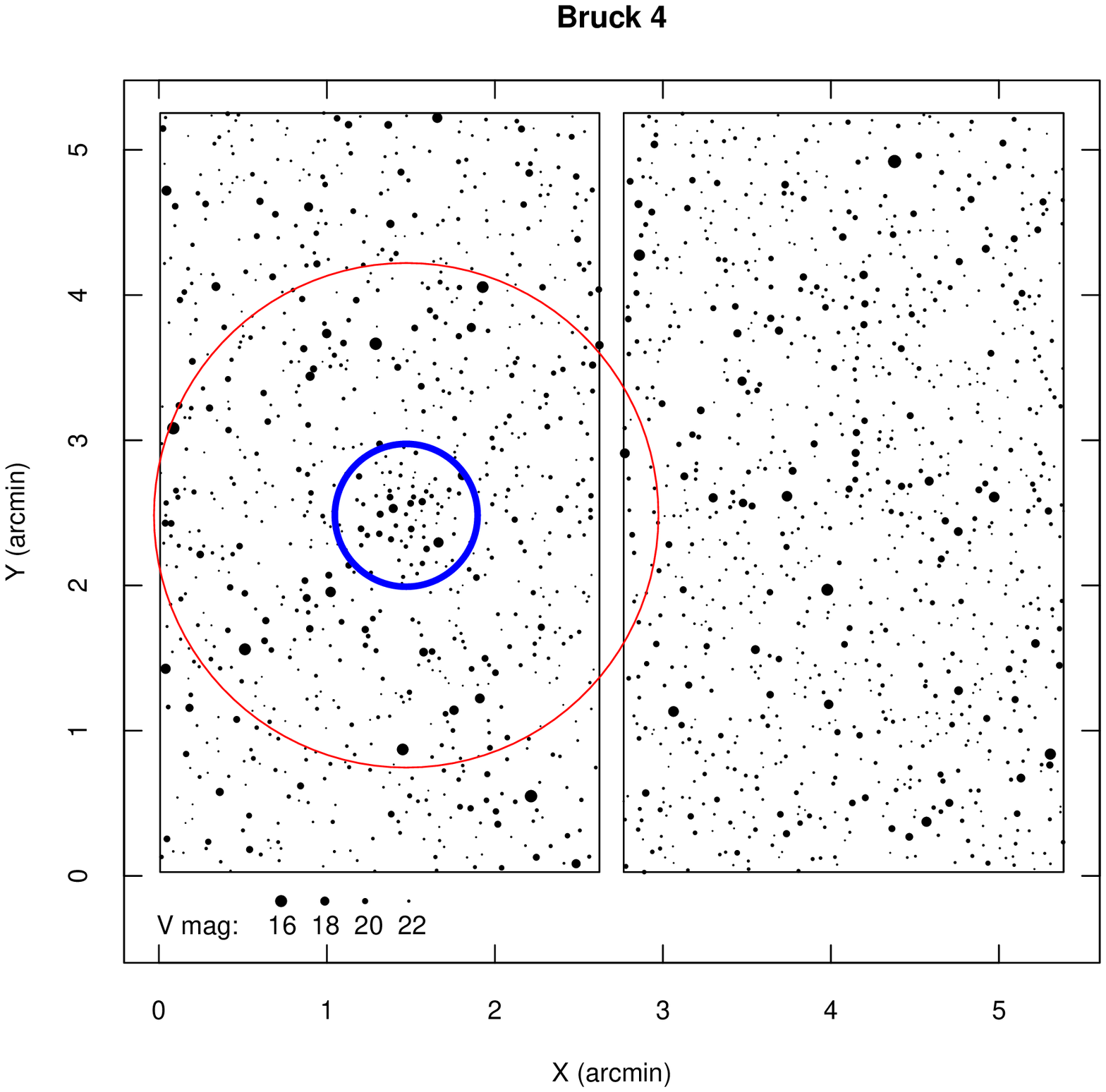}
   \includegraphics[width=0.32\textwidth]{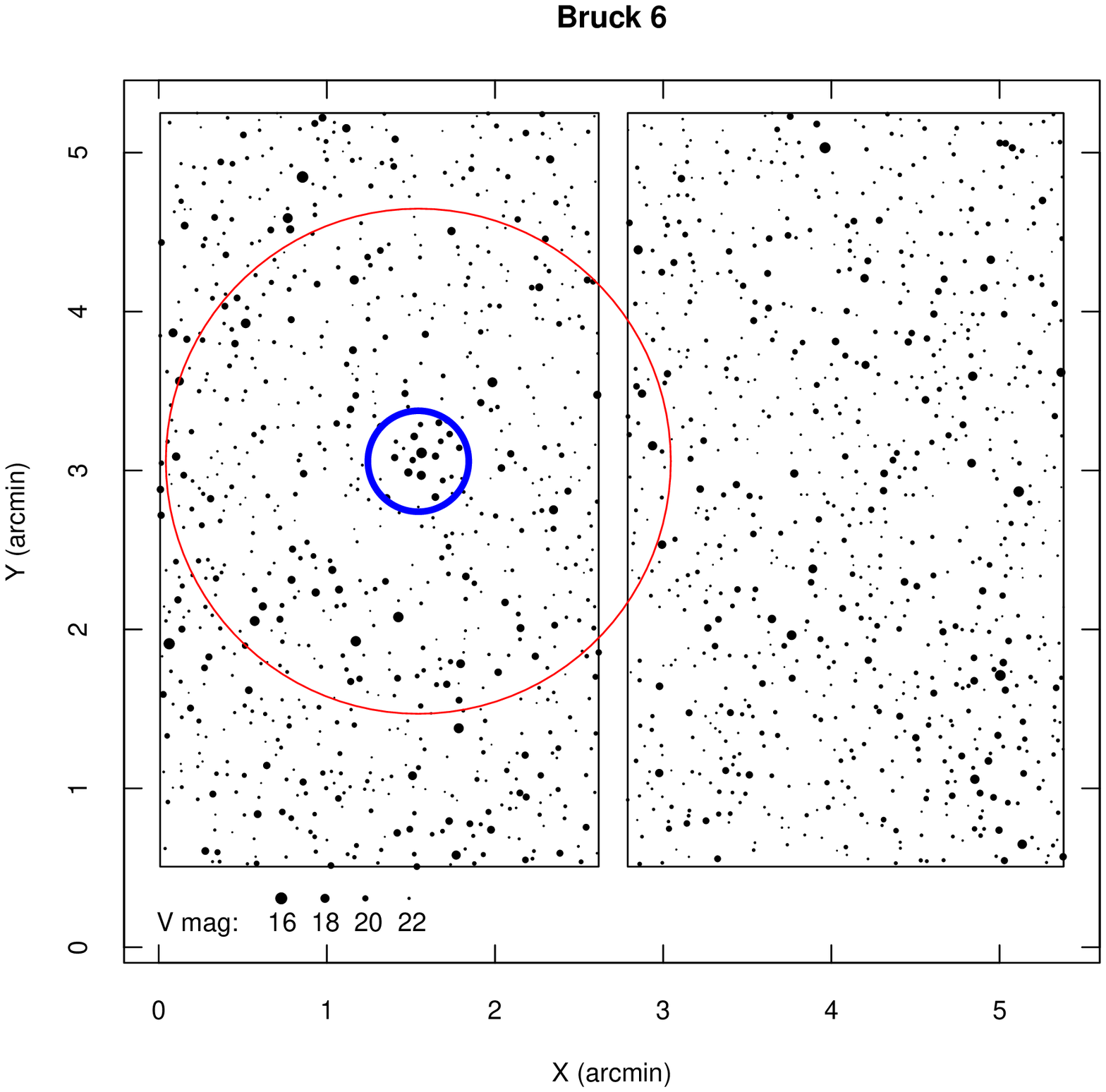}
   \includegraphics[width=0.32\textwidth]{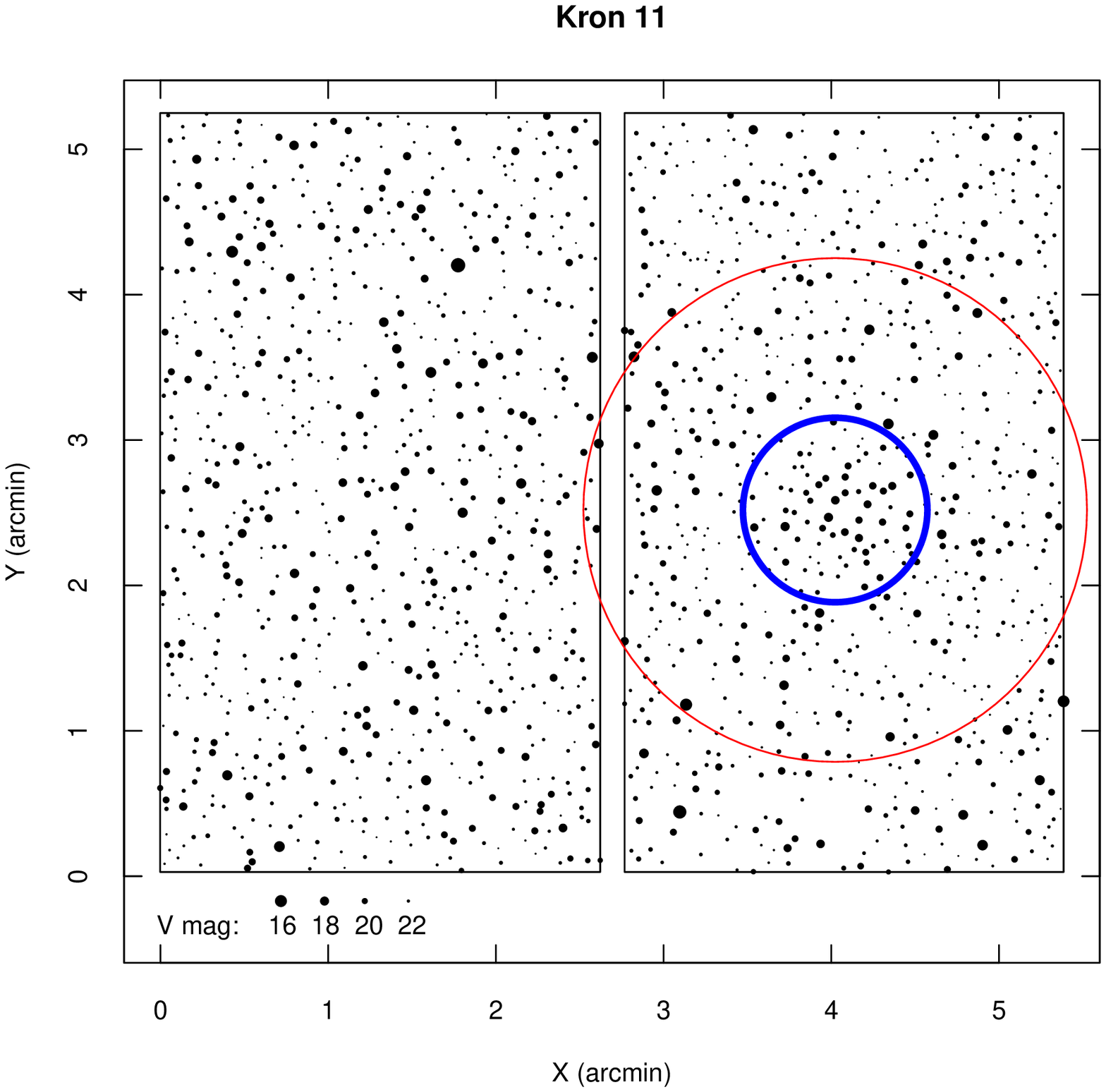}
   \caption{Sky maps of all nine clusters. All plots have the same
     contrast scale for magnitudes to allow a fair visual
     comparison. Blue circles indicate the adopted cluster size based on the catalogue of \cite{bica+08a} 
     that were increased in some cases to include more stars for the statistical field-star decontamination.
     Red circles are the smallest limit around the cluster considered to select field stars, which has
     a radius of 2.5$\arcmin$ for NGC~152, and 1.5$\arcmin$ for the others. The values
 are provided     in Table \ref{stats-fits}.}
 \label{skymaps-A}
   \end{figure*}

Optical images of the SMC WH clusters were obtained using the SOAR
Optical Imager (SOI) in the 4.1m Southern Astrophysical Research (SOAR)
telescope, under the project SO2013B-019 using the same setup as in Paper
I. 
B and V filters were used to
detect stars in the magnitude range 16 $<$ V $<$ 24, which covers all
required CMD features to derive age, metallicity, distance, and
reddening, that is, a few magnitudes below the main-sequence
turnoff (MSTO) and few magnitudes above the red clump (RC).
While the MSTO is a crucial CMD feature to determine age,
the RC and the red giant branch (RGB) are particularly important
to constrain distance and metallicity.
 SOI/SOAR has a field of view of
$5.26\arcmin\times5.26\arcmin$ and is composed of two CCDs that
are separated by gap
of 7.8$\arcsec$. Because of the gap, we centred the clusters
in the centre of one of the CCDs to avoid losing important
information on the cluster. The sky maps of all observed
clusters show their positions in the SOI FOV in Fig.
\ref{skymaps-A}. The pixel 
scale of 0.077$\arcsec$/pixel was converted into
0.154$\arcsec$/pixel because we used a setup of 2x2 binned pixels.
Observations were carried out during two nights, the first had
poor seeing (1.5 - 2.0$\arcsec$) ,
 the second good seeing (0.8 - 1.2$\arcsec$). The
only cluster good enough from the first night is Kron~11, all other
eight clusters analysed in this work were observed in the second night.
The observation log is presented in Table \ref{log}.

\begin{table}[!htb]
\caption{Observation log. The CCDs were displaced by $\sim
  20\arcsec$ from the cluster centre to avoid the gap
  between the set of two E2V CCDs in SOI, as shown in Fig.
  \ref{skymaps-A}. The ($\alpha$, $\delta$)
  coordinates and the (D)iameters are from \cite{bica+08a}.}
\label{log}
\centering
\small
\begin{tabular}{l@{ }c@{  }cc@{ }c@{ }c@{ }c@{ }c}
\hline \hline
\noalign{\smallskip}
Name(s) & $\alpha$ (2000)& $\delta$ (2000) & D & Filter  & Exp. & Airmass & seeing  \\
 & h:m:s & $\degr:\arcmin:\arcsec$ & $\arcmin$ &  & sec. &   & $\arcsec$ \\
\noalign{\smallskip}
\hline
\noalign{\smallskip}
\multicolumn{8}{c} {2013-09-12} \\
\noalign{\smallskip}
\hline
\noalign{\smallskip}
\object{Br\"uck\,2}  & 00:19:18 & -74:34:28 & 0.45 & B & 3x600 & 1.41 & 1.1 \\
                  &                 &                  &         & V & 3x200 & 1.42 & 1.1  \\
\object{Br\"uck\,4}  & 00:24:54 & -73:01:50 & 0.85 & B & 3x600 & 1.57 & 1.2 \\
                  &                 &                  &         & V & 3x200 & 1.54 & 1.1 \\
\object{Br\"uck\,6}  & 00:27:57 & -74:24:02 & 0.60  & B & 3x600 & 1.49 & 1.2 \\
                  &                 &                  &          & V & 3x200 & 1.52 & 0.8 \\
\object{Kron\,8}      & 00:28:01 & -73:18:15  & 1.10  & B & 3x600 & 1.42 & 0.9 \\
                  &                 &                  &          & V & 3x200 & 1.41 & 0.9  \\
\object{HW\,5}        & 00:31:03 & -72:20:35  & 0.55  & B & 3x600 & 1.35 & 0.9 \\
                  &                 &                  &          & V & 3x200 & 1.35 & 0.8 \\
\object{Lindsay\,14} & 00:32:41 & -72:34:53 & 1.20 & B & 3x600 & 1.47 & 1.1 \\
                  &                 &                  &          & V & 3x200 & 1.44 & 0.9 \\
\object{NGC\,152}  & 00:32:56 & -73:06:58  & 3.00  & B & 3x600 & 1.42 & 1.1 \\
                  &                 &                  &          & V & 3x200 & 1.43 & 1.0  \\
\object{HW\,6}         & 00:33:04 & -72:39:07 & 0.80  & B & 3x600 & 1.37 & 1.2 \\
                  &                 &                  &          & V & 3x200 & 1.36 & 0.9 \\
\noalign{\smallskip}
\hline
\noalign{\smallskip}
\multicolumn{8}{c} {2013-09-11} \\
\noalign{\smallskip}
\hline
\noalign{\smallskip}
\object{Kron\,11}             & 00:36:27 & -72:28:41  & 1.10  & B & 3x600 & 1.53 & 1.8 \\
                           &                 &                  &          & V & 3x200 & 1.50 & 1.4 \\
\noalign{\smallskip}
\hline
\end{tabular}
\end{table} 
\normalsize


\subsection{Data reduction and photometry}

   \begin{figure}[!htb]
   \centering
   \includegraphics[width=\columnwidth]{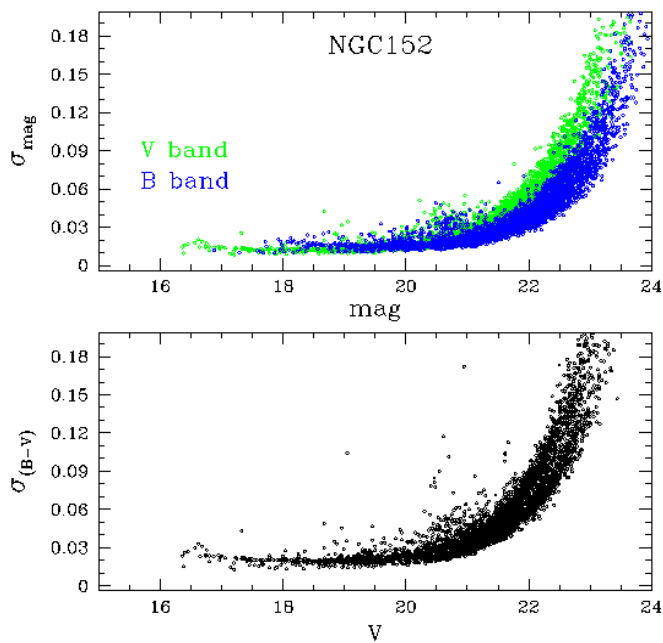}
   \caption{Photometric errors from IRAF for the reference cluster NGC~152 in bands B
     and V, and colour B-V.}
 \label{fig:photerr}
   \end{figure}

\begin{figure*}[!htb]
\centering
\includegraphics[height=0.8\textwidth,angle=-90]{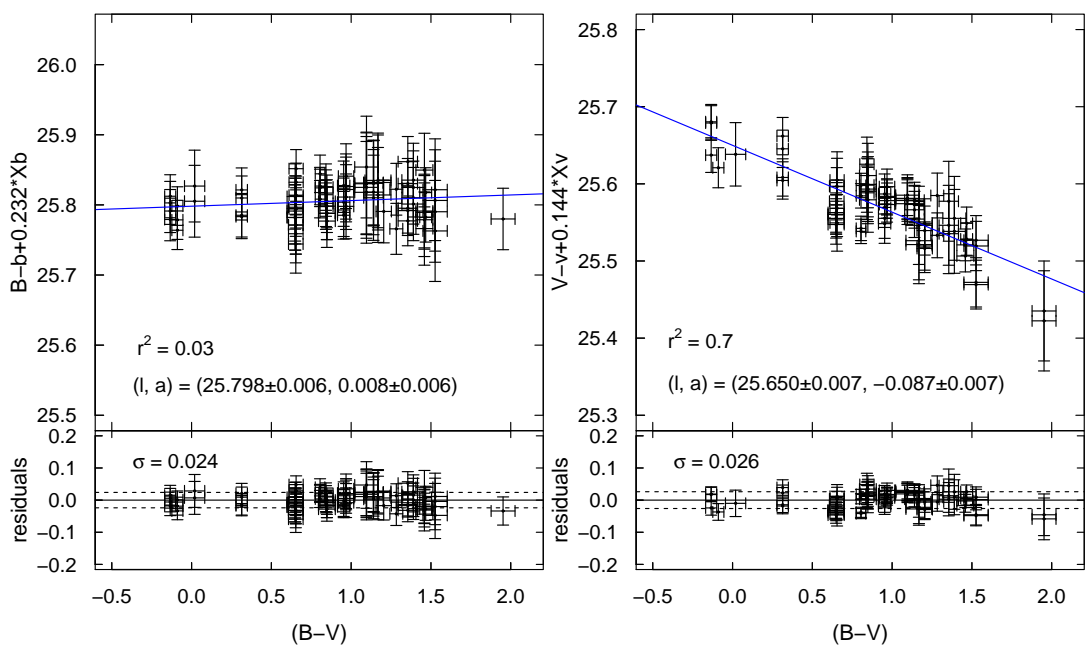}
\caption{Standard star calibration curve fit, derived with Eq.
  \ref{calib-cur}. The linear fit coefficients are displayed in the
  plots and listed in Table \ref{calibcoef} together with the standard
  deviation of the residuals $\sigma$ and the coefficient of
  determination $r^2$.} 
\label{stdfit}
\end{figure*}

The data were reduced with dedicated
SOAR/IRAF
packages\footnote{http://www.soartelescope.org/observing/documentation/soar-optical-imager-soi/image-reduction}
to properly handle the SOI mosaic images. 
Aperture and PSF photometry were performed using the DAOPHOT/IRAF package
\citep{stetson87}. Photometric errors are displayed in
Fig. \ref{fig:photerr} for the reference cluster NGC~152 as an
example, a similar plot for all clusters is presented in Appendix
\ref{app}. Some of these plots show
two stripes for a given set of stars, for example the bottom panel
of Fig. \ref{fig:photerr}. Each stripe corresponds to one of the two
chips of the instrument (as shown in Fig. \ref{skymaps-A}). The
difference of the photometric errors between the two chips is very
small and larger for fainter stars, which does not affect the
analysis in this paper.


The same fields of standard stars as were used in Paper I were
observed here. We used the selection of \cite{sharpee+02}. However,
instead of using only a few reference stars listed in \cite{sharpee+02},
we matched all stars in the fields with the stars in the
Magellanic Clouds Photometric Survey by \citet[MCPS][]{zaritsky+02}. Only
bright stars with V $<$ 17.5 (above the red clump) were considered, which
avoids most of possible variable stars in the field and guarantees
smaller photometric errors. We also excluded outliers with large
errors from MCPS sample. Using MCPS magnitudes as references, we
fitted Eq. \ref{calib-cur} to find the zero point $\beta$ and colour
coefficient $\alpha$: 

\begin{center}
\begin{equation}
  M-m=\alpha\cdot (B-V)+ \beta{\rm ,}
  \label{calib-cur}
\end{equation}
\end{center}

\noindent where $M$ corresponds to either B or V, and $m$ corresponds
to the respective instrumental magnitudes 
(given by -2.5~$\times$~log(counts/exptime) already corrected by
airmass effects). Airmass coefficients used to correct the magnitudes
were 0.22$\pm$0.03~mag/airmass and 0.14$\pm$0.03~mag/airmass for the B
and V bands, respectively (as can be found at the CTIO
website\footnote{http://www.ctio.noao.edu/noao/content/13-m-photometric-standards}).
Using the residuals of the fit, we made three
1-$\sigma$ clippings to exclude outliers and selected a
narrow distribution of well-behaved stars in a range of colours of
-0.3 $<$ (B-V) $<$ 2.0. The final fitting is displayed in Fig.
\ref{stdfit} for filters B and V, with the respective residual plots.
Fitted coefficients are listed in Table \ref{calibcoef} together
with quality factors $r^2$ and $\sigma$.


\begin{table}[!htb]
\caption{Coefficients of Eq. \ref{calib-cur} from the fits of the
  2007 and 2008 standard stars. }
\label{calibcoef}
\centering
\begin{tabular}{l|cc}
\hline \noalign{\smallskip}
Coef.          & B                      & V                  \\
\noalign{\smallskip}
\hline 
\noalign{\smallskip}
$\alpha$    & 0.008 $\pm$ 0.006   & -0.087 $\pm$ 0.007  \\
$\beta$ (mag) & 25.798 $\pm$ 0.006 & 25.650 $\pm$ 0.007  \\
r$^2$         & 0.03                     & 0.7  \\
$\sigma_{\rm residuals}$ (mag) & 0.024                     & 0.026  \\
\noalign{\smallskip}
\hline
\end{tabular}
\end{table} 


\subsection{Photometric completeness and cluster membership}

\begin{figure}[!htb]
   \centering
   \includegraphics[width=\columnwidth]{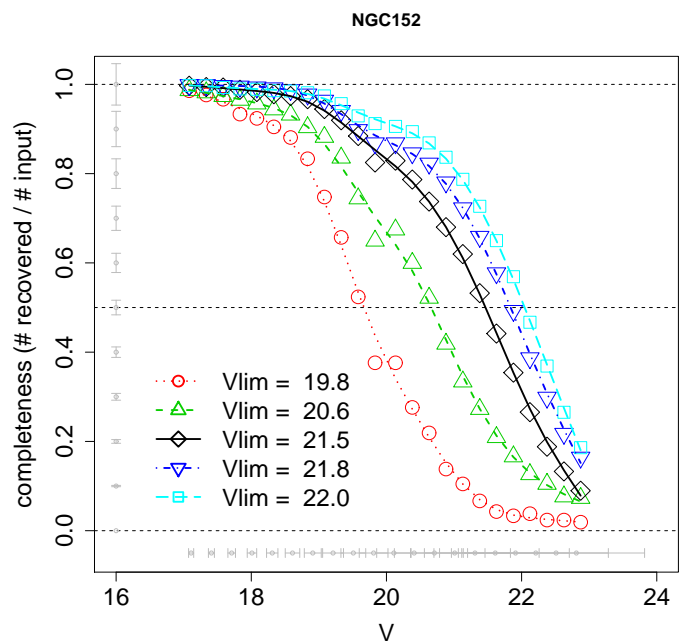}
   \caption{Completeness curves for the reference cluster NGC~152. Different
     curves represent different annuli around the centre of the
     cluster in steps of 22$\arcsec$ (from 0$\arcmin$ to
     1.8$\arcmin$). The curves in a crescent distance from the
     cluster centre are represented by red circles and a dotted line,
     green triangles and a dashed line, black diamonds and solid lines,
     blue inverted triangles and a dot-dashed line, and cyan squares and
     a long-dashed line. Uncertainty bars from the artificial star
     tests are presented in grey. The horizontal black dashed line at
     completeness level 0.5 marks the intersection with each line whose
     magnitudes are shown in the legend in each panel.}
 \label{fig:complete}
   \end{figure}

Artificial star tests (ASTs) were made using the task addstar in IRAF and additional
scripts to determine photometric completeness. We generated stars with
magnitudes 17 $<$ V $<$ 23 in steps of 0.25~mag to cover all features
in the CMD of target clusters. We randomly chose V magnitudes within
each interval of magnitude and distributed the stars in radial
symmetry around the cluster centre. We verified that the smallest
distance between any pair of stars was $\gtrsim 3.5 \cdot <{\rm
  FWHM}>$ to avoid introducing artificial crowding (e.g. Paper
I; \citealp{rubele+11}). For NGC~152 we generated 2301 stars for each
magnitude bin within a radius of 1.8$\arcmin$ and 467 stars for the
other clusters within 0.8$\arcmin$ around the cluster centre. To
calculate B magnitudes for these stars that were to be used to include
artificial stars in the B-band images, we tried to follow approximate
colours of the CMD structures for each cluster and applied this
difference to the V magnitudes, keeping the same positions as
were calculated
for V images. For each magnitude bin of 0.25~mag we introduced
artificial stars as described above in B and V bands for a given
cluster and performed the photometry and calibration exactly as was done
before for the cluster. After this, we counted the percentage of
recovered artificial stars. For each magnitude bin we repeated this
procedure twice to obtain different random positions for the artificial
stars and avoid biases. In Fig. \ref{fig:complete} we show the
completeness curves for the reference cluster NGC~152 for different
radii and V magnitudes; the curves for the other clusters are
presented in Appendix \ref{app}.

\begin{figure}[!htb]
   \centering
   \includegraphics[width=\columnwidth]{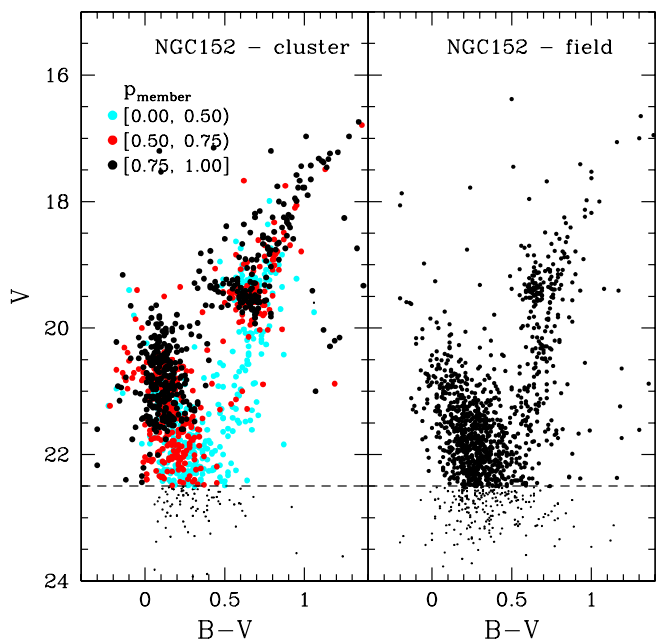}
   \caption{V, (B-V) CMD for the reference cluster NGC~152 ($R <
     R_{\rm{clus}}$, left panel) and the control field ($R >
     R_{rm{field}}$, right panel). The point colours depend on the
     membership probability ($p_{\rm{member}}$) for each star in the
     cluster direction. The horizontal dashed line corresponds to the
     magnitude limit, derived from the completeness curves around
     $R_{\rm{clus}}$.}
 \label{fig_membership}
   \end{figure}

Cluster member stars were estimated statistically following the method
of Paper I developed by \cite{kerber+05}. The idea is to compare the
CMD of stars in the direction of the cluster with another made of
nearby field stars. For each bin in colour and magnitude the number of
stars in the two CMDs are counted and normalized by the area in the sky
that is covered by the cluster and field region. This information is combined with
the completeness calculated for each star interpolating the curves
from Fig. \ref{fig:complete} in magnitude and position. At the end,
we determine a  membership probability for each star in the
cluster CMD. For a detailed explanation we refer to Paper I. We
show the CMD for cluster and field stars for NGC~152 in
Fig. \ref{fig_membership}, where the membership probability is
indicated by colour scale. Similar plots for the other clusters are
displayed in Appendix \ref{app}.

\begin{table}[!htb]
\caption{Stellar counts after the corrections for incompleteness and
field contamination. Columns correspond to the cluster name, the adopted
cluster radius ($r_{\rm{clus}}$), the distance from the cluster centre
to define the control field ($r_{\rm{field}}$), the magnitude limit
($V_{\rm{lim}}$), the number of observed stars in the cluster
direction before any treatment for selection effects
($N_{\rm{clus}}^{\rm{(obs)}}$) and after the corrections for
incompleteness ($N_{\rm{clus}}^{\rm{(comp)}}$) and field contamination
($N_{\rm{clus}}^{\rm{(clean)}}$).}
\label{stats-fits}
\centering
\begin{tabular}{l@{}c@{ }c@{ }c@{ }c@{ }c@{ }c}
\hline \hline
Target  & $r_{\rm{clus}}$ & $r_{\rm{field}}$ & $V_{\rm{lim}}$ &
$N_{\rm{clus}}^{\rm{(obs)}}$ & $N_{\rm{clus}}^{\rm{(comp)}}$ &
$N_{\rm{clus}}^{\rm{(clean)}}$ \\
~~~~    & arcsec & arcsec & ~~~~  & ~~~~  &~~~& \\
\hline
NGC\,152 & 90 & 150 & 22.5 & 1149 & 2931 & 1887 \\
Br\"uck\,6 & 18 & 90 & 22.5 & 50 & 129 & 105 \\
Kron\,11 & 33 & 90 & 22.5 & 62 & 194 & 167 \\
Kron\,8  & 33 & 90 & 23.0 & 187 & 491 & 367 \\
HW\,6  & 24 & 90 & 22.7 & 50 & 104 & 66 \\
Lindsay\,14  & 36 & 90 & 23.0 & 167 & 471 & 349 \\
Br\"uck\,2 & 20 & 90 & 22.0 & 23 & 66 & 62 \\
Br\"uck\,4 & 25 & 90 & 22.5 & 50 & 129 & 105 \\
HW\,5  & 16 & 90 & 22.5 & 41 & 95 & 87 \\
\hline
\end{tabular}
\end{table}

%

\section{Statistical isochrone fitting}
\label{isot_fit}

\subsection{Method}

To determined age, metallicity, distance modulus, and
reddening in an objective and self-consistent way, we fit the CMD
of each cluster against a set of synthetic CMDs using a numerical-statistical isochrone fitting.
This method was extensively explained in previous works of our group,
where it was applied to LMC clusters observed with HST/WFPC2 \citep{kerber+02,kerber+05,kerber+07} and
Galactic open clusters from 2MASS \citep{alves+12}.
In Paper I we analysed
the CMDs of five SMC stellar clusters observed with SOAR/SOI using
the same setup as in this paper.
We therefore limit the description here and refer to the previous
works for further details on the method.

   \begin{figure}[!htb]
   \centering
   \includegraphics[width=\columnwidth]{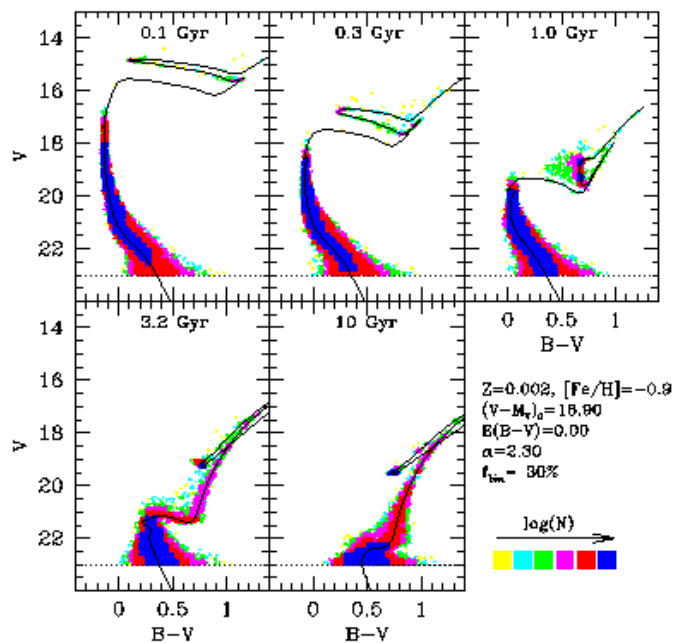}
   \caption{Synthetic CMDs for different ages keeping other typical
     SMC parameters fixed as indicated in the plot. The colour scale
     indicates the logarithm of the number of stars to compose a Hess diagram
     used to fit the observer CMDs.}
\label{cmd_modelling}
   \end{figure}

The first step is to make a visual isochrone fitting to have \textup{\textup{{\it \textup{a
  priori}}} }values for age, metallicity, distance and reddening. A grid
  of synthetic CMDs was constructed using the \textup{{\it \textup{a priori}} }information.
  All synthetic CMDs were simulated based on PARSEC
isochrones \citep{bressan+12}, assuming a binary fraction of 30\%, a
Salpeter IMF slope of $\alpha = 2.30$, and photometric errors from the
observations (Fig. \ref{fig:photerr}).
Figure \ref{cmd_modelling} illustrates five synthetic CMDs for ages
varying from 0.10 Gyr to 10 Gyr; the other parameters were kept
fixed.
They reproduce the observed features well,
including the spread of points that is due to the photometric uncertainties and
unresolved binaries.

The observed CMD was fitted against the grid of synthetic CMDs
by applying
the maximization of the likelihood statistics. The likelihood of
each comparison was calculated as the
product of the probabilities for the observed stars to be
reproduced by each synthetic CMD -- an implicit combination of age,
metallicity, distance modulus, and reddening.
The set of synthetic CMDs that maximize the likelihood were
identified as the best, and their parameters were
averaged out to obtain the final parameters and uncertainties
for a given cluster.
To avoid local maxima, we explored a wide range of values in
the parameter space, considering virtually all solutions among the best ones.
For further details we refer to the aforementioned works.

\subsection{Results}

\begin{figure*}[!htb]
   \centering
   \includegraphics[width=0.32\textwidth]{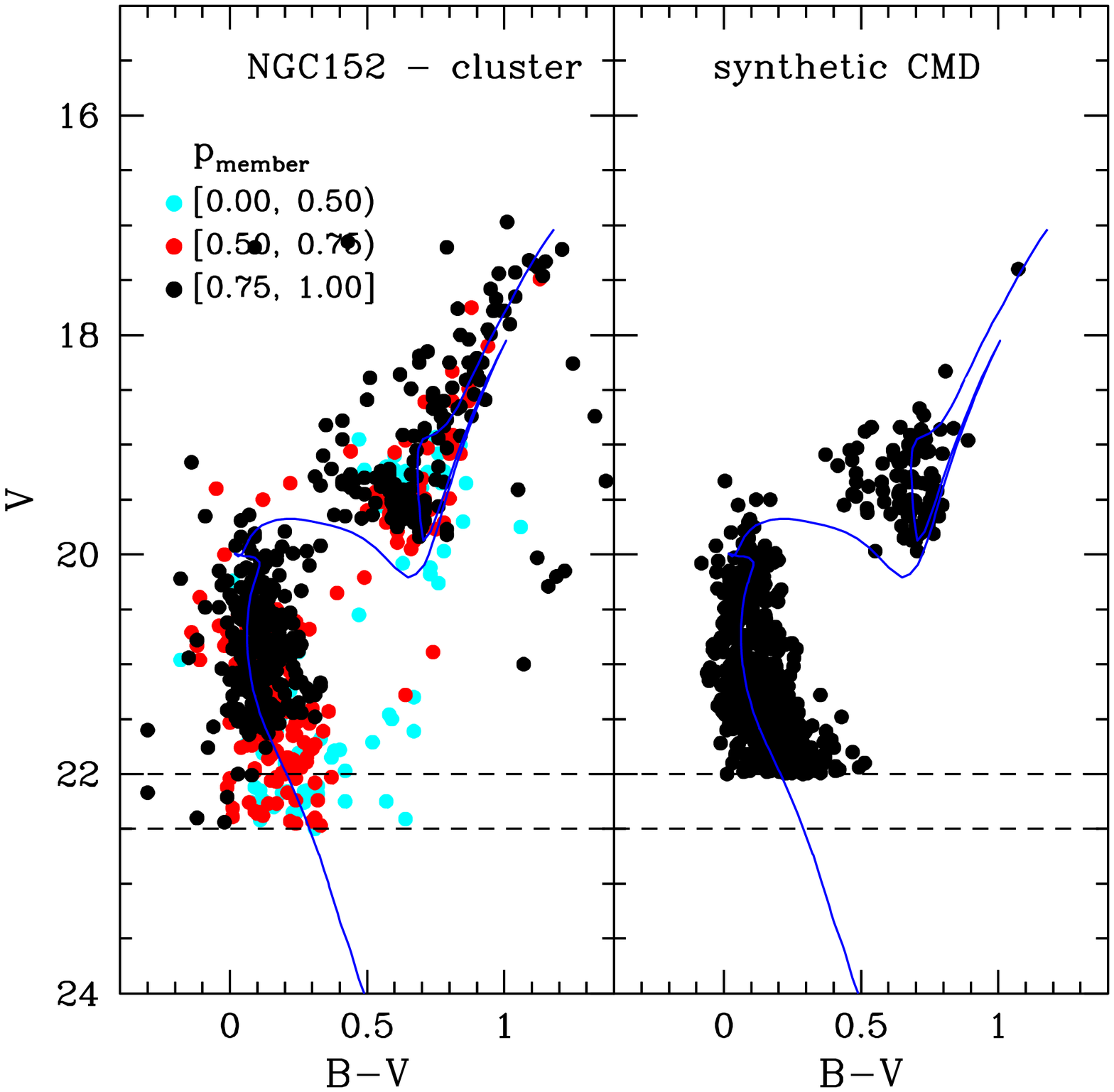}
   \includegraphics[width=0.32\textwidth]{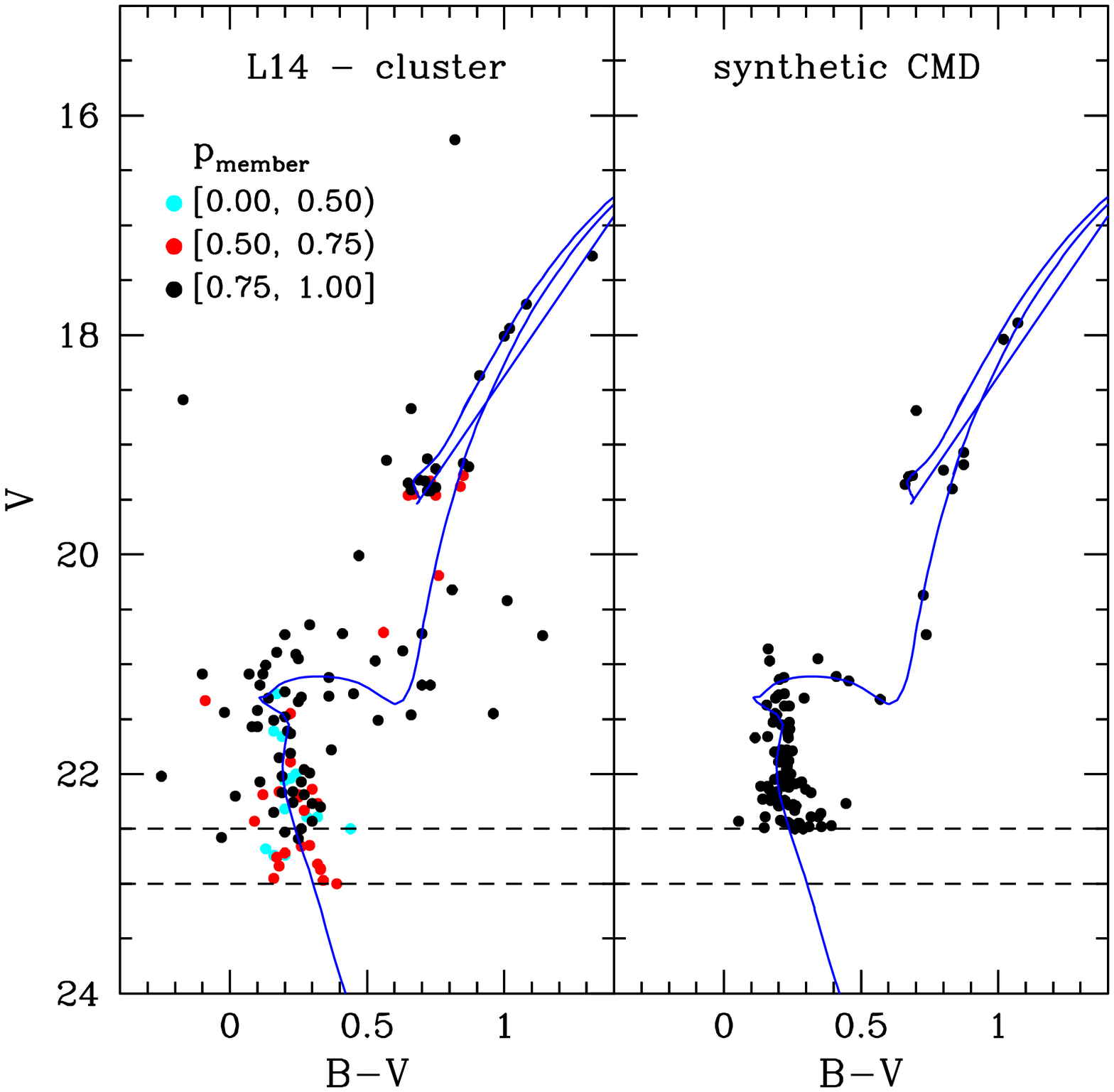}
   \includegraphics[width=0.32\textwidth]{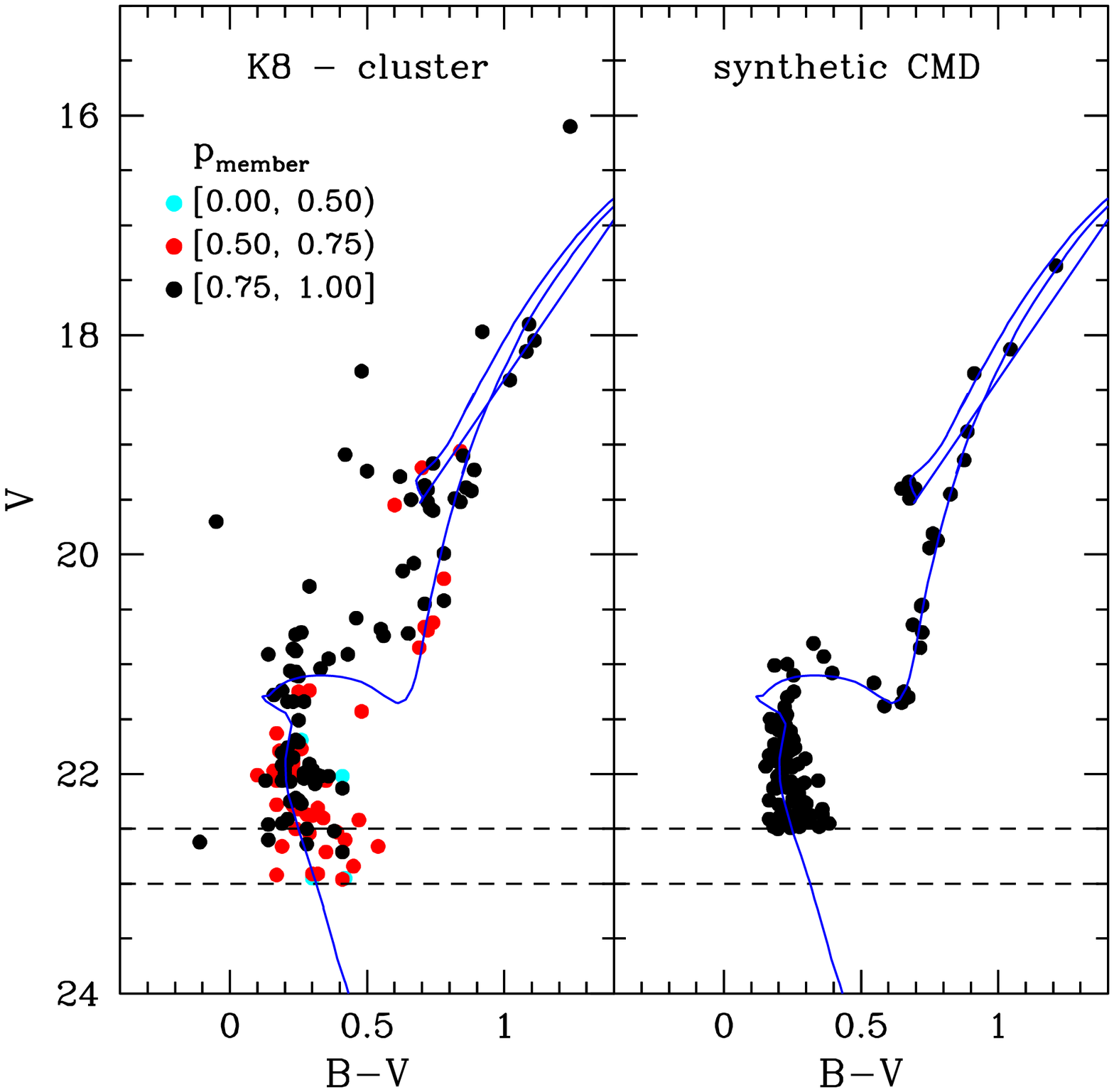}
   \includegraphics[width=0.32\textwidth]{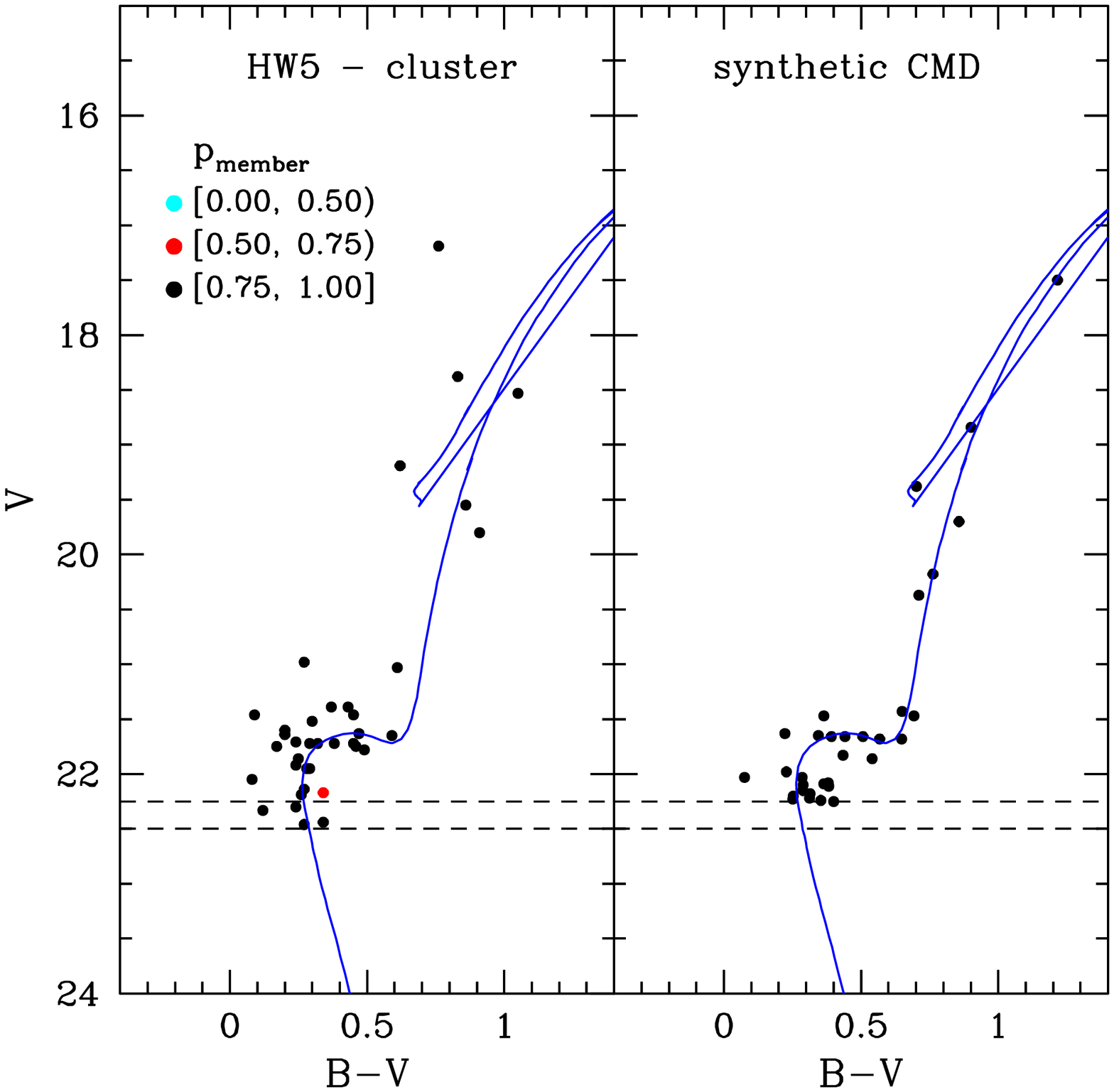}
   \includegraphics[width=0.32\textwidth]{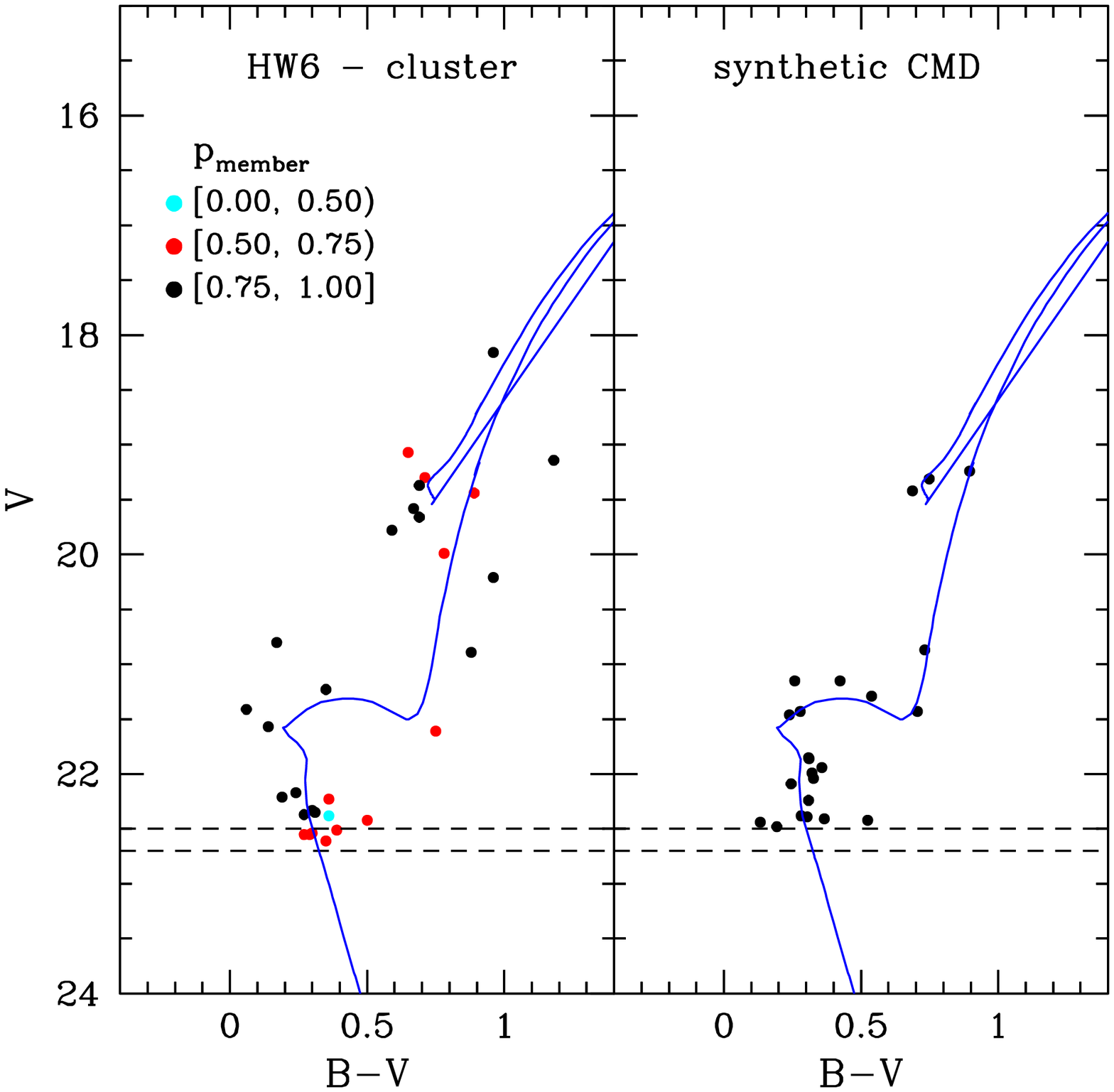}
   \includegraphics[width=0.32\textwidth]{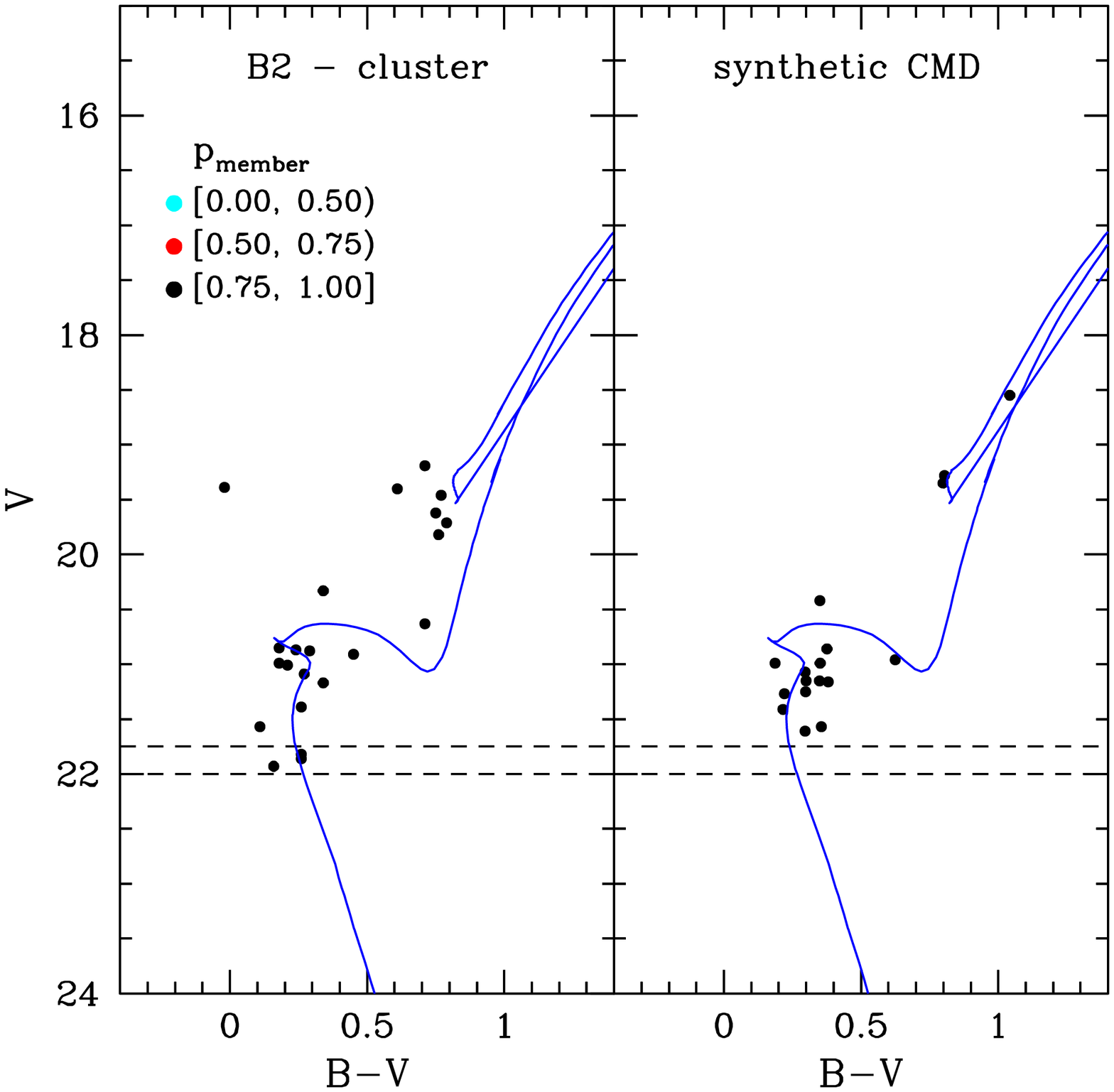}
   \includegraphics[width=0.32\textwidth]{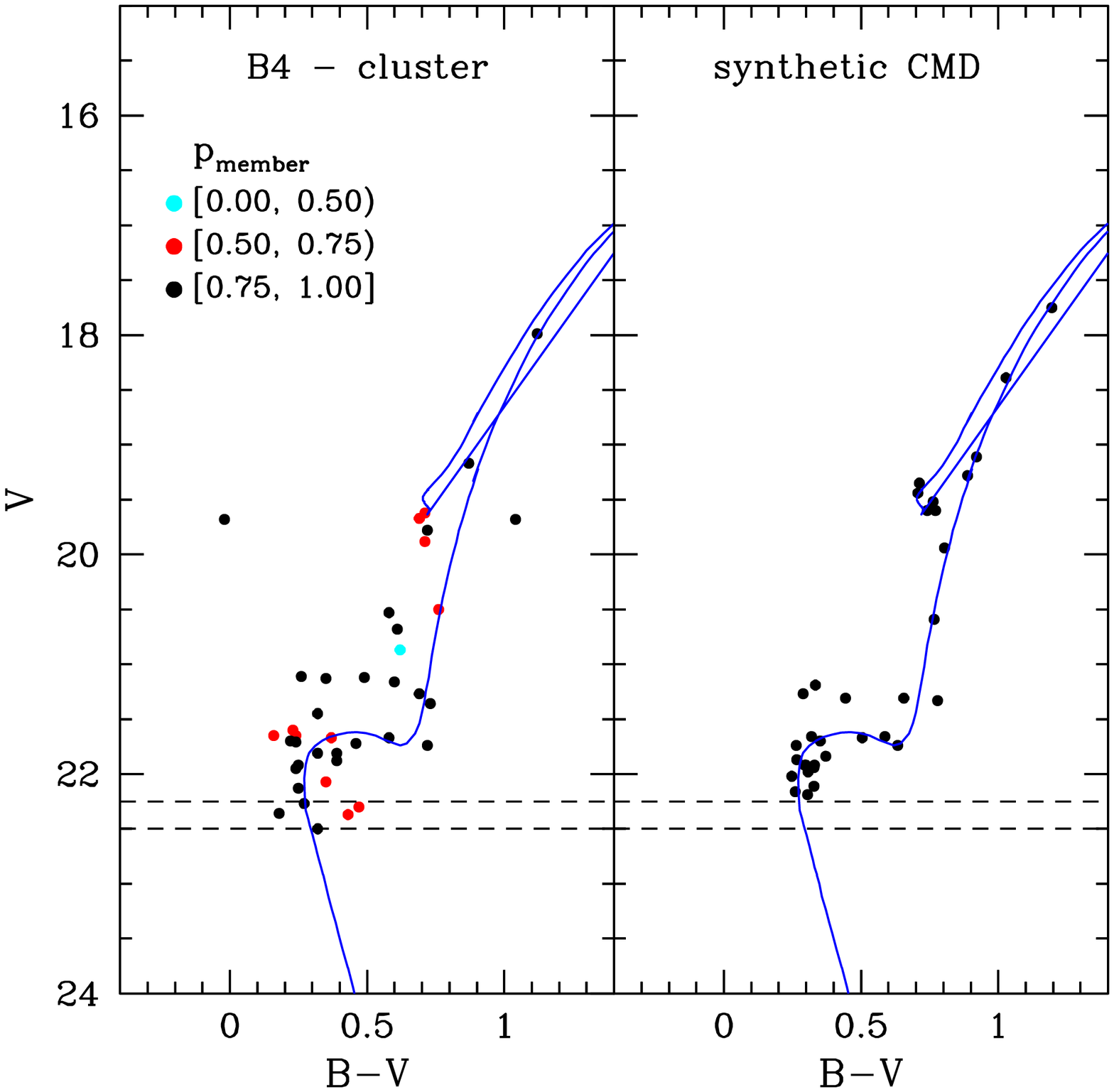}
   \includegraphics[width=0.32\textwidth]{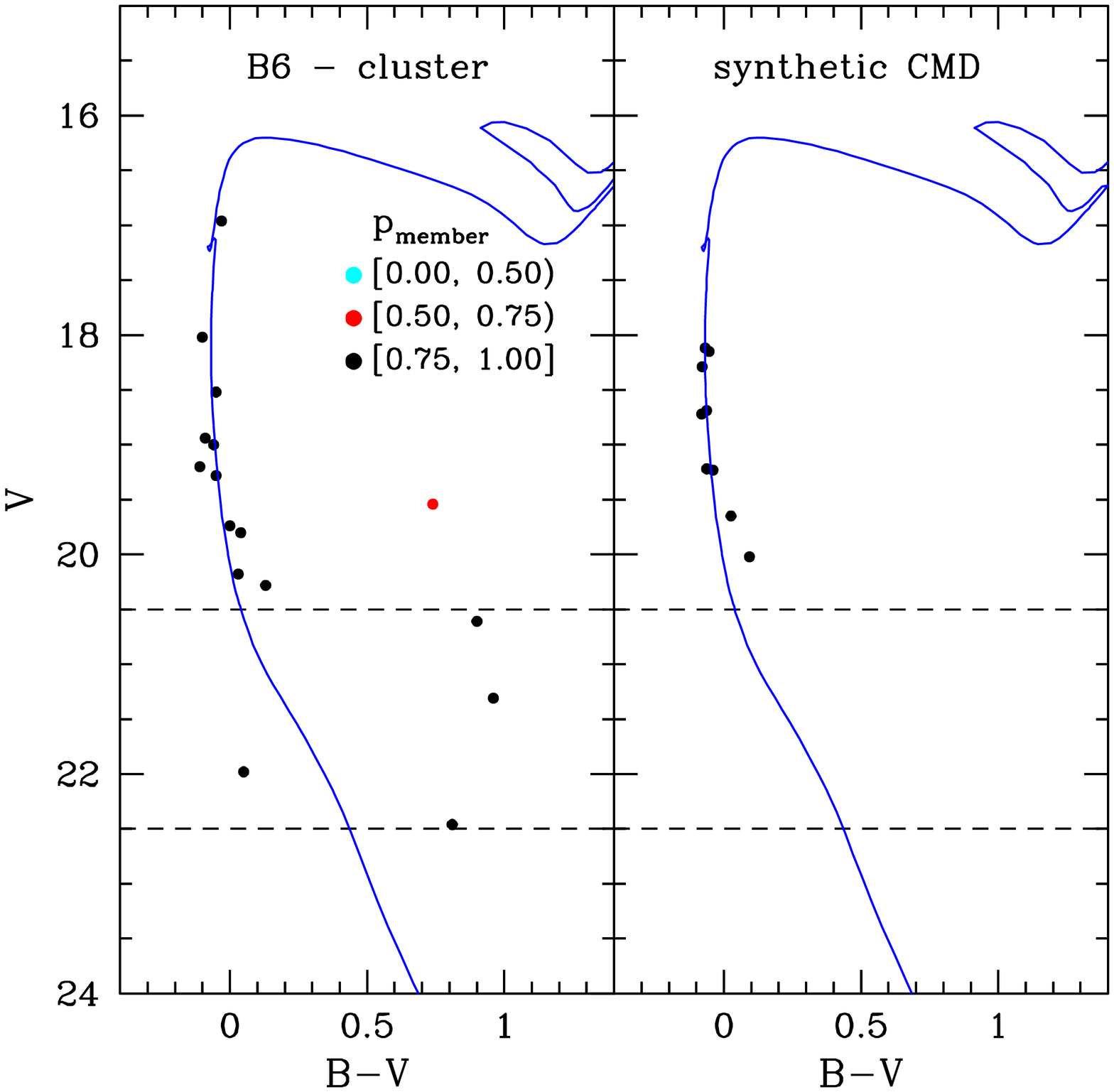}
   \includegraphics[width=0.32\textwidth]{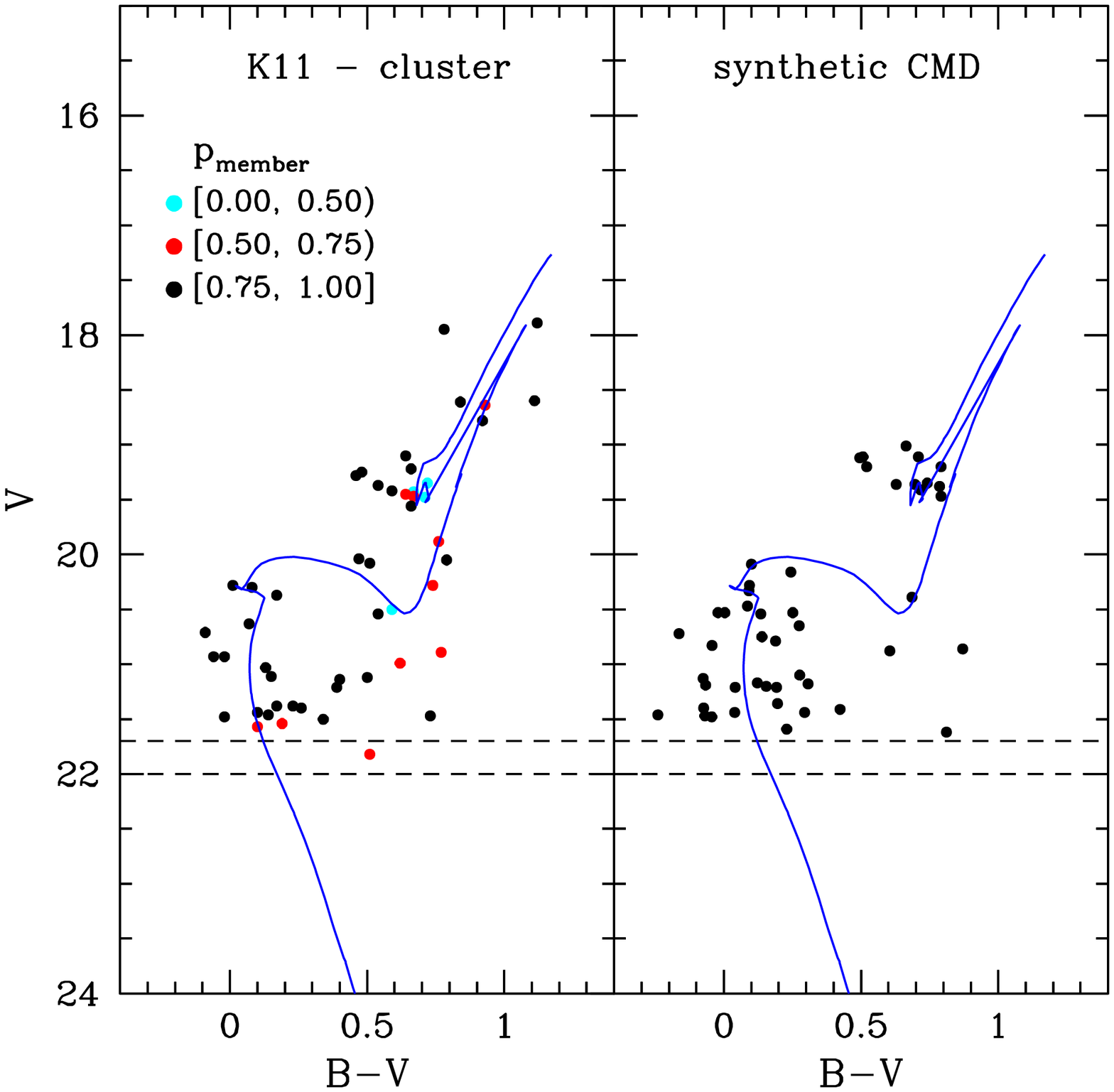}
   \caption{Best isochrone fittings for all clusters. {\it Left panels}: cluster stars according to the membership probabilities ($p_{\rm{member}}$).
{\it Right panels}: synthetic CMD-generated parameters found for the best solution. The number of points is equal to the observed CMDs within the same magnitude limits. The horizontal dashed lines correspond to the magnitude limits used to compute the likelihood (brighter mag) and $p_{\rm{member}}$ (fainter mag). 
The solid blue line correspond to the PARSEC isochrones with the parameters 
found in Table \ref{tab_bestfit}.}
\label{fig_bestfit}
   \end{figure*}

We present the observed CMDs for the clusters and the respective
synthetic CMD based on the best-fit parameters in
Fig. \ref{fig_bestfit}. To present a cleaned CMD for each
cluster, we statistically removed stars in accordance to their
probability of being a non-cluster member ($1- p_{\rm{member}}$) and the
expected number of field stars within $R_{\rm clus}$. The same
isochrone is overplotted in the panels of observed and synthetic CMDs
to guide the eye. For all nine clusters the similarity between
synthetic and observed CMDs is reflected in the quality of the derived
parameters shown in Table \ref{tab_bestfit}. We note that
all CMDs present MSTO and RC, which is required to derive the parameters. The only
exception is the young cluster Br\"uck~6, where its well-defined
main sequence constrains the fit. Kron~11 is the only cluster observed
in the night with the poorer seeing, but still its CMD presents all
features, in particular a clear MSTO and RC.

\begin{table*}[!htb]
\caption{Physical parameters determined in this work. Columns list cluster name, age, metallicity (assuming
  Z$_{\odot}$=0.0152, \citealp{caffau+11}), distance modulus,
  distance, reddening, and semi-major axis corresponding to the
  distance of the cluster to the centre of the SMC as done in Paper I.}
\label{tab_bestfit}
\centering
\renewcommand{\arraystretch}{1.5}
\begin{tabular}{lcccccc}
\hline \hline
Name & Age(Gyr)  & [Fe/H] & (m-M)$_{0}$ & d(kpc) &
E(B-V)  & $a$($^\circ$) \\
\hline
NGC\,152       &   1.23$\pm$0.07 & -0.87$\pm$0.07 &  18.89$\pm$0.10  &  60.0$\pm$2.9   &   0.03$\pm$0.01  & 2.0   \\
Br\"uck\,6     &   0.13$\pm$0.04 & -0.04$\pm$0.06 &  18.88$\pm$0.19  &  60.0$\pm$5.1   &   0.06$\pm$0.03  & 2.3   \\
Kron\,11       &   1.47$\pm$0.11 & -0.78$\pm$0.19 &  19.11$\pm$0.14  &  66.5$\pm$4.1   &   0.02$\pm$0.02  & 2.3   \\
Kron\,8        &   2.94$\pm$0.31 & -1.12$\pm$0.15 &  19.22$\pm$0.07  &  69.8$\pm$2.3   &   0.04$\pm$0.03  & 2.4   \\
HW\,6          &    3.2$\pm$0.9  & -1.32$\pm$0.28 &  19.07$\pm$0.12  &  65.2$\pm$3.6   &   0.08$\pm$0.05  & 2.5   \\
Lindsay\,14    &    2.8$\pm$0.4  & -1.14$\pm$0.11 &  19.24$\pm$0.05  &  70.6$\pm$1.6   &   0.03$\pm$0.02  & 2.6   \\
Br\"uck\,2     &    1.8$\pm$0.7  &  -1.0$\pm$0.5  &  18.91$\pm$0.18  &  60.8$\pm$4.9   &   0.11$\pm$0.05  & 2.9   \\
Br\"uck\,4     &    3.8$\pm$0.6  & -1.19$\pm$0.24 &  19.11$\pm$0.13  &  66.6$\pm$3.7   &   0.05$\pm$0.04  & 3.0   \\
HW\,5          &    4.3$\pm$0.9  & -1.28$\pm$0.32 &  19.15$\pm$0.10  &  67.7$\pm$3.0   &   0.03$\pm$0.03  & 3.1   \\
\hline
\end{tabular}
 \end{table*}

The calibration cluster NGC~152 is 1.23$\pm$0.07~Gyr old, with [Fe/H] = -0.87$\pm$0.07 from our
fitting procedure. Correnti et al. (2016, in prep.) detected an extended the main-sequence turnoff for this 
cluster assuming the same metallicity [Fe/H] = -0.6 and ages = 1.25, 1.4, and 1.6 Gyr. Our derived age agrees well with their findings, but our metallicity is more metal-poor. \cite{rich+00}
derived 1.3$<$age$<$2.0~Gyr assuming a metallicity of [Fe/H] = -0.7. From comparing our CMD to that of 
Rich et al., it is possible to see that their CMD is redder by 0.1-0.15 mag, which allows them to
fit with a fainter turnoff by about 0.3-0.4 mag and hence to obtain an older age. Because of the possible problems
with the magnitude system conversion by Rich et al. that we mentioned before and also based
on the compatibility of our ages with those from Correnti et al., we can say that all ages derived in
this work and in Paper I are reliable. Neither Correnti et al. nor Rich et al. used
a statistical fitting for the metallicities as we did to derive photometric abundances, therefore we can use them as
references. In Paper I, however, we compared our metallicity derivation for HW~40 with that
derived using CaII triplet of individual stars by \cite{parisi+15}, and they were compatible. Therefore
we can assume that our age and metallicity scales are satisfactory.

%

\section{Age and metallicity gradients}

Radial distributions of age and metallicity for the SMC seem to be
tangled. With a significant dispersion, \cite{parisi+14}
have found an increasing trend on age distribution for $a$
$\lesssim$ 4.5$^{\circ}$ and decreasing above that. The opposite
pattern is found for metallicity \citep{parisi+15}, that is, values
decrease until $a < 4.5^{\circ}$ and increase beyond this. In Paper I
(Fig. 10) these trends were already detectable (see also
\citealp{dobbie+14}).
Figure \ref{agemetgrad} shows age and [Fe/H] vs. distance from
the SMC centre in terms of semi-major axis $a$.
\cite{parisi+15} explained the
V shape by splitting their sample into two parts on $a$ = 4$^{\circ}$ and
averaging the metallicities in each bin in $a$. From this they found a very low
gradient for clusters located at $a$ $<$ 4$^{\circ}$, in agreement with
other findings (e.g. \citealp{dobbie+14}). However, they were
unable to justify the behaviour of the metallicity distribution above $a$ $>$
4$^{\circ}$. We proposed a different explanation for the radial
distribution in age and metallicity in Paper I and we endorse it
here. 

To discuss the gradients based on a large sample, we compiled ages
and metallicities of SMC clusters available in the literature. We took
weighted averages of the parameters, where the weights were attributed
depending on the technique used, as follows. Ages from resolved
photometry received weight 5, and those from integrated
photometry or spectroscopy received weight 2. Metallicities derived using
resolved spectroscopy received weight 5, those from resolved
photometry received weight 3, and integrated spectroscopy received 2.
This is not a complete catalogue from the literature, but it covers the
most relevant works with large homogeneous samples:
\cite{dias+10,parisi+15,parisi+09,parisi+14,piatti+07a,piatti+08,piatti+11,piatti11a,piatti12b,glatt+08b,glatt+08a,glatt+10,DCH98,MSF98,rafelski+05}. Based
on the catalogue of \cite{bica+08a}, we selected 637 SMC clusters. We
found ages for 346 of them and metallicities for 58 of them. The
catalogue will be published separately.

   \begin{figure*}[!htb]
   \centering
   \includegraphics[height=\columnwidth,angle=-90]{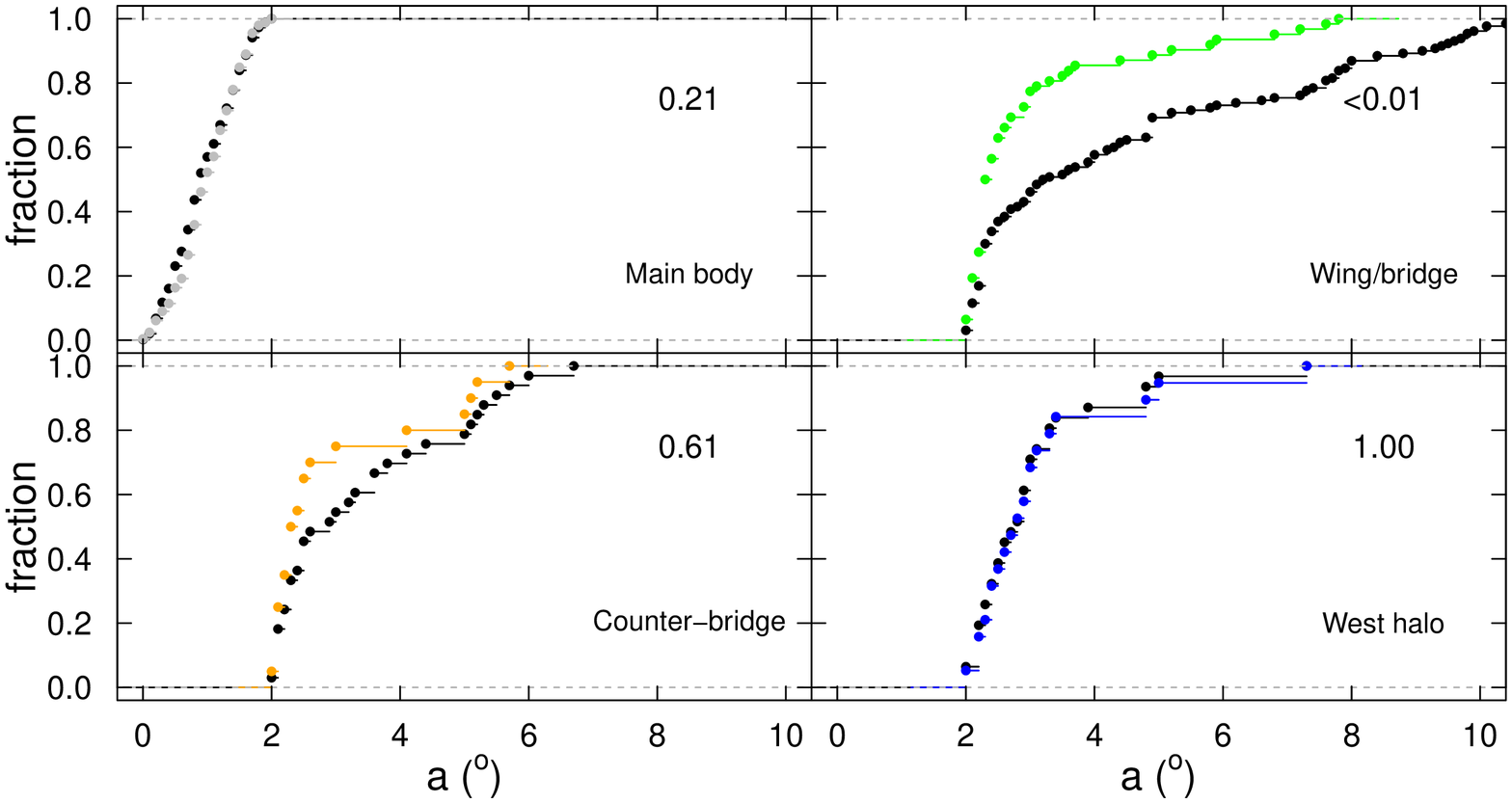}
      \includegraphics[height=\columnwidth,angle=-90]{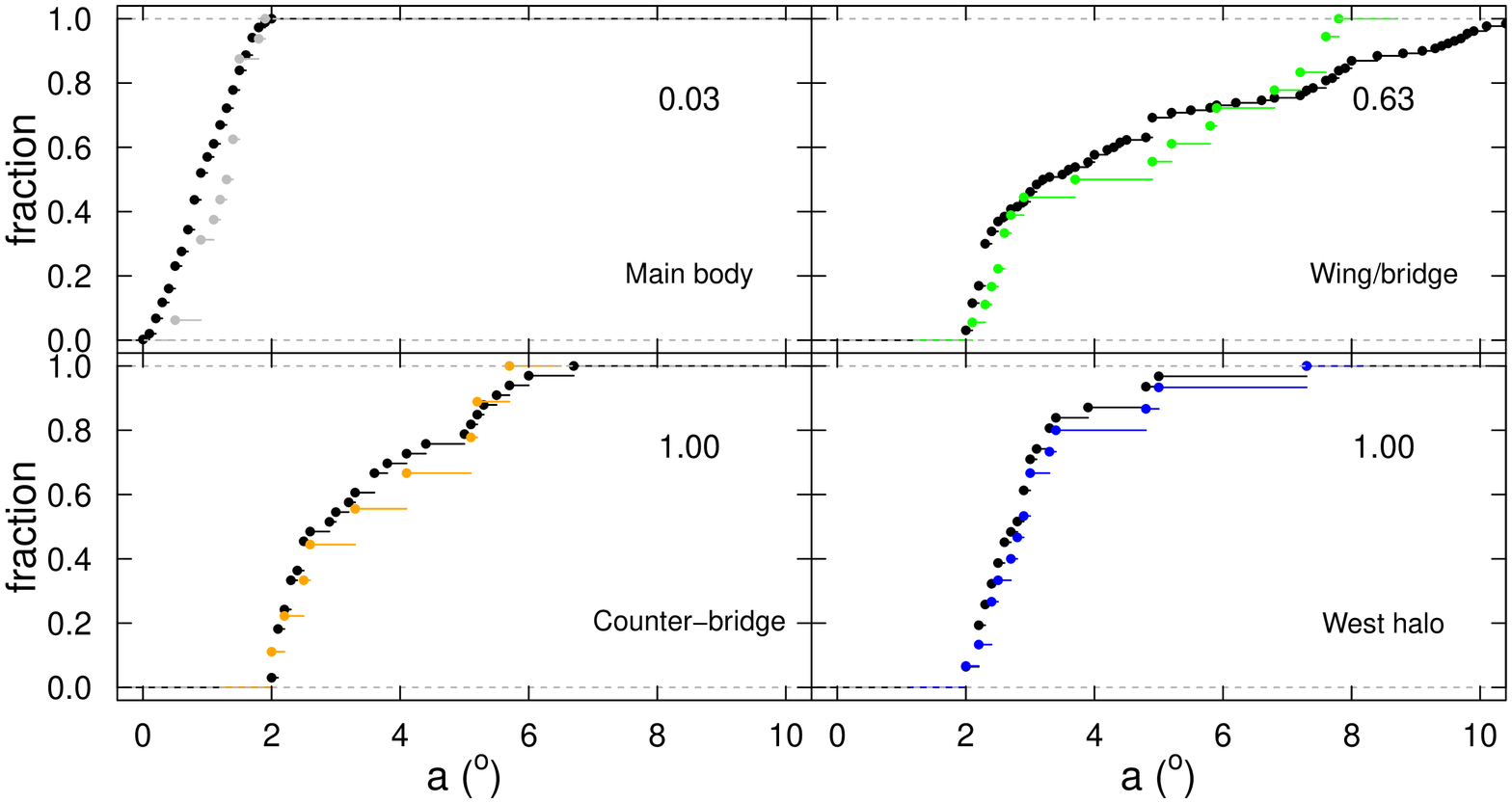}
   \caption{Cumulative distributions of distance to the centre {\it a} for each
   group of the SMC star cluster population: main body, wing
and bridge,
   counter-bridge, and west halo. Left panels show ages, right panels
   metallicities available in the literature compilation described in the text. Black
   curves are the distributions of all clusters, and the colour curves represent the
   clusters with available parameters. The numbers in the panels are the p-values
   obtained by applying the Kolmogorov-Smirnov test to the black and coloured
   curves.}
 \label{sampling}
   \end{figure*}

The literature compilation of ages and metallicities for SMC clusters is
representative for most of the cases. In Fig. \ref{sampling} we show the cumulative
distributions of distance to the centre {\it a} for each group of the SMC
star cluster population. The reference curves in black were derived from the 637
clusters from the \cite{bica+08a} catalogue split into the four regions. The coloured curves
represent ages and metallicities available for these clusters that are used in 
Fig. \ref{agemetgrad} and the following discussions. We ran 
Kolmogorov-Smirnov tests to check whether the samples are representative of their 
respective SMC component. For the main body, the age distribution is representative,
but the metallicity distribution is missing for some clusters at around $a \sim 1^{\circ}$. 
Wing and bridge clusters have a non-uniform distribution of ages, with more information
available for inner clusters; the metallicity distribution is well covered. The counter-bridge and west halo 
samples are well spread over at all possible distances from the
centre, and age and metallicity distributions represent this group well.

   \begin{figure*}[!htb]
   \centering
   \includegraphics[height=\columnwidth,angle=-90]{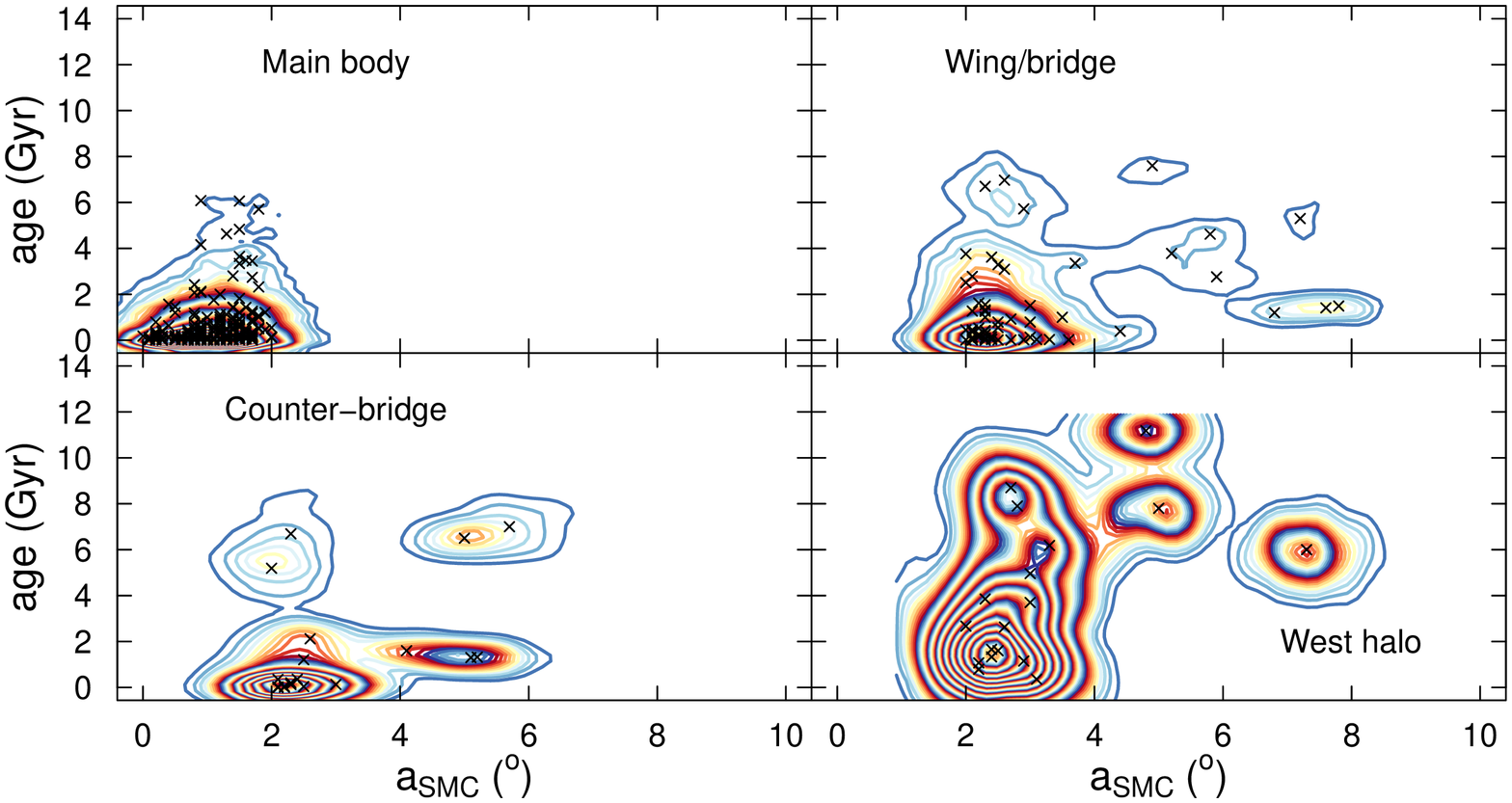}
      \includegraphics[height=\columnwidth,angle=-90]{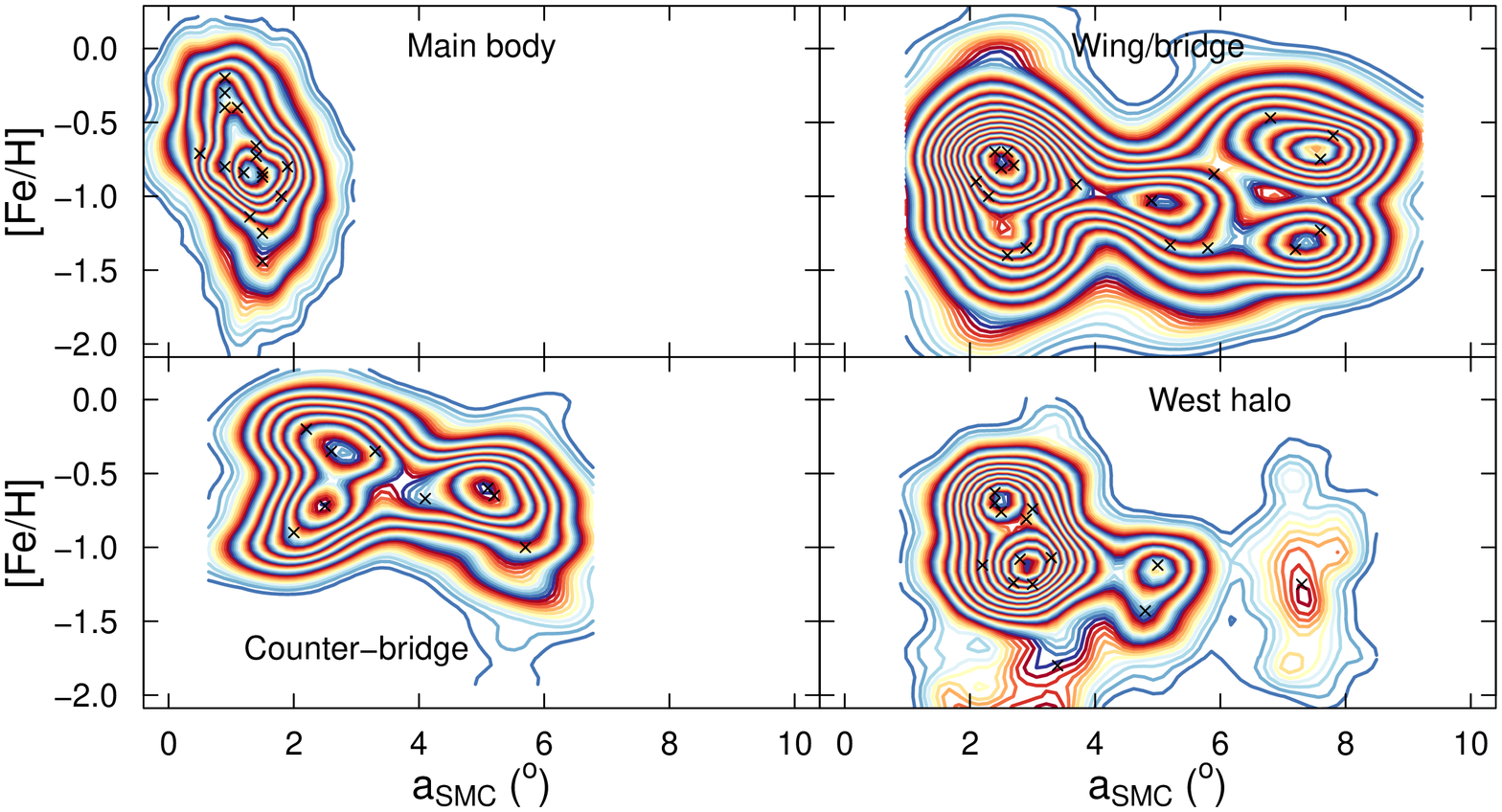}
   \caption{Age and metallicity radial distributions in terms of the
     semi-major axis {\it a} in degrees, as done in Paper I. Age distributions are shown separately for the four components in the left panel, and the equivalent for metallicity is displayed in the right panel. Ages and metallicities are weighted averages from literature data as explained in the text. No results from this work are included here. Contours are the density maps considering a two-dimensional Gaussian around each point using their uncertainties. See text for details. The left panel shows the radial distribution.}
 \label{agemetgrad}
   \end{figure*}

In Fig. \ref{agemetgrad} we show the average
age and metallicity for the clusters from the
literature as described above,
indicating the different regions in the galaxy: main body, wing
and bridge,
counter-bridge, and west halo following the definitions of
Sect. \ref{selection}. 
For each point we assumed an uncertainty in {\it a} of 0.2$^{\circ}$, and 
uncertainties in age and metallicity from the literature. When uncertainties
in metallicity were not available, we assumed 10\% of [Fe/H]. These uncertainties
were used to generate two-dimensional Gaussian distributions to populate the 
plots and trace the density curves shown in Fig. \ref{agemetgrad}. 
The dispersion in the distributions may be caused in part by 
old and metal-poor clusters from the outskirts projected into the direction of the
inner parts of the SMC. Moreover, the absolute value of the gradients can change 
if the distance scale {\it a} takes into account the three-dimensional shape of the SMC;
the slope varies with $sin(i)$, where $i$ is the angle between the line of sight and 
the sky plane. However, if a gradient is detected in the projected distribution, it can be converted into the deprojected distribution if the angle $i$
is known. The three-dimensional distribution is beyond the scope of this work and does
not affect our conclusions here. Therefore we focus on the projected distributions below.

The SMC is
classified as a dwarf irregular galaxy, which means that it is still forming stars.
\cite{glatt+10} showed that the bulk of recent star formation in the SMC occurs
in its central bar and the older clusters are spread across the galaxy. If we assume
that this was always the case for star formation in the SMC, we would expect
age and metallicity gradients, with younger and metal-rich clusters in the innermost
regions of the galaxy. In Fig. \ref{agemetgrad} cluster ages increase with {\it a} in the 
main body, although there is a large group of young clusters from \cite{glatt+10}.
For metallicities the trend is cleaner and shows decreasing [Fe/H] with radial 
distance {\it a}. For the main body we confirm the findings from Glatt et al.
We note that the discussions in Glatt et al. and in this work are based on the
projected distribution of clusters. 

For the wing and brigde, ages increase rapidly between $2^{\circ} < a < 3^{\circ}$ and 
there are sparse clusters beyond this distance that seem to have decreasing
ages. Metallicities present larger uncertainties, nevertheless the distribution
shows a valley at around $a \approx 4.5^{\circ}$. The combination of both 
distributions indicates that traces of the V-shape distribution described by 
\cite{parisi+15} can be detected even in non-homogeneous average literature 
data. This deserves further investigation.

The counter-bridge has only a handful of clusters with available
parameters, and
it seems to be a tidal counterpart of the wing and bridge \citep{beslaPHD,diaz+12}.
Therefore its composition might be more complex. The age distribution seems 
to be increasing slowly with {\it a} but with a double trend. On the other hand,
the metallicity distribution is monotonic, but with a large spread. More data are needed
to confirm whether the distribution is double or not, and to
confirm our first classification of
counter-bridge clusters.

Finally, west halo clusters have monotonically increasing ages and monotonically
decreasing metallicities until $a < 6^{\circ}$. Beyond this lies only the peculiar cluster
AM-3, which we discuss in Sect. \ref{sec:am3}.

   \begin{figure}[!htb]
   \centering
   \includegraphics[width=\columnwidth]{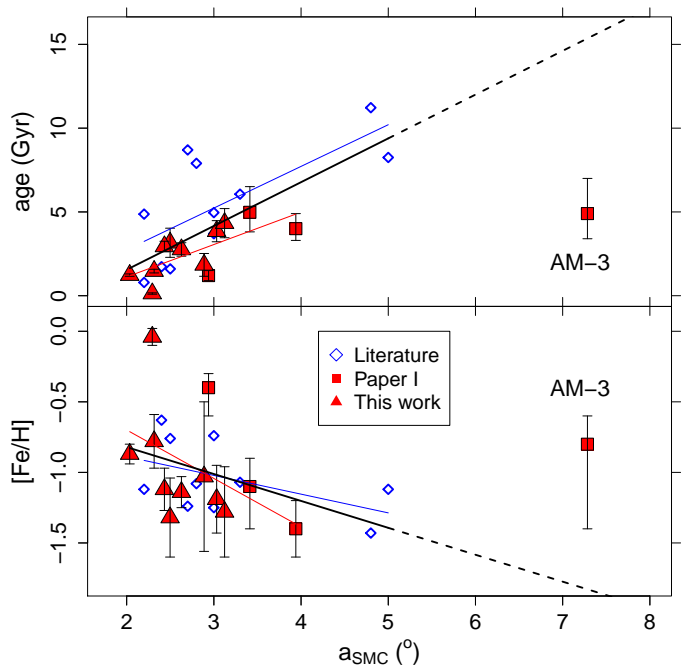}
   \caption{Same as Fig. \ref{agemetgrad}, but only with WH
     clusters. Blue diamonds are the same literature values as in the
     previous figure, excluding the clusters in common with our sample
     to avoid duplicates. Red squares are the results from Paper I for
   the WH, except AM-3. Red triangles are the results from this
   work. Blue lines are the linear fits to the literature points, red
   lines are the linear fits to our data points, and black lines are
   the linear fits to all points together. The parameters of these
   fits are presented in Table \ref{tab:WHgrad}.}
 \label{fig:WHgrad}
   \end{figure}

To these distributions from literature, we now add our
results for WH clusters in
Fig. \ref{fig:WHgrad}.
Blue points are
the same as those from Fig. \ref{agemetgrad}, and red points represent
the clusters analysed in Paper I and in this work. For duplicate
clusters we removed the points from literature from the plot. Linear
fits were made to the literature points (blue line), to our results
(red line), and to both samples together (black line). The parameters
from the fits are detailed in Table \ref{tab:WHgrad}. The age gradient
for WH clusters is found to be -3.7$\pm$1.8~Gyr/$^{\circ}$, with a
coefficient of determination $r^2 = 0.5$. For metallicity, the
gradient is -0.19$\pm$0.09~dex/$^{\circ}$ with $r^2 = 0.2$. The
coefficients $r^2$ indicate that a linear fit is a good representation
of the data distribution. The gradients are different from zero by more
than 2-$\sigma$. Therefore the radial gradients of age and metallicity
for WH clusters are confirmed. \cite{dobbie+14} derived a metallicity
gradient of -0.075$\pm$0.011~dex/$^{\circ}$ for field stars in the SMC
for all stars internal to $r < 5^{\circ}$ from the centre. This region
encompasses most of the star clusters, and we have demonstrated that it
does not make sense to analyse the metallicity gradient of all
clusters together. This explains why we derived a steeper gradient
than that found by Dobbie et al.

\begin{table}[!htb]
\caption{Age and metallicity gradients for WH clusters. Results from
  the fits presented in Fig. \ref{fig:WHgrad}. Columns named
  ``literature'', ``our results'', and ``all WH'' refer to the blue,
  red, and black lines in Fig. \ref{fig:WHgrad}.}
\label{tab:WHgrad}
\centering
\begin{tabular}{lccc}
\hline \hline
\noalign{\smallskip}
Coefficient & Literature & Our results & All WH \\
\noalign{\smallskip}
\hline
\noalign{\smallskip}
\multicolumn{4}{c}{Age gradients}\\
\noalign{\smallskip}
\hline
\noalign{\smallskip}
linear (Gyr) & -2.2$\pm$2.6 & -2.7$\pm$1.8 & -3.7$\pm$1.8\\
angular (Gyr/$^{\circ}$) & 2.5$\pm$0.8 & 1.9$\pm$0.6 & 2.6$\pm$0.6\\
r$^2$& 0.5 & 0.5 & 0.5\\
$\sigma$ (Gyr) & 2.5 & 1.1 & 2.1\\
\noalign{\smallskip}
\hline
\noalign{\smallskip}
\multicolumn{4}{c}{Metallicity gradients}\\
\noalign{\smallskip}
\hline
\noalign{\smallskip}
linear (dex) & -0.62$\pm$0.27 & -0.01$\pm$0.59 & -0.44$\pm$0.27\\
angular (dex/$^{\circ}$) & -0.13$\pm$0.08 & -0.34$\pm$0.21 & -0.19$\pm$0.09\\
r$^2$& 0.3 & 0.2 & 0.2\\
$\sigma$ (dex) & 0.23 & 0.37 & 0.31\\
\noalign{\smallskip}
\hline
\end{tabular}
 \end{table}

\begin{figure}[!htb]
   \centering
   \includegraphics[width=\columnwidth]{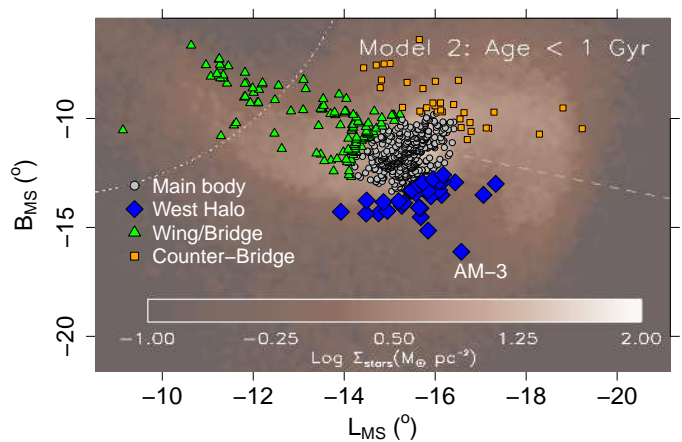}
   \caption{Same as Fig. \ref{2Dpos}, but using Magellanic stream
     coordinates as defined by \cite{NMB08}. Colours are the same
   as in Fig. \ref{2Dpos}, and the clusters are overplotted on a model by
   \cite{beslaPHD} (adapted figure from Besla).}
 \label{2D-models}
   \end{figure}

\cite{beslaPHD} described the final distribution of gas and stars of
the SMC after interaction with the LMC, following the scenario where the two
galaxies are on their first close encounter with the Milky Way. In their
Fig. 7.5 it becomes clear that stars in the external regions of the
SMC present two large structures similar to spiral arms, one in the
region of the wing and bridge, and the other starting in the region of the
counter-bridge, which extends to the region of the WH. Although
the aim of the simulations was to derive the general evolution of the
Magellanic system and not to determine the exact final shape of the
SMC, Fig. \ref{2D-models} shows that the distribution of the clusters
qualitatively matches the star distributions from the simulation
well.
We note that the stars represented by the simulation in the figure are
younger than 1~Gyr, but they represent the consequences of the tidal
forces from the LMC-SMC interaction, which may have stripped the star
clusters of all ages that were once in the main body. In this process,
the gas movement could have formed some star clusters as well. 

The WH is not part of the two arms, but corresponds to a fading
stellar distribution outwards of the main body that appears to be a
consequence of the tidal forces. The same is true for
some counter-bridge clusters at the top of the plot in
Fig. \ref{2D-models}. This could indicate that the distribution of WH
clusters is a sparse counterpart, a slice, of the main body in the outer region
of the SMC that was stripped by the tidal force of the LMC.
This is also indicated by the age and metallicity trends.
The observations show that the trends of the main
body and WH are similar, but the WH group is displaced by $a =
2^{\circ}$ (Fig. \ref{agemetgrad}). More specifically, main-body
clusters reach older ages of up to about 6~Gyr at $a=2^{\circ}$. WH
clusters at this distance are younger with ages around 1~Gyr and
increase with distance from the SMC centre. The metallicity of main-body clusters decreases to [Fe/H]~$\approx -1.4$ at $a=2^{\circ}$
and WH clusters have [Fe/H]~$\approx -0.8$ decreasing with
distance. We cannot rule out other scenarios to explain these breaks
in age and metallicity trend between main body and west
halo. Nevertheless, we were able to confirm and derive the gradients in the WH
based on homogeneous results. 
In the model of \cite{beslaPHD} the interaction that originated the WH
probably took place at about 100-160 Myr, which would agree with
the strong common peaks in the cluster formation history of the clouds
with OGLE \citep{pietrzynski+00}. Brück~6, with 130$\pm$40~Myr, appears
to be a legacy of that enhancement, while the remaining clusters in
Table \ref{tab_bestfit} possibly show the results of tidal effects.

Moreover, the very old and metal-poor clusters that were
located in the outskirts of the main body before the interaction were
sent to the outer region of the WH group. 
In particular, the most metal-poor globular cluster NGC~121 is located in this region.
 Its metallicity, previously obtained from low-resolution spectra and CMDs, 
estimated to be of [Fe/H]$\sim$$-$1.5 \citep{dias+10}, was recently confirmed
from analysis of spectra obtained with FLAMES at the VLT by Mucciarelli et
al. (priv. comm.) as [Fe/H]$\sim$-1.4. This metallicity is 
compatible with the mean metallicity of field giants of [Fe/H]$\sim$-0.9/-1.0, as
analysed by \cite{mucciarelli+14}.
We do not exclude
the possibility of other scenarios to explain the gradients when
splitting the clusters into the groups mentioned above. Dedicated
models
need to be discussed on the WH to address whether it is a tidal
structure and to describe the processes involved.

\begin{figure}[!htb]
   \centering
   \includegraphics[width=\columnwidth]{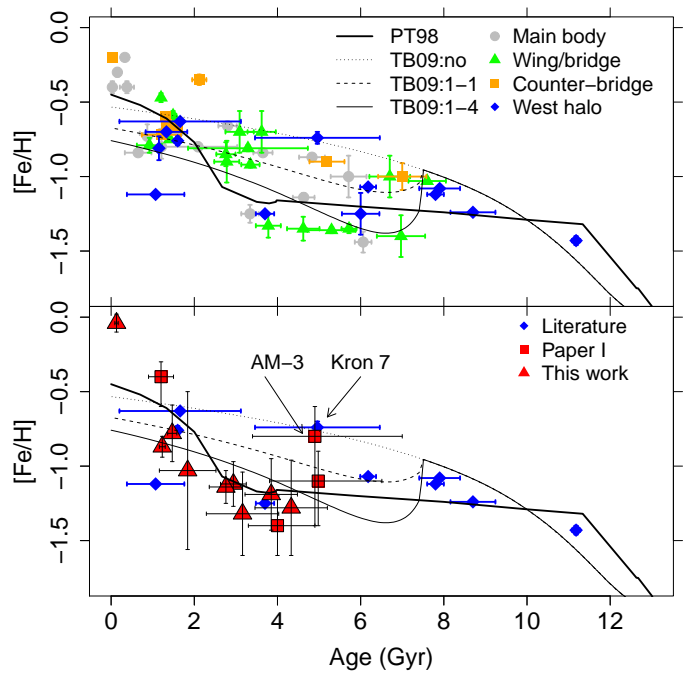}
   \caption{Age-metallicity relation for SMC clusters. The upper panel
     shows the literature average, as in
     Fig. \ref{agemetgrad}. The lower panel shows the WH clusters from
the     literature, Paper I, and this work. In both panels we
     overplot the burst chemical evolution model of \cite{PT98} and
     the merger models of \cite{tsujimoto+09}, which correspond to no-merger
and      merger of progenitors with mass ratios of 1-1 and 1-4.}
 \label{fig:amr}
   \end{figure}

Figure \ref{fig:amr} shows the age-metallicity relation of SMC clusters with 
particular attention on the WH objects. We overplot the burst model of \cite{PT98} 
and the major merger model of \cite{tsujimoto+09} for reference. 
In the upper panel we show average parameters from the literature using colour and 
symbols to identify WH, main body, wing and bridge, and counter-bridge clusters, as 
in Fig. \ref{agemetgrad}. In general, 
the clusters tend to follow the two chemical evolution models, but the dispersion in metallicity at a 
given age is as high as $\sim0$.5~dex. When we select only WH objects (avoiding duplications as in Fig. \ref{fig:WHgrad}), the clusters in the bottom panel show a lower dispersion in [Fe/H] for a given age around the model of \cite{PT98}. 
There are two exceptions: AM-3 and Kron~7. The first is probably not from the WH, 
but from the counter-bridge, as we discuss in the next section. The second has
parameters derived mostly from integrated light, which are less accurate because of degeneracies. 
The only study based on six individual stars has been published
by \cite{DCH98}, who derived equivalent widths 
for CaII triplet lines. If we adopt the updated calibration of \cite{saviane+12}, the metallicity obtained 
for Kron~7 is [Fe/H] = -0.9 instead of the average $<$[Fe/H]$>$ = -0.7 shown in the plot. This 
indicates that in fact WH clusters follow the burst model of \cite{PT98} with a lower dispersion in metallicity. An alternative scenario would be the major merger 1-1 proposed by \cite{tsujimoto+09}
but 5~Gyr, not 7.5~Gyr ago. This would follow the argument
of the model of  \cite{PT98} with a burst
at about 4~Gyr ago.

A third interesting cluster is the most metal-rich and youngest of the sample, Br\"uck~6, which 
was pointed out above as a possible product of the formation of the WH at the same epoch of
the formation of the Magellanic bridge. 
Finally, we note that the dispersion in metallicity is not clearly explained by
the radial distribution alone, since it mixes the external groups and the dispersion
remains. \cite{parisi+15} compared their data with five
different chemical evolution models and also split the clusters in
distance bins of a = 2$^{\circ}$, and the dispersion remained. 
In summary, the age-metallicity relation appears to be different for
different groups, as we also stated for age and metallicity gradients.
More investigations on the other regions should therefore be carried out to solve this open question.

%

\subsection{AM-3: west halo or counter-bridge?}
\label{sec:am3}

Based on Fig. \ref{2Dpos}, we classified AM-3 as a WH cluster because of its
position in the sky plane. It is the outermost WH cluster at distance
$a=7.3^{\circ}$ from the SMC centre, while all other WH clusters are
located within $a < 5^{\circ}$. This cluster does not follow the
gradients in age and metallicity, as shown in
Fig. \ref{agemetgrad}. \cite{dacosta99} has pointed out that this
cluster was the most distant form the SMC centre and still within the
limits of the SMC field star and HI distribution. Recently,
\cite{beslaPHD} and \cite{diaz+12} have shown evidence for a stellar
counterpart of the counter-bridge. In particular, \cite{beslaPHD}
showed the predicted stellar distribution of the SMC after the tidal
interaction with the LMC. We overplot the SMC clusters in the
predicted stellar distribution and find a qualitatively good
agreement of the main body, wing and bridge, counter-bridge, and west halo
clusters. AM-3 is located in the tail of the counter-bridge structure
predicted by the model (see Fig. \ref{2D-models}). AM-3 age and metallicity seem to
agree well with the extrapolation of the radial trends shown in
Fig. \ref{agemetgrad}. Moreover, AM-3 does not follow the general trend
of the chemical evolution of the WH, as revealed by the age-metallicity
relation in Fig. \ref{fig:amr}. Instead it seems to follow a trend together with 
counter-bridge clusters.
Therefore it is possible that AM-3 is a
counter-bridge cluster that was stripped to its current position
following the counter-bridge arm described by the model.

%

\section{Conclusions}

We presented photometric parameters  age, metallicity, distance, and reddening
 for nine clusters located in the WH of the SMC. Eight
of them were studied for the first time, and NGC~152 was used as a
reference cluster. We proposed to split the clusters in groups according to their 
position in the galaxy for a more enlightening analysis of the SMC history: main
body, wing and bridge, counter-bridge, and WH. We focused on
the last group. A detailed study of WH clusters has confirmed that to study 
the complex star formation and dynamical history of the SMC, it is crucial to
analyse the galaxy region by region. The main results for the WH that led
to this conclusion were as follows.

\begin{itemize}
\item{The age gradient of WH clusters is 2.6$\pm$0.6~Gyr/$^{\circ}$ with a 
negative linear coefficient equal to -3.7$\pm$1.8~Gyr, indicating that the gradient
is not compatible with the main body of the SMC. Moreover, in the transition from the
main body to the WH at $a = 2^{\circ}$, main-body clusters reach 6~Gyr, while WH
have ages around 1-2~Gyr, indicating a discontinuity of radial gradients in cluster ages
from the two groups.}
\item{The metallicity gradient of WH clusters is more subtle than the age gradient, but the
same comparisons are valid. The gradient is -0.19$\pm$0.09~dex with a linear coefficient
equal to -0.44$\pm$0.27~dex, which is lower than the metallicities of the innermost clusters
in the main body. In the boundary between main body and WH at $a = 2^{\circ}$, main-body
clusters extend to [Fe/H] $\sim$ -1.3, while WH clusters are around [Fe/H] $\sim$ -0.8, with
an outlier with solar metallicity close to the main body, Br\"uck~6, which might have formed during
the tidal formation of the WH. There is no continuity in the metallicity gradient from the main body
to the WH.}
\item{The age-metallicity relation of all SMC clusters presents a high dispersion in metallicity
at a given age, but if only WH are plotted, the dispersion is significantly reduced and the
distribution agrees with the burst chemical evolution model of \cite{PT98}.}
\item{The dynamical model from Besla et al. (2011) releases the stellar distribution in the SMC
after tidal interactions with the LMC. Qualitatively, the cluster distribution agrees well with the 
predictions; this is very clear for main body, wing and bridge, and counter-bridge. WH clusters are 
located over a fading stellar distribution in the outer ranges of the main body between the two arms that 
represent the wing and bridge and counter-bridge.}
\item{The cluster AM-3 seems to be an important key to separating WH and counter-bridge
because it is located in the region of the WH, but is far away, and it is also located in the tail of the extended arm shape
of the counter-bridge. It follows the age and metallicity gradient and the age-metallicity relation of
the counter-bridge, not of the WH.}
\item{The cluster Br\"uck~6 seems to be an important witness of the recent epoch of interaction
that created the Magellanic bridge about 100~Myr ago. During the possible tidal disruption that
generated the west halo, Br\"uck~6 would have been formed.}
\end{itemize}

It is crucial to have homogeneous and precise ages and metallicities for SMC clusters in its 
four groups to understand the complexities in the history of our neighbour dwarf irregular 
galaxy. Spectroscopic metallicities and ages from CMDs from future 
observations are highly desired.

%

\begin{acknowledgements}
BB, BD, EB and LK acknowledge partial financial support from
FAPESP, CNPq, CAPES, and the LACEGAL project,
and the CAPES/CNPq for their financial support with
the PROCAD project number 552236/2011-0. BD acknowledges ESO for the
one-year studentship. SO acknowledges financial support of the University of
Padova. BD acknowledges discussions and comments from
G. Besla and C. Parisi. BD acknowledges T. Tsujimoto and G. Besla for kindly
providing their models, and A. Mucciarelli for providing his
preliminary results on NGC~121. LK acknowledges Matteo Correnti and Paul Goudfrooij 
for providing their CMD and results for NGC~152.
\end{acknowledgements}

%

\bibliographystyle{aa} 
\bibliography{smc} 

\appendix

\section{Extra plots}
\label{app}

   \begin{figure*}[!htb]
   \centering
   \includegraphics[width=0.32\textwidth]{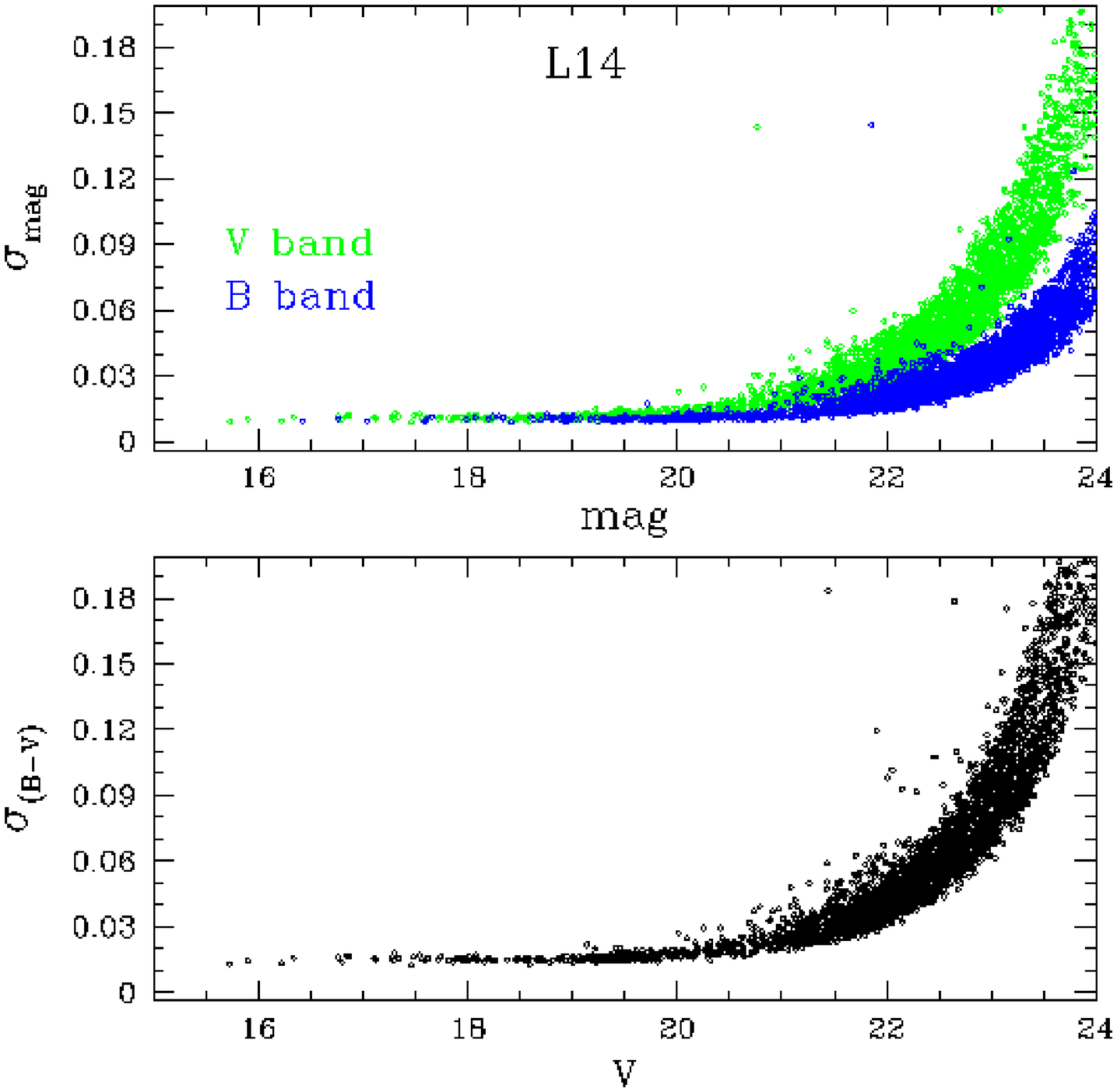}
   \includegraphics[width=0.32\textwidth]{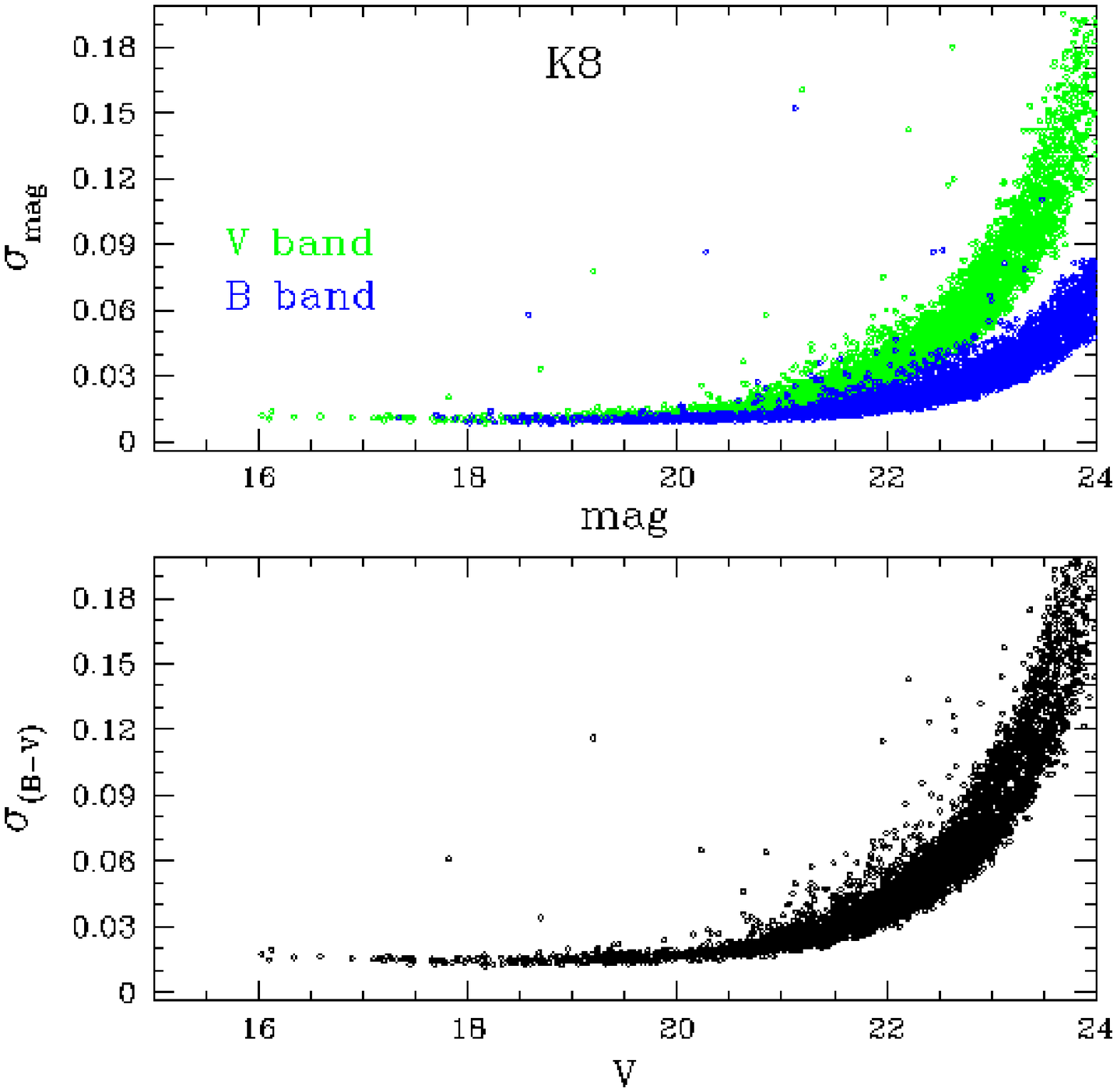}
   \includegraphics[width=0.32\textwidth]{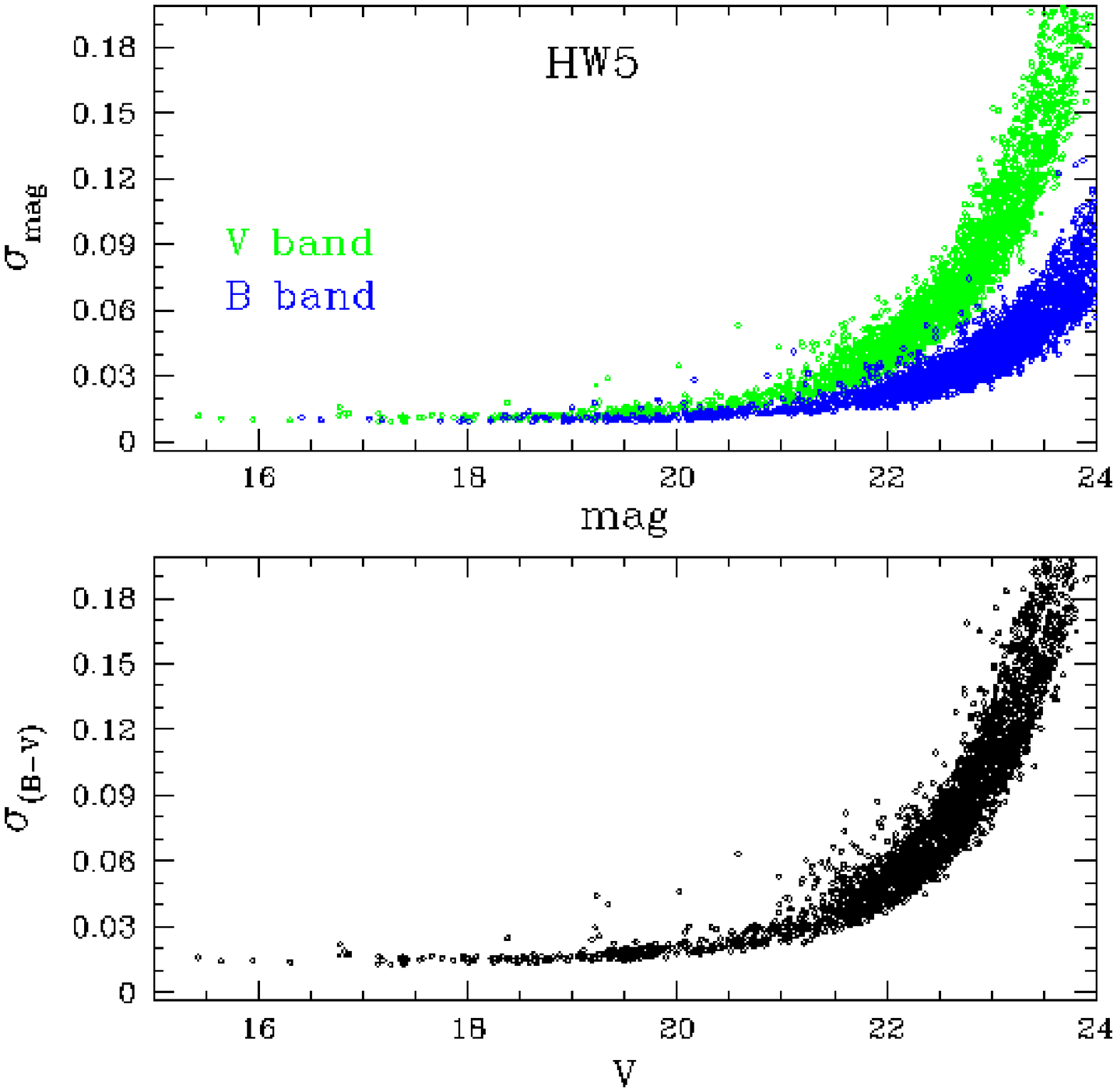}
   \includegraphics[width=0.32\textwidth]{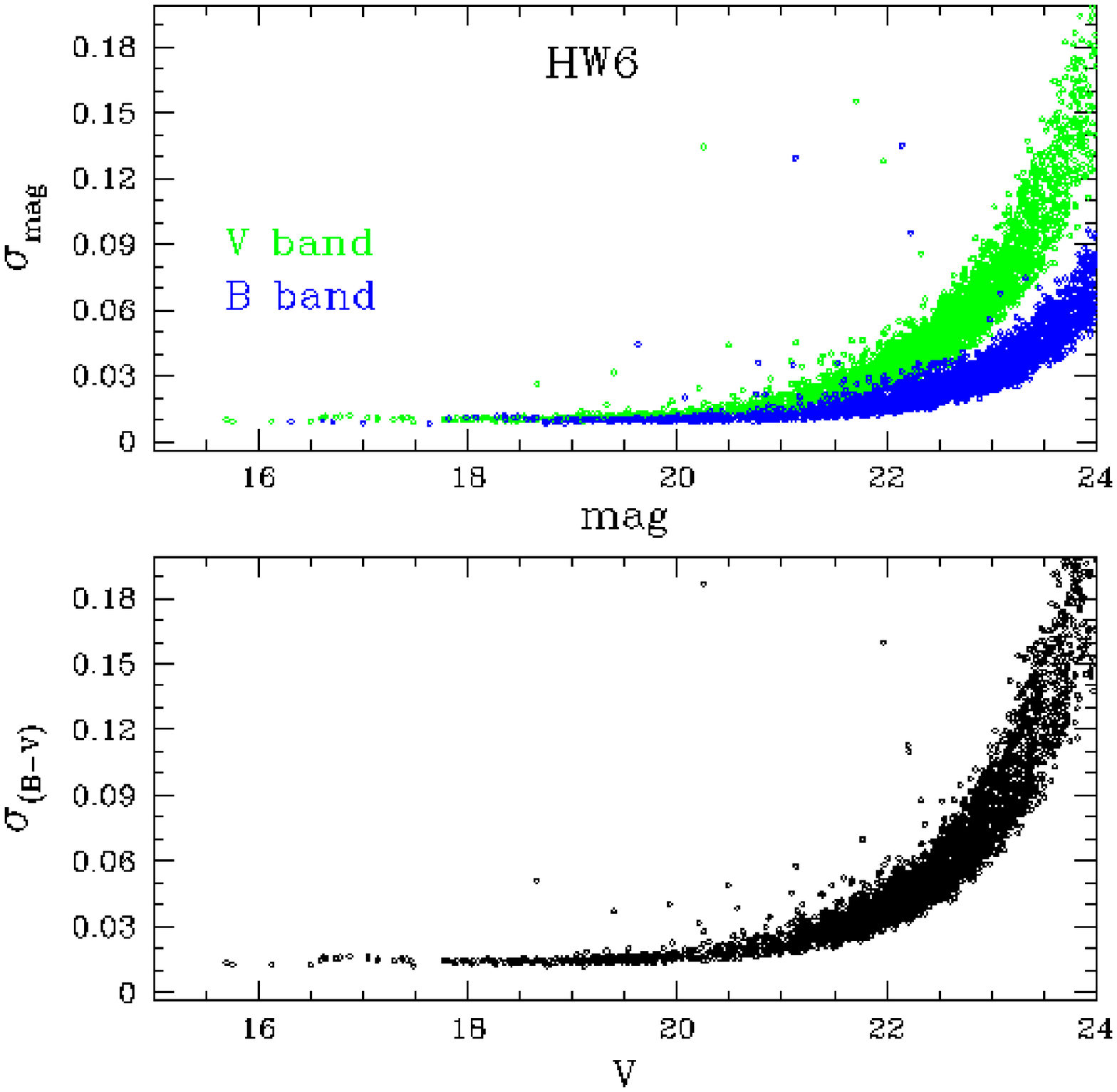}
   \includegraphics[width=0.32\textwidth]{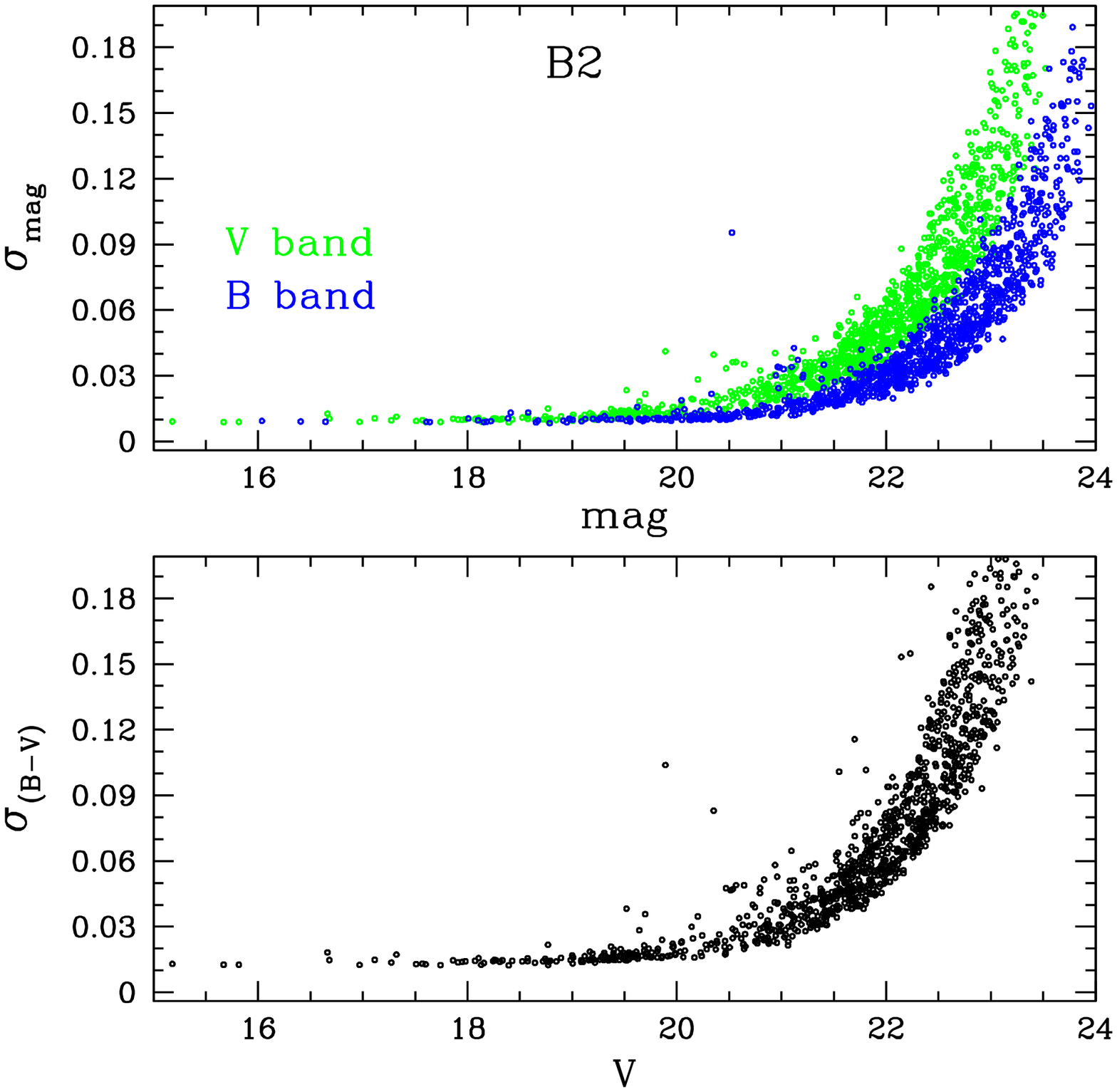}
   \includegraphics[width=0.32\textwidth]{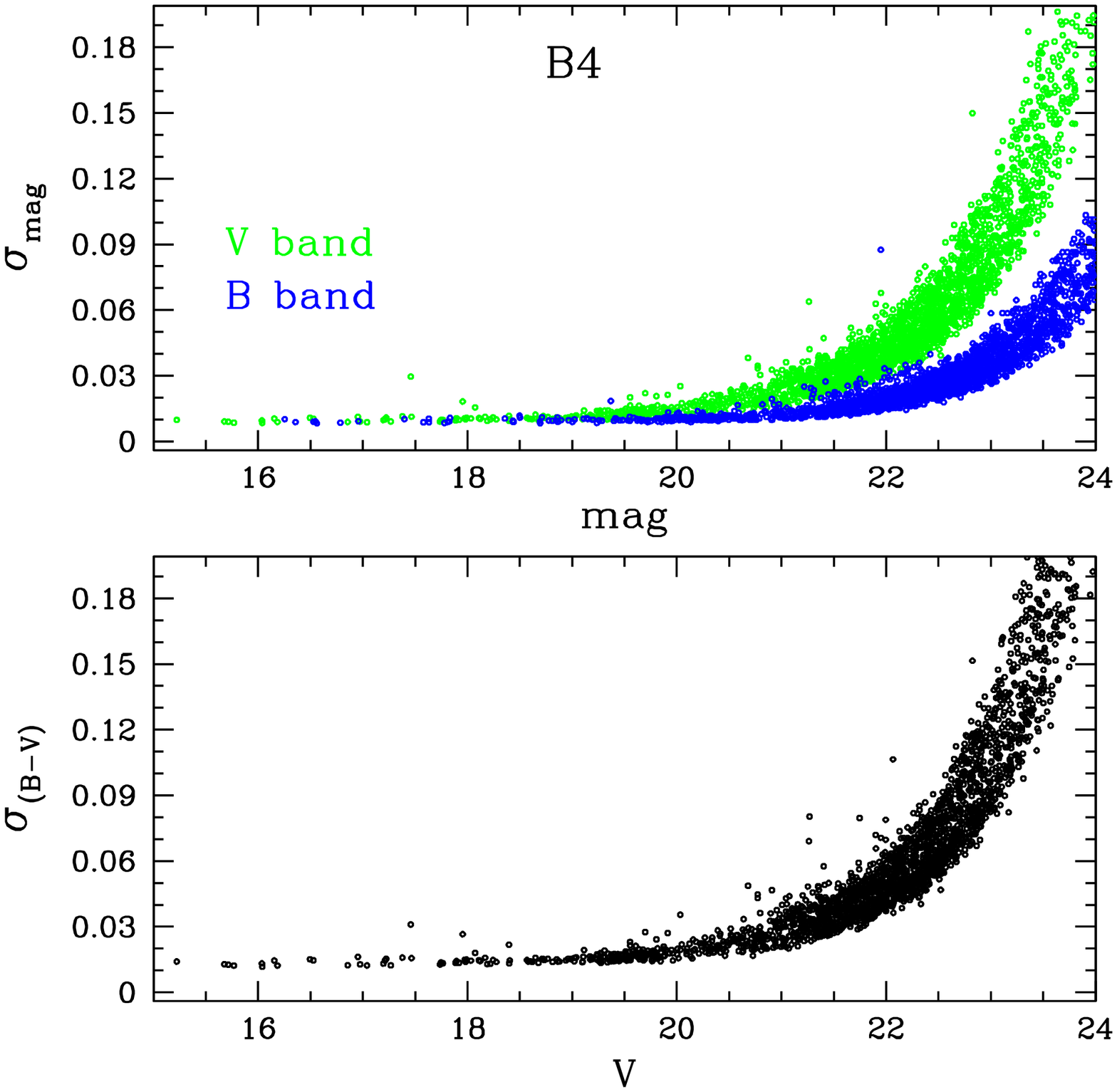}
   \includegraphics[width=0.32\textwidth]{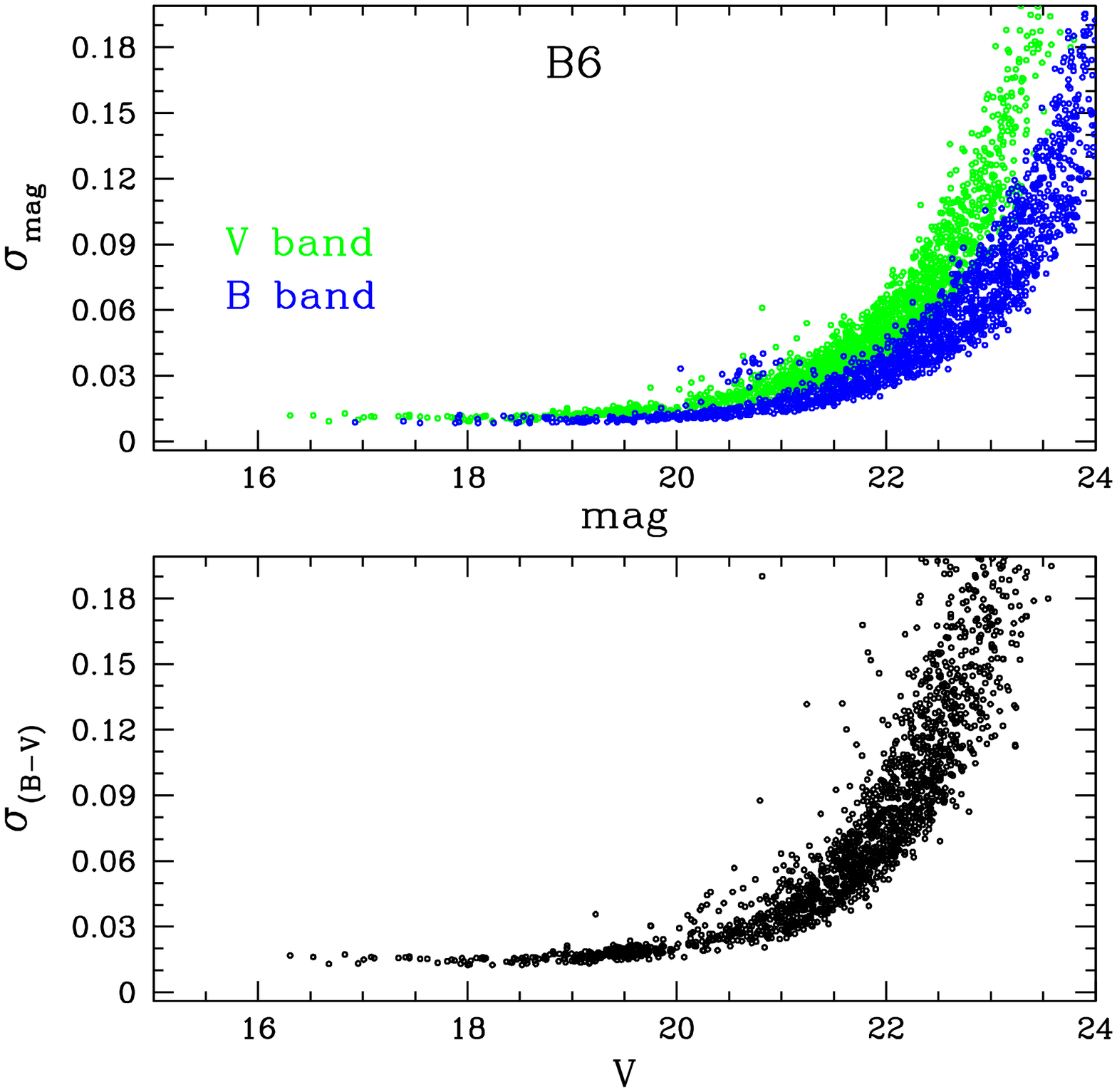}
   \includegraphics[width=0.32\textwidth]{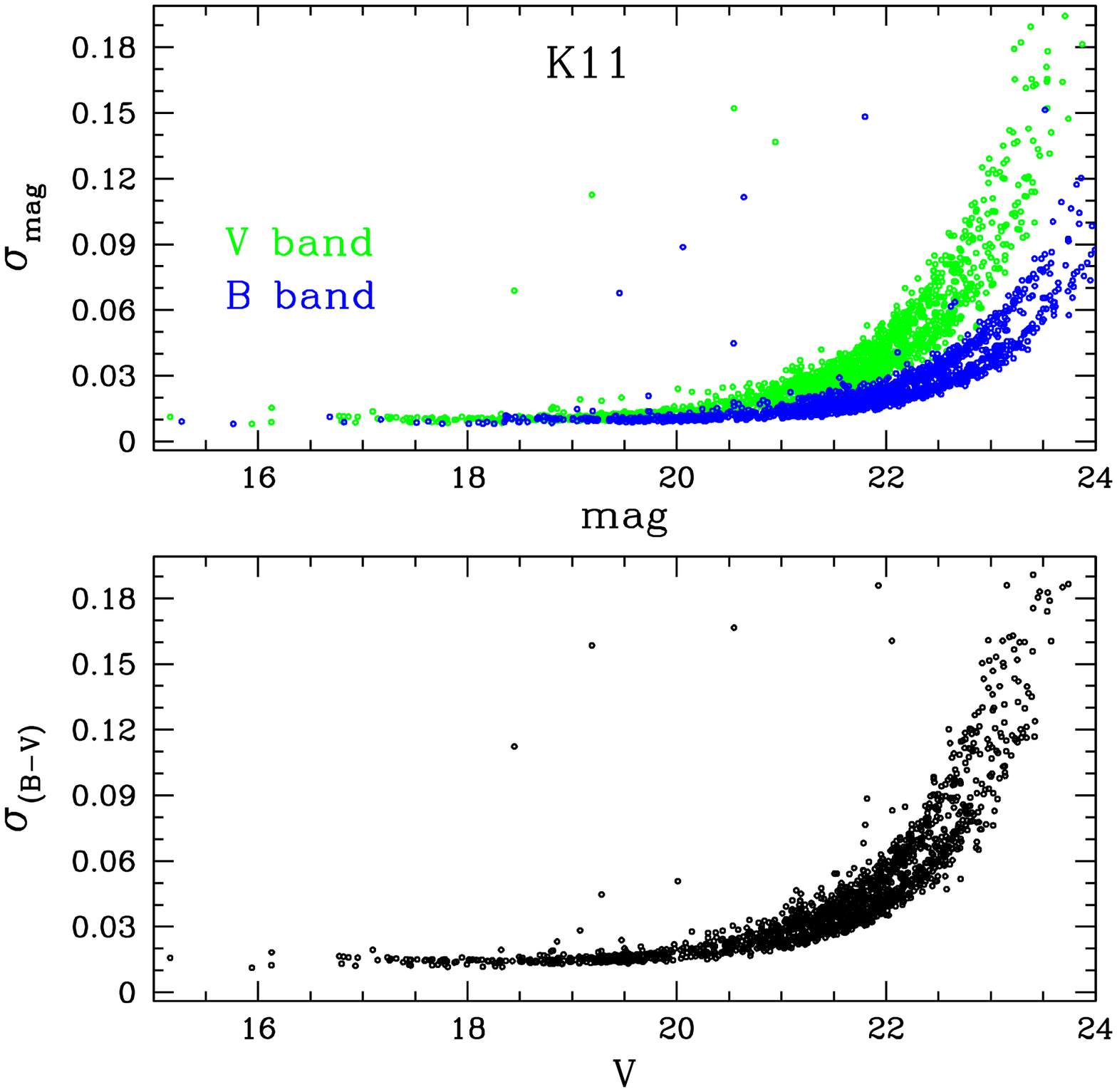}
   \caption{Photometric errors from IRAF for all clusters in bands B
     and V.}
 \label{figapp:photerr}
   \end{figure*}

   \begin{figure*}[!htb]
   \centering
   \includegraphics[width=0.32\textwidth]{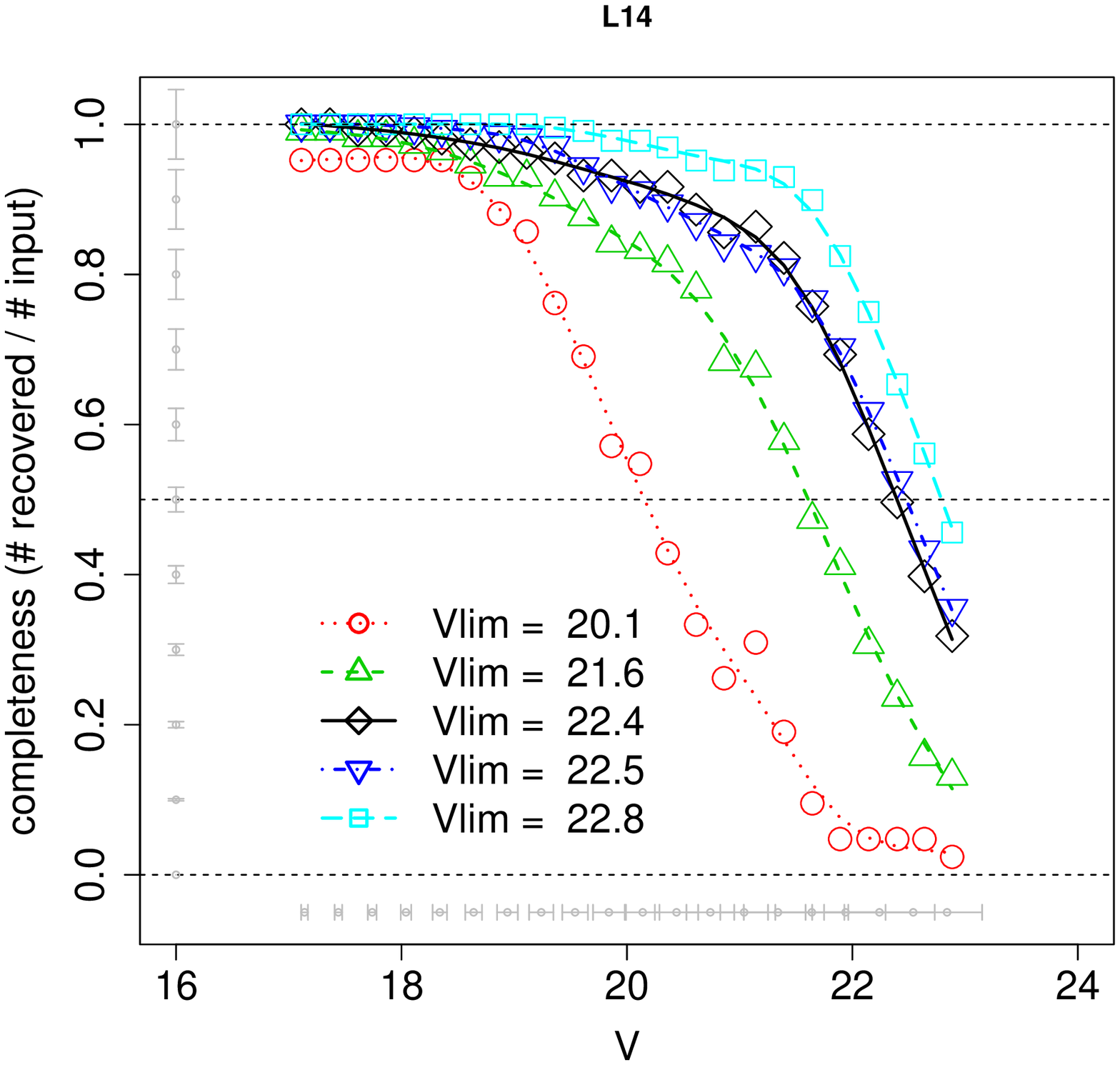}
   \includegraphics[width=0.32\textwidth]{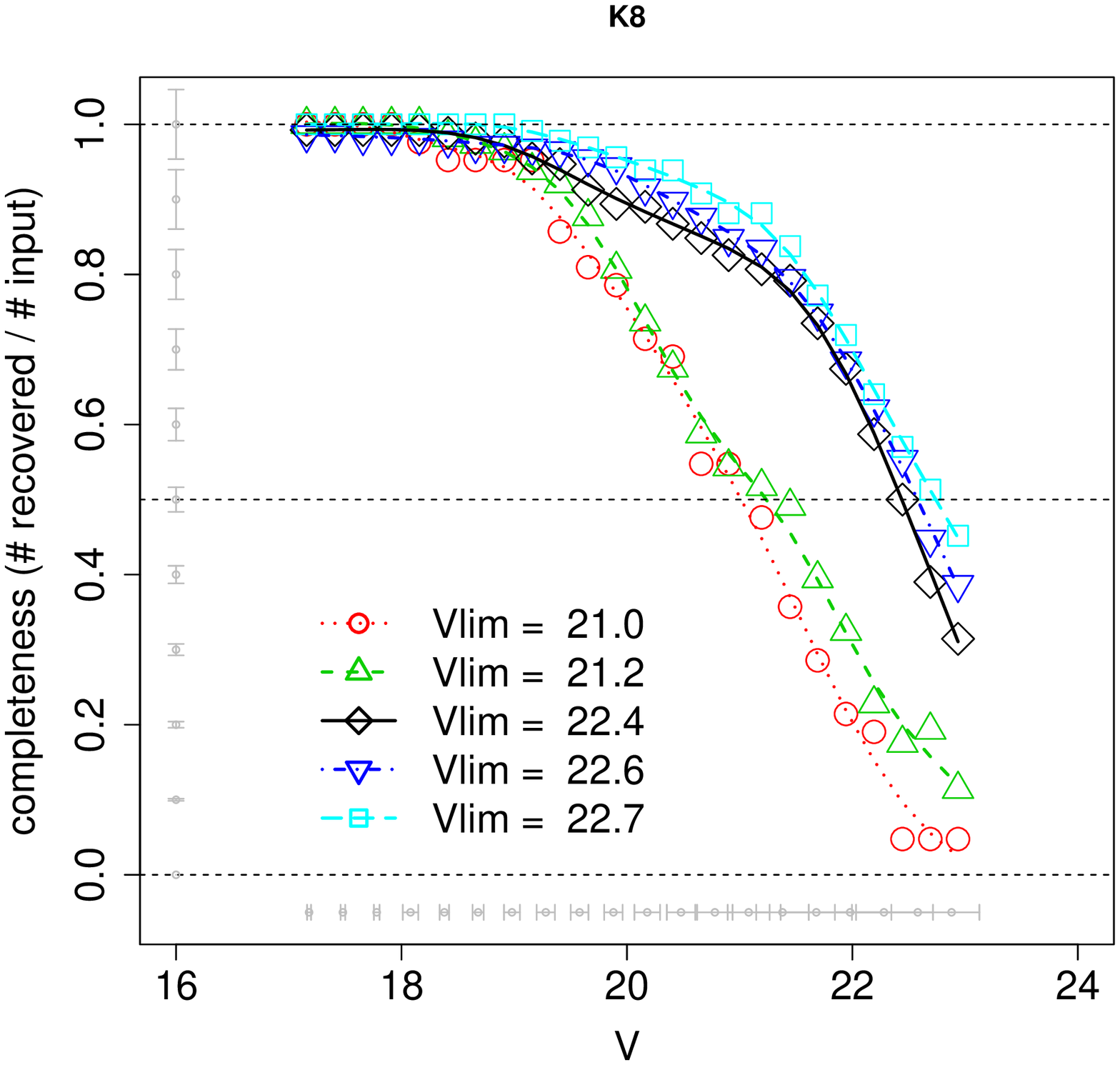}
   \includegraphics[width=0.32\textwidth]{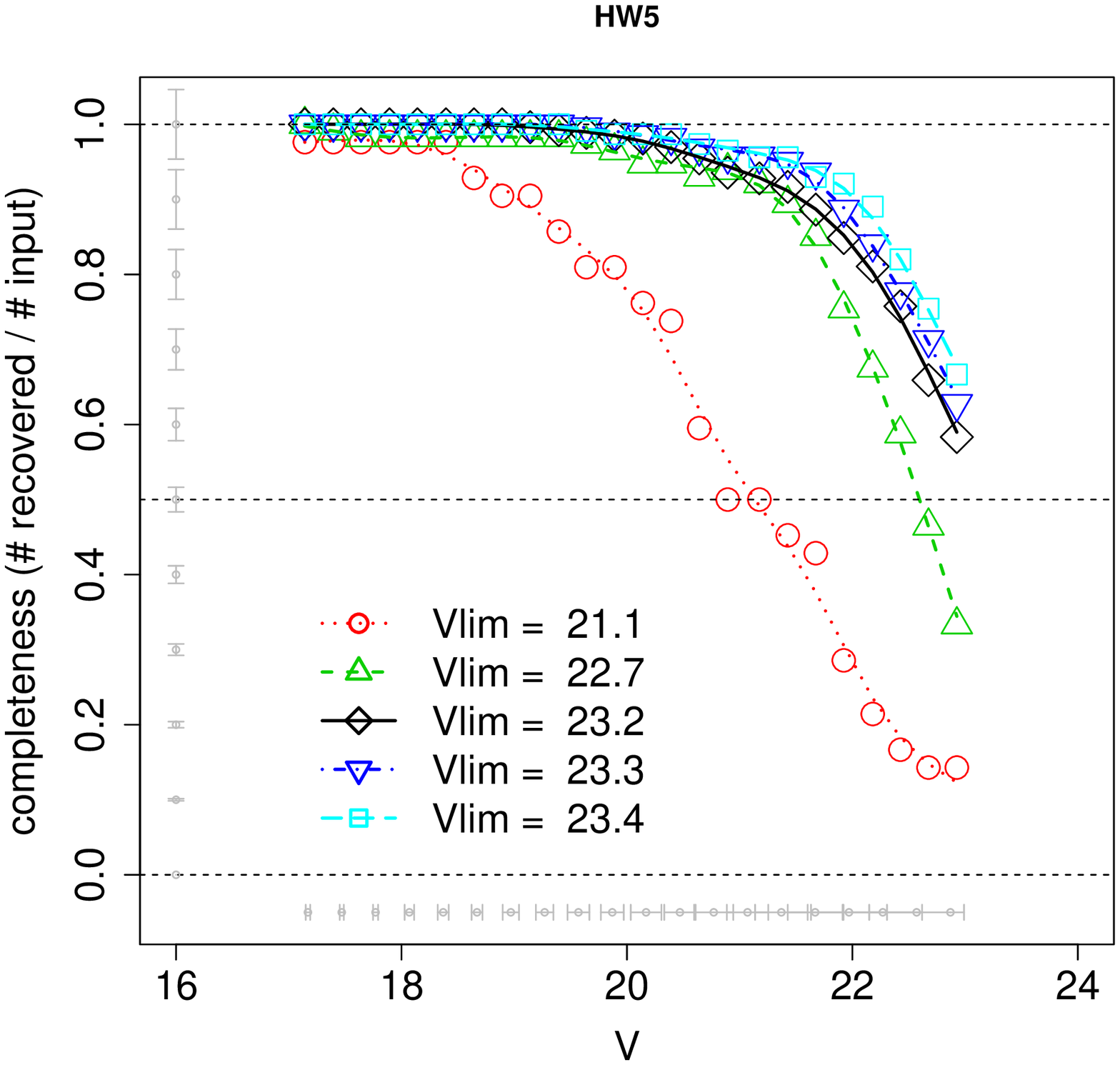}
   \includegraphics[width=0.32\textwidth]{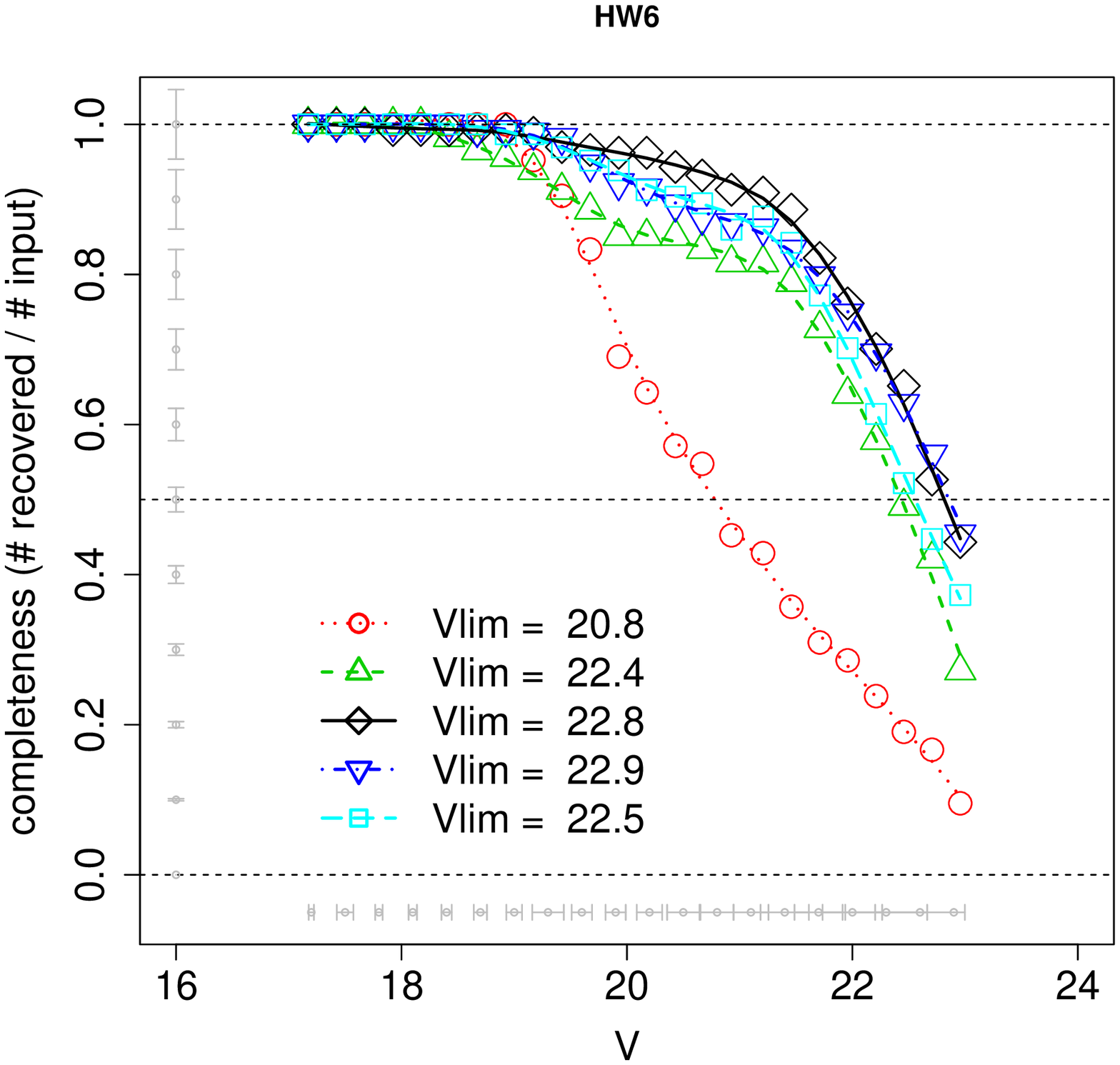}
   \includegraphics[width=0.32\textwidth]{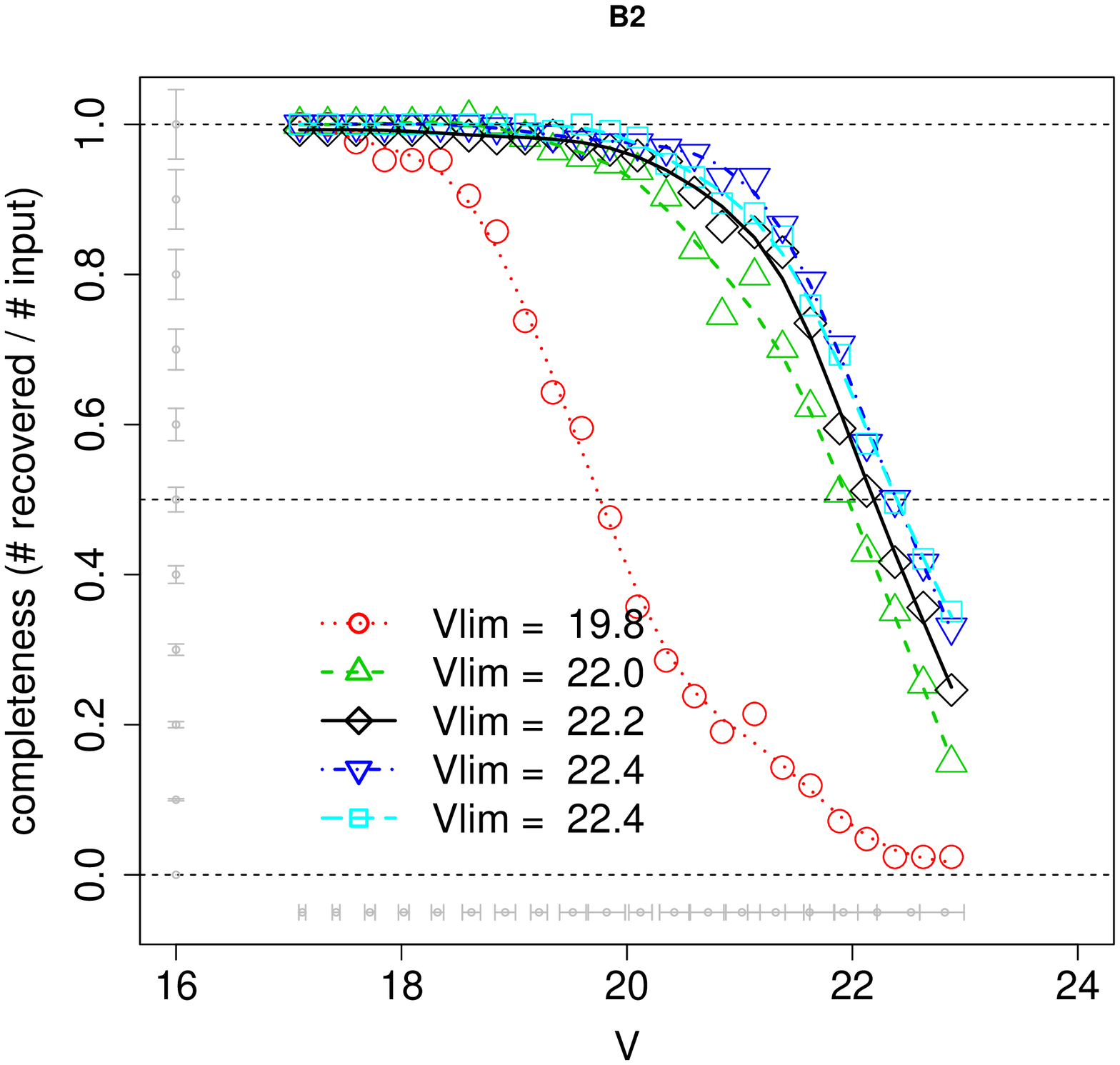}
   \includegraphics[width=0.32\textwidth]{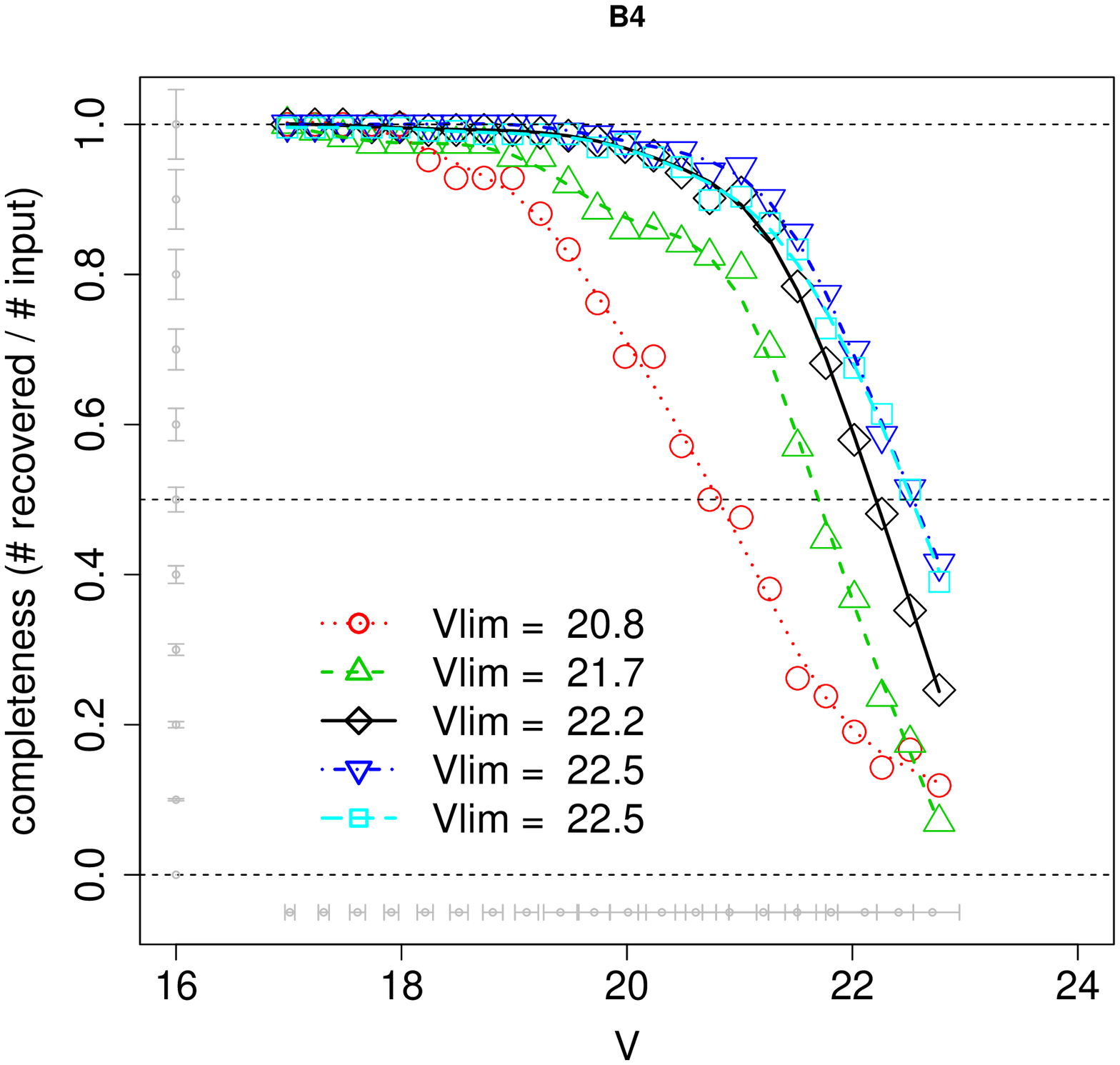}
   \includegraphics[width=0.32\textwidth]{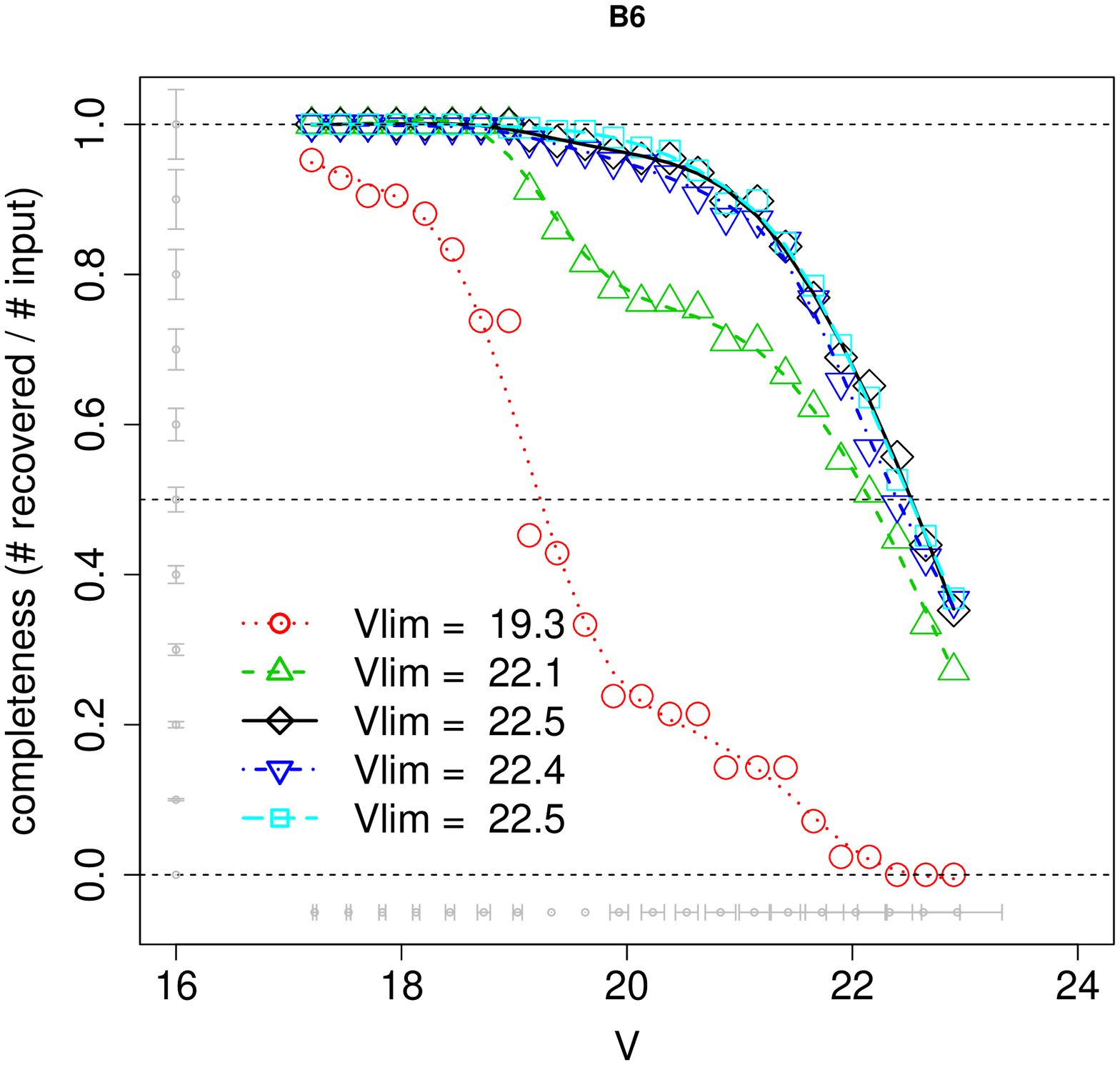}
   \includegraphics[width=0.32\textwidth]{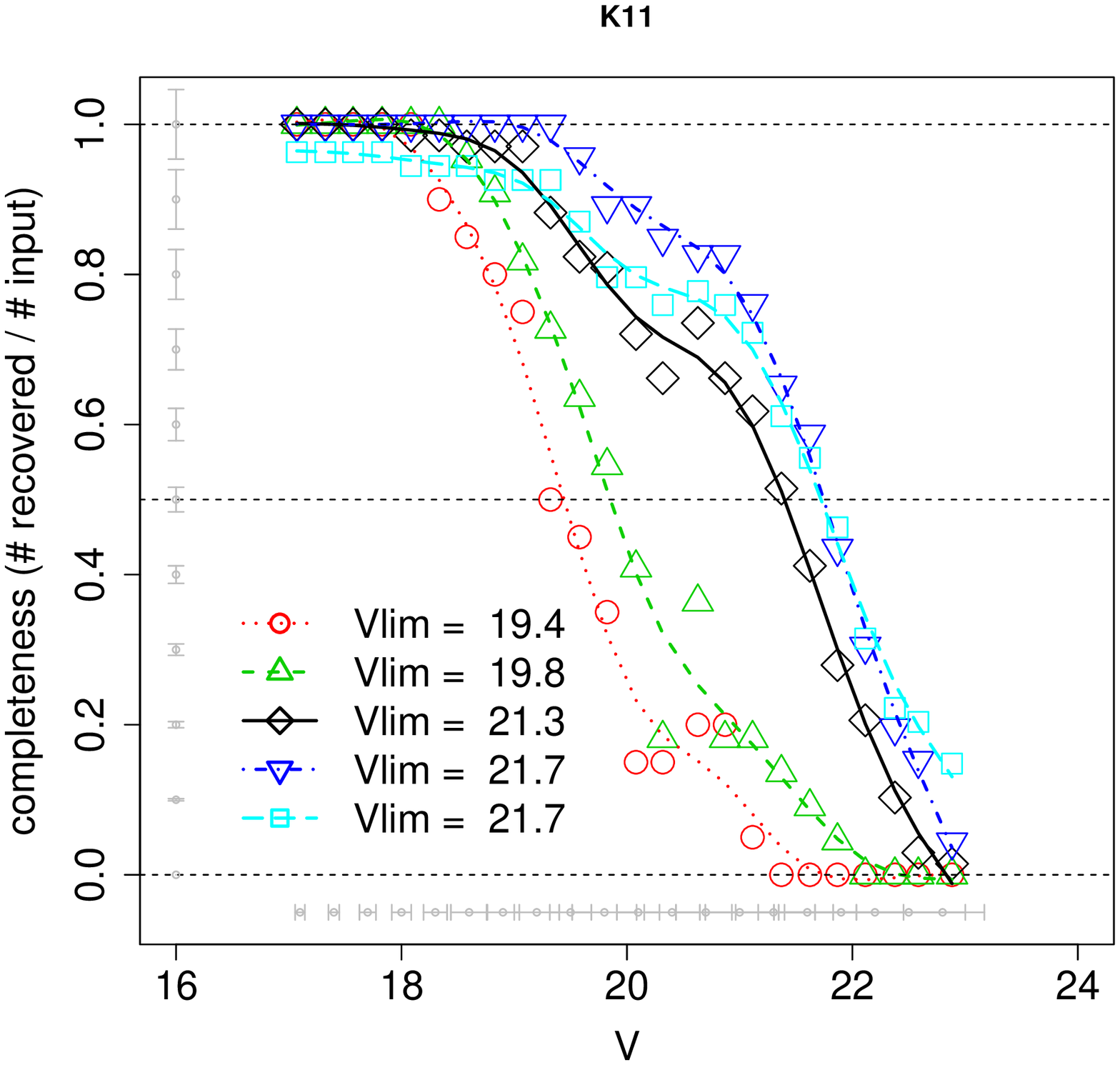}
   \caption{Completeness curves for the nine clusters. Different
     curves for a same cluster represent different annuli around the
     centre of the cluster in steps of 22$\arcsec$ for NGC~152 (from 0$\arcmin$
     to 1.8$\arcmin$) and of 10$\arcsec$ for the other clusters (from 0$\arcmin$
     to 0.8$\arcmin$). The curves in a crescent distance from the
     cluster centre are represented by red circles and a dotted line,
     green triangles and a dashed line, black diamonds and solid lines,
     blue inverted triangles and a dot-dashed line, and cyan squares and
a     long-dashed line. Uncertainty bars from the artificial star
     tests are presented in grey. The horizontal black dashed line at
     a completeness level of 0.5 marks the intersection with each line, whose
     magnitudes are shown in the legend in each panel.}
 \label{figapp:complete}
   \end{figure*}

   \begin{figure*}[!htb]
   \centering
   \includegraphics[width=0.32\textwidth]{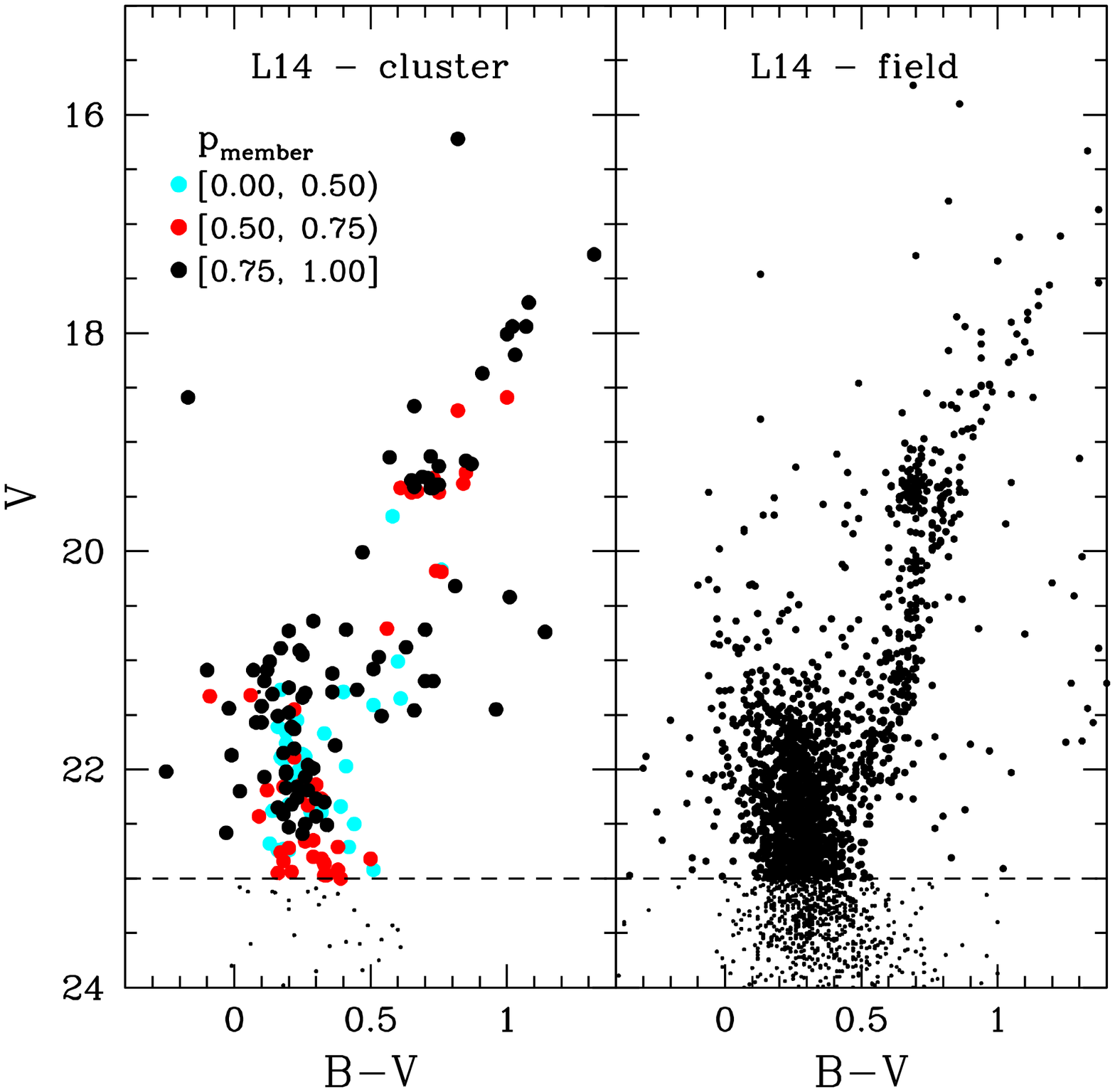}
   \includegraphics[width=0.32\textwidth]{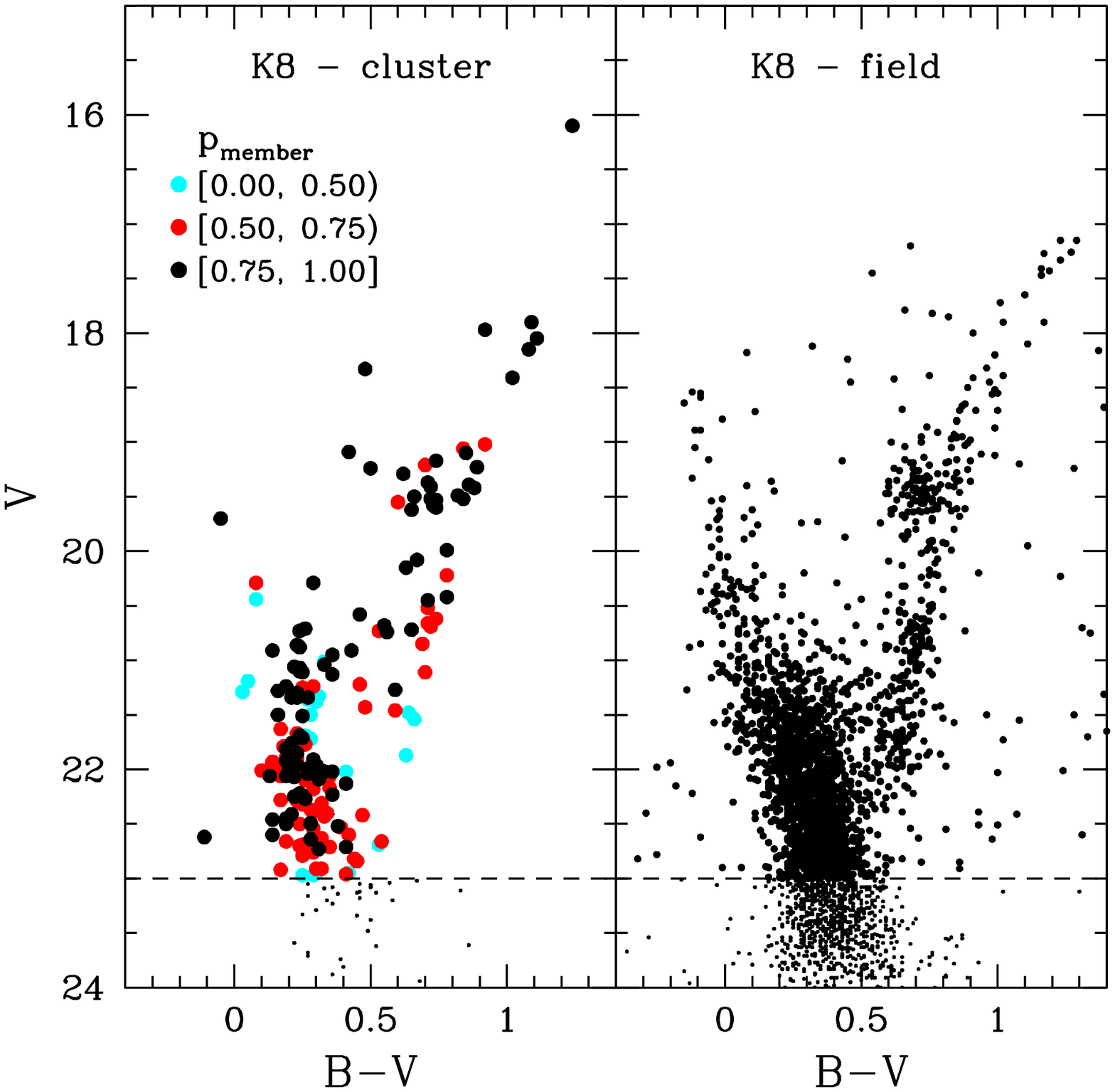}
   \includegraphics[width=0.32\textwidth]{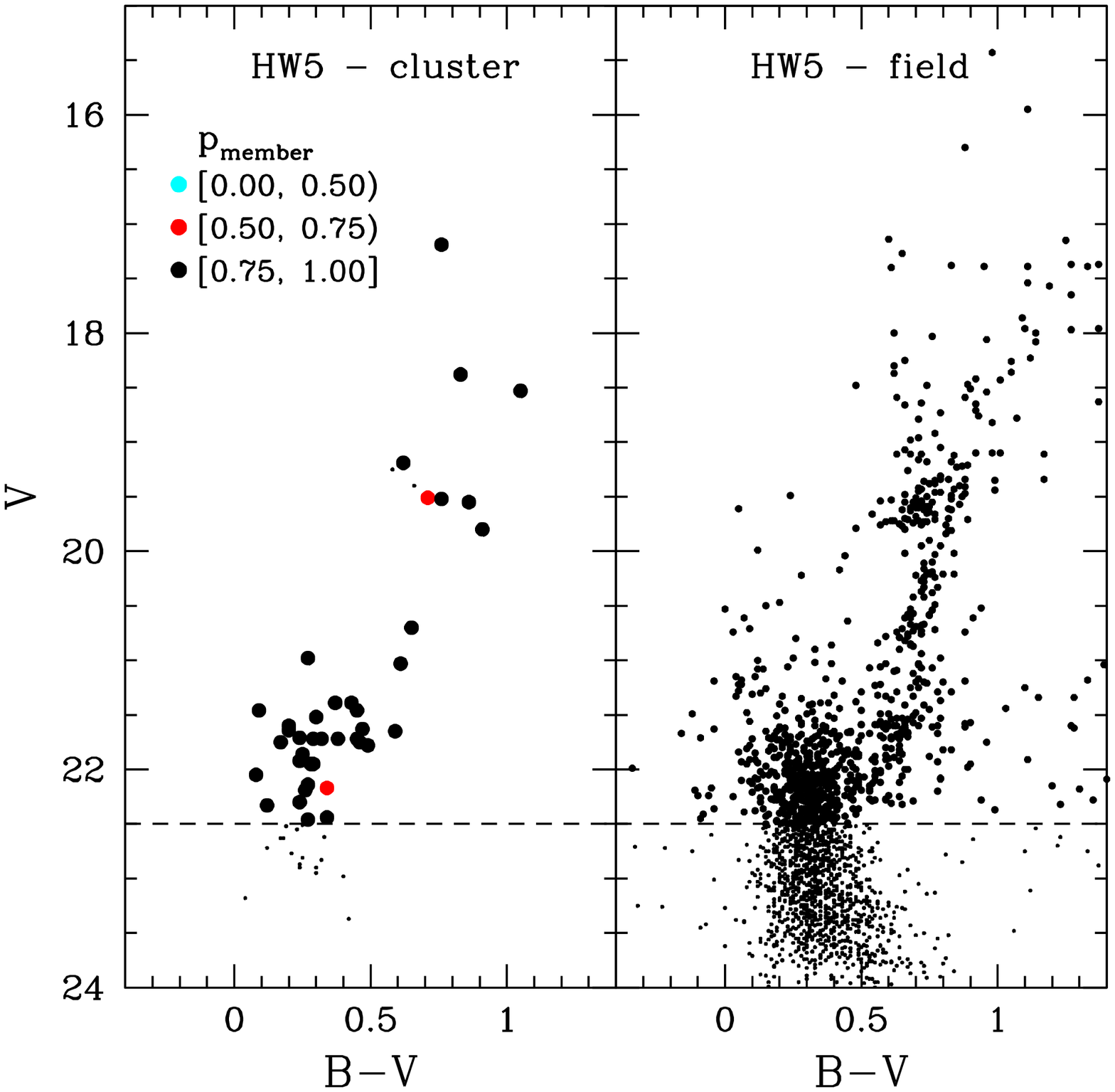}
   \includegraphics[width=0.32\textwidth]{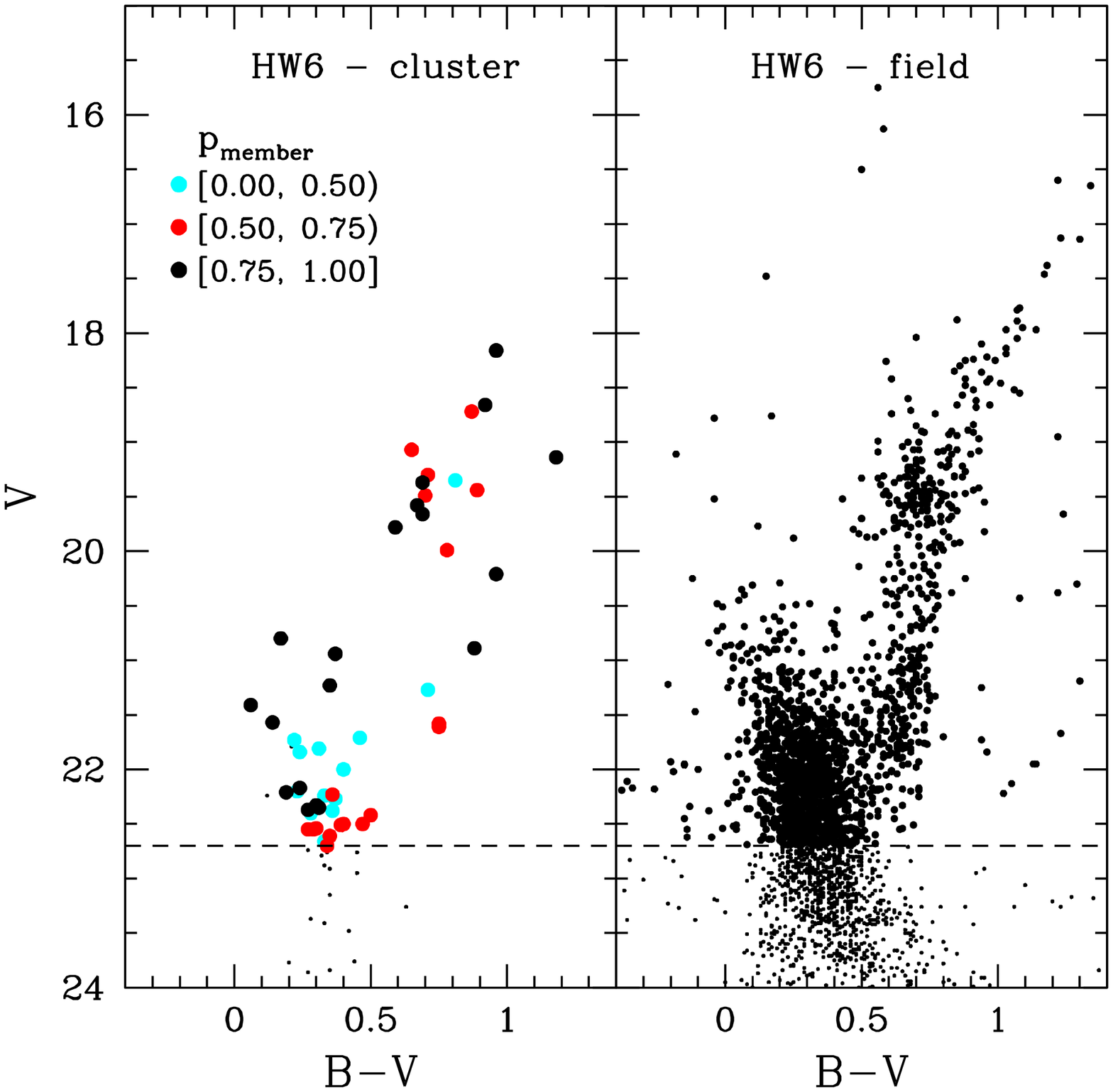}
   \includegraphics[width=0.32\textwidth]{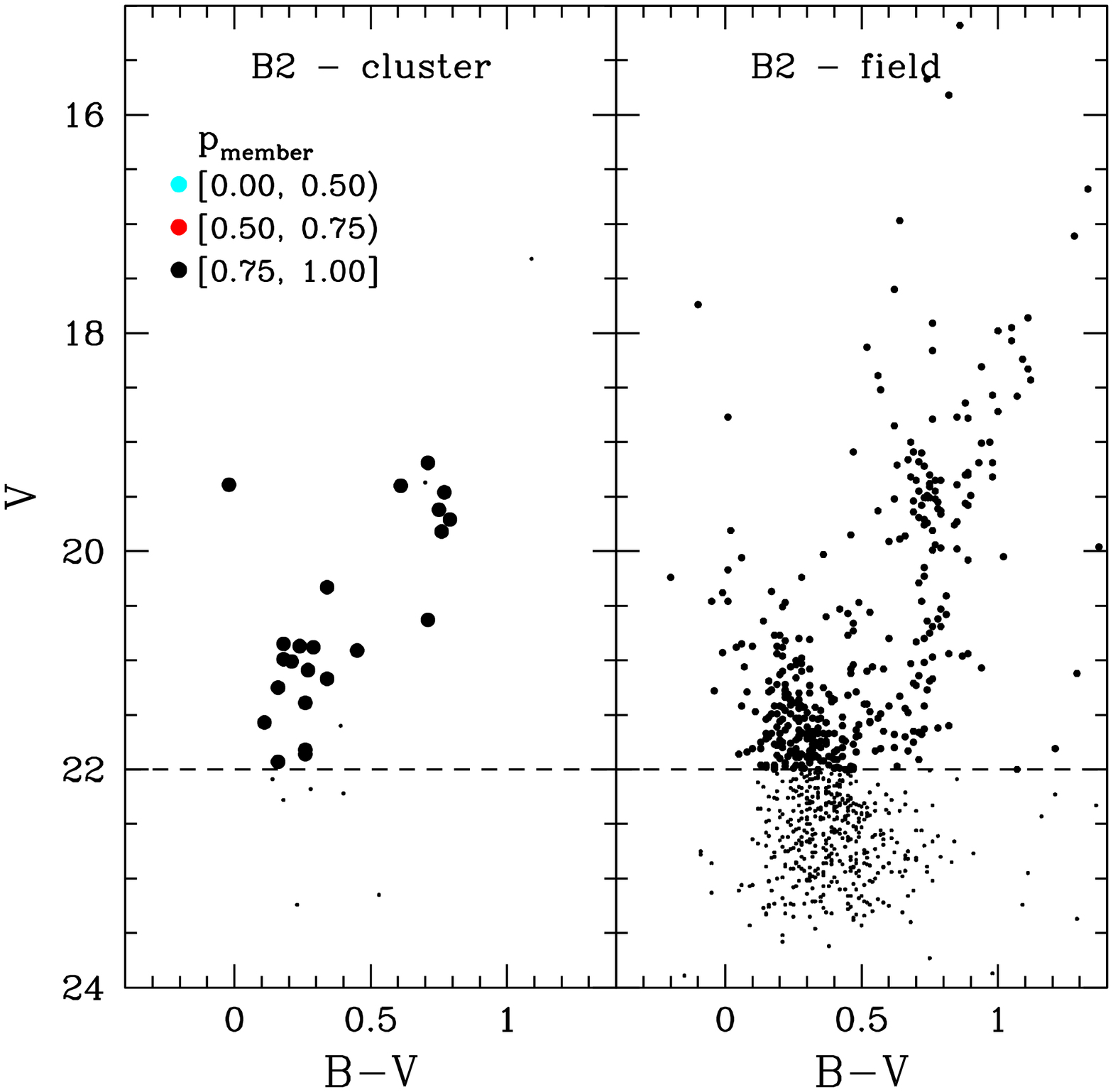}
   \includegraphics[width=0.32\textwidth]{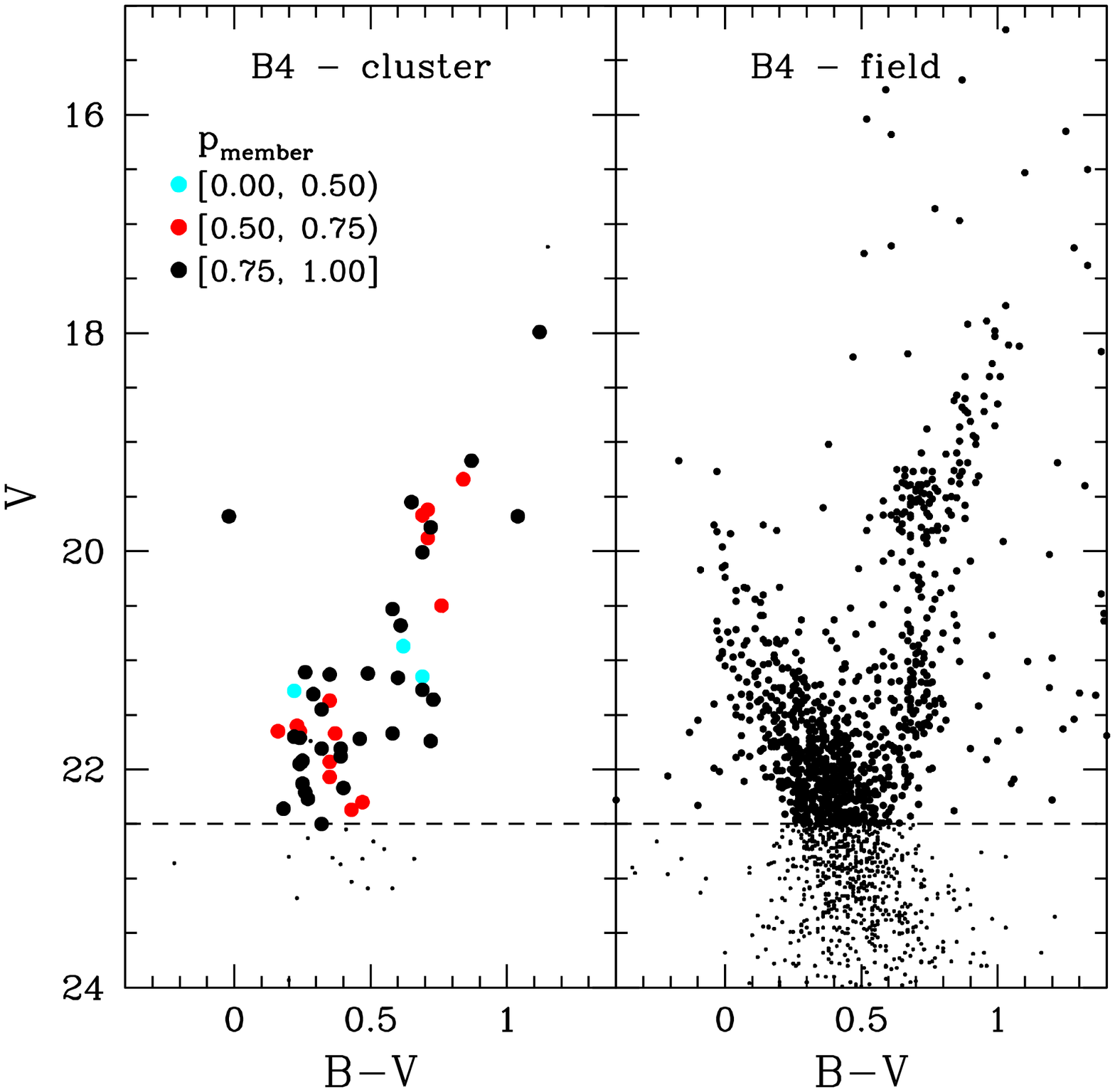}
   \includegraphics[width=0.32\textwidth]{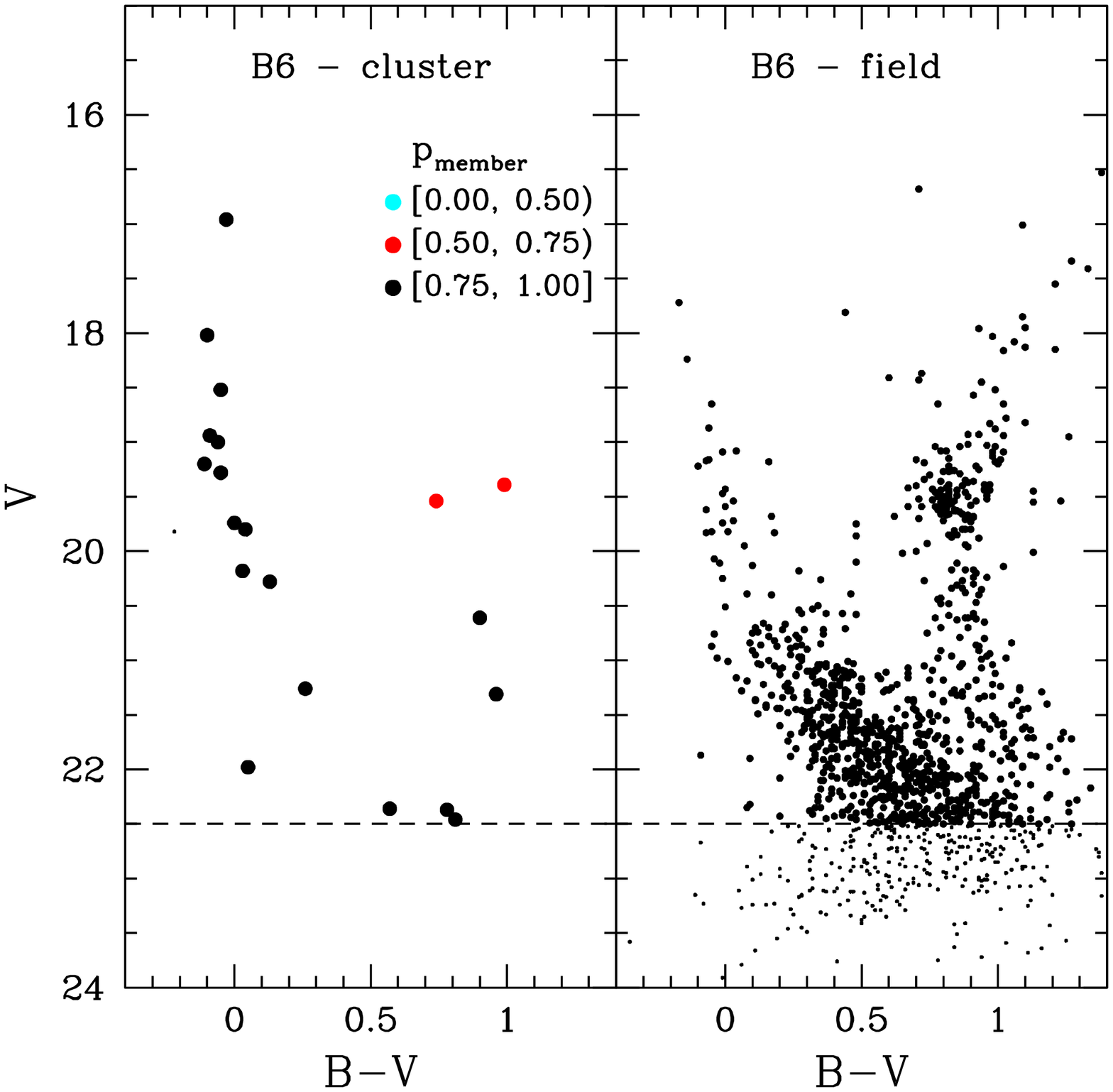}
   \includegraphics[width=0.32\textwidth]{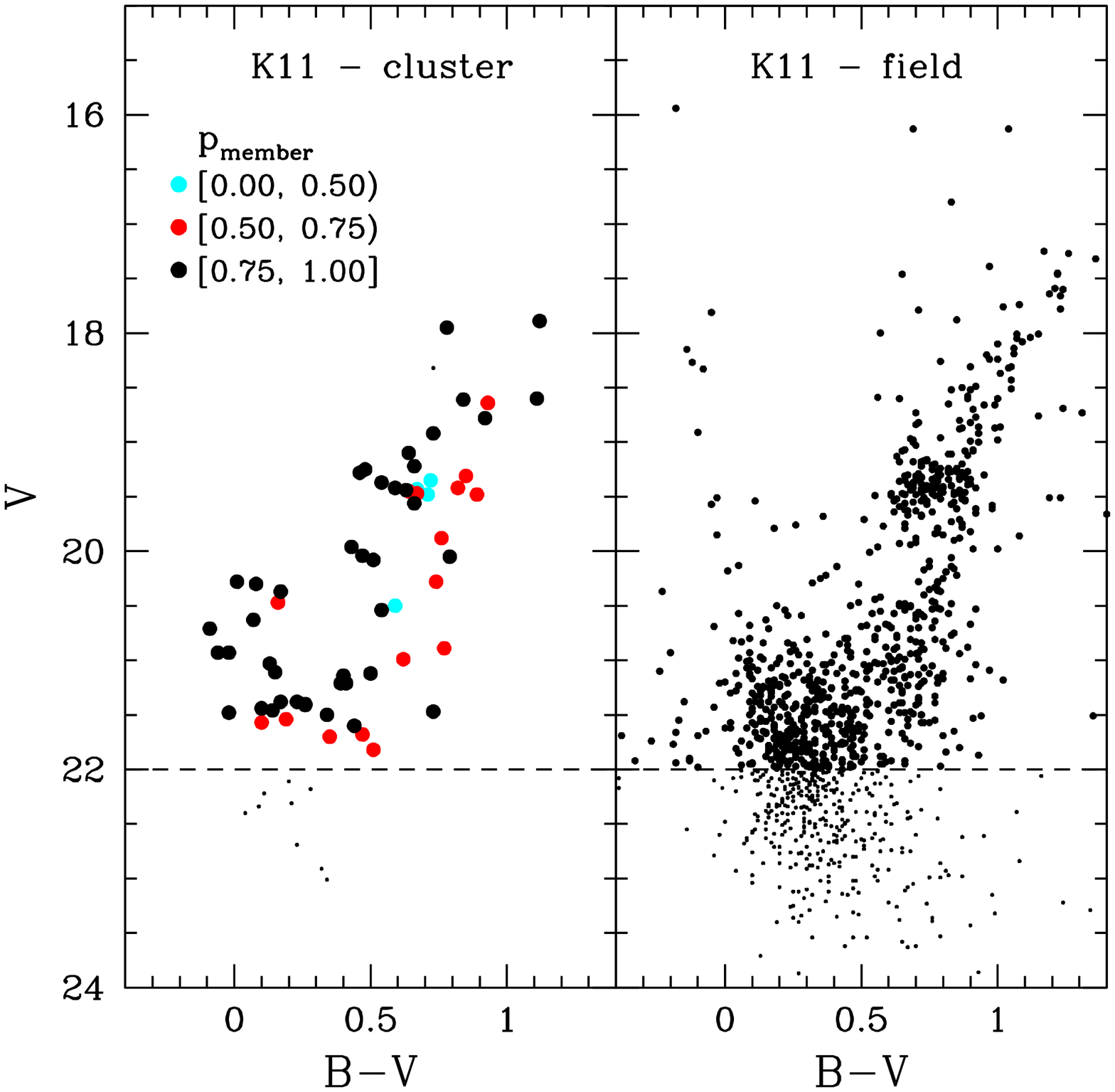}
   \caption{Same as Fig. \ref{fig_membership} for all other clusters.}
 \label{figapp_membership}
   \end{figure*}

\end{document}